\documentclass[pra,showpacs,showkeys,floatfix,amsmath,amssymb,superscriptaddress,reprint]{revtex4-1}
\usepackage{graphicx,exscale,bm,color}

\begin{document}
\title{Multiple re-encounter approach to radical pair reactions and the role of
nonlinear master equations}
\date{\today}
\author{Jens Clausen}
\affiliation{Institute for Theoretical Physics, University of Innsbruck,
             Technikerstr. 25, A-6020 Innsbruck, Austria}
\affiliation{Institute for Quantum Optics and Quantum Information,
             Austrian Academy of Sciences, Technikerstr. 21a,
             A-6020 Innsbruck}
\author{Gian Giacomo Guerreschi}
\affiliation{Institute for Quantum Optics and Quantum Information,
             Austrian Academy of Sciences, Technikerstr. 21a,
             A-6020 Innsbruck}
\affiliation{Department of Chemistry and Chemical Biology,
             Harvard University,
             Cambridge MA 02138 USA}
\author{Markus Tiersch}
\affiliation{Institute for Theoretical Physics, University of Innsbruck,
             Technikerstr. 25, A-6020 Innsbruck, Austria}
\affiliation{Institute for Quantum Optics and Quantum Information,
             Austrian Academy of Sciences, Technikerstr. 21a,
             A-6020 Innsbruck}
\author{Hans J. Briegel}
\affiliation{Institute for Theoretical Physics, University of Innsbruck,
             Technikerstr. 25, A-6020 Innsbruck, Austria}
\affiliation{Institute for Quantum Optics and Quantum Information,
             Austrian Academy of Sciences, Technikerstr. 21a,
             A-6020 Innsbruck}
\begin{abstract}
We formulate a multiple-encounter model of the radical pair mechanism that is
based on a random coupling of the radical pair to a minimal model environment.
These occasional pulse-like couplings correspond to the radical encounters and
give rise to both dephasing and recombination. While this is in agreement with
the original model of Haberkorn and its extensions that assume additional
dephasing, we show how a nonlinear master equation may be constructed to
describe the conditional evolution of the radical pairs prior to the detection
of their recombination. We propose a nonlinear master equation for the evolution
of an ensemble of independently evolving radical pairs whose nonlinearity
depends on the record of the fluorescence signal. We also reformulate
Haberkorn's original argument on the physicality of reaction operators using the
terminology of quantum optics/open quantum systems. Our model allows one to
describe multiple encounters within the exponential model and connects this with
the master equation approach. We include hitherto neglected effects of the
encounters, such as a separate dephasing in the triplet subspace, and predict
potential new effects, such as Grover reflections of radical spins, that may be
observed if the strength and time of the encounters can be experimentally
controlled.
\end{abstract}
\maketitle
\section{
\label{sec1}
Introduction}

It has been shown that certain birds such as the European Robin use the
geomagnetic field for orientation \cite{Wil72,*Wil96,*Mou01,*Joh08}. The
presently leading hypothesis for this magnetic sense is the radical pair
mechanism (RPM) \cite{Sch78,*Sch82,Rit00,*Rod09}, which is a model of a
light-induced chemical reaction whose products depend on an external magnetic
field. Two paired electrons undergo photo-induced separation and evolve due to
their coupling with surrounding nuclear spins until they recombine, giving rise
to a biological signal which depends on their spin state and hence on the
geomagnetic field. The field sensitivity is thus based on unpaired electron
spins that constitute the radical pair (RP).
Such reactions have been studied extensively in the context of spin chemistry
\cite{Ste89,*Sal84,*Nag98}, and avian magnetoreception
\cite{Wil02,*Rit04,*Mou04,*Mae08,*Hen08,*Rit09,*Sol10,*Hei11}.

This work is in part motivated by the recent controversy on the proper reaction
operators \cite{Jon10,Jon11a,Kom11c,Jon11,Kom11m,Del11}. Our goal is to
encompass and enlarge the variety of possible descriptions but, at the same
time, to clearly distinguish between parameters which depend on the specific
setup and that ultimately must be measured, and the requirements of a formal
description with a consistent interpretation.
~
Despite its potential role in bird navigation, we here restrict ourselves to the
RPM itself, leaving open its role in quantum biology. The reason is that it is
often easier to design experiments in order to test a given theory in vitro than
in vivo.
An example where the influence of the strength of an external magnetic field on
the recombination fluorescence has been investigated experimentally in solution
is Pyrene-Dimethylaniline (Py-DMA) as acceptor-donor pair \cite{Rod07,*Ste89}.
In contrast, the \emph{directional} sensitivity of the chemical compass requires
some anisotropy of either the initial state, or the molecule or its
surroundings, which must be aligned in the bird
\cite{Hog12,Sch78,*Sch82,Rit00,*Rod09}. 
Which molecules are involved remains unknown to date, although the photopigment
cryptochrome has been found in the retina of the bird's eye.
It is obvious that the details of these reactions depend on the respective
setup, such as the molecules involved, the way the radicals are formed, their
environment, and the whereabouts of the reaction products, which suggests a
large variety of possible scenarios. It is therefore reasonable to consider a
simplified model that captures their commonalities and allows one to explain the
relevant experiments. In particular, we skip intermediate reaction steps
associated with short-lived transient states.

We assume that the RPM takes place in a chemical system which has a stable
ground state with two paired electrons. This system can
consist of a donor-acceptor-pair of two separate molecules or a single molecule
that can undergo a photo-induced dissociation or conformational change.
~
We distinguish between the external kinetics of the chemical system which is
assumed to be describable as a diffusion and its internal dynamics. While the
internal dynamics involves quantum processes which depend on the external
kinetics, the diffusion itself is here regarded as a classical process that is
independent on what is going on internally in the chemical system. Our model
assumes that the system's diffusion, during which the internal state evolves
unitarily, is occasionally interrupted by encounters (collisions) of the
radicals, which have on the internal state the effect of a generalized
measurement. Such measurements may be completely unread, or they may be more or
less perfectly read out, typically, by means of a fluorescence detector.


In this work, we derive a general reaction operator describing the recombination
of the RP by an interaction of the chemical system with a model environment.
This allows for varying degrees of relaxation and dephasing and contains the
models of Haberkorn \cite{Hab76} as well as Jones and Hore \cite{Jon10} as
special cases corresponding to either pure or balanced relaxation and dephasing,
respectively. We also compare this with a different approach put forward by
Kominis \cite{Kom09,Kom11} who suggests the use of a nonlinear master equation.
We can rule out the latter approach based on general considerations; at the same
time, we explain how a (different) nonlinear equation can be applied to describe
a RP-evolution conditioned on a record of measurements.
Note that there exist alternative models based on an electric dipole field
caused by accumulation of molecules in a long-lived triplet signaling state
\cite{Sto12} or models based on quantum criticality \cite{Sun11}. In our
approach, we do not distinguish between different chemical reaction products,
i.e., we assume one single ground state. This distinction is made in the
different orthogonal final states of our model environment.

We further develop a Markovian diffusion model for the mechanical motion of the
RP, presupposing encounters of the radicals as a necessary condition for their
recombination. The encounters are described as a transient interaction with the
model environment and form part of a quantum measurement process. We consider a
read-out of these measurements with given efficiency, thus allowing for
conditional evolution and feedback. The master equation approach to the reaction
operator can be obtained from this encounter model as a limiting case. In this
way, we show how different re-encounter models \cite{Ped77,Bur04,Gor02} affect
the master equation. This is relevant, since a large variety of scenarios is
conceivable for the molecular motion, such as a one- two-, or three-dimensional
free diffusion, caged diffusion, where the radicals are confined within a
micelle or connected by a flexible chain. Alternatively, they could be located
on different parts of a stiff molecule which might be excited by vibrational
degrees of freedom and in addition undergo conformational changes.

An analysis of the RPM will provide a better understanding of the processes
involved in magnetoreception \cite{cai10,Cai11,Cai11a}, which in turn may help
to develop future applications inspired by natural processes. For example, it
has been demonstrated experimentally that a RP-reaction can be used to map
nanomagnetic fields \cite{Lee11a}.
An artificial chemical compass may potentially allow for a nanoscale mapping of
the direction of the magnetic fields, solely by distributing suitable chemicals
onto or into a substrate or organism, followed by an optical excitation and
analysis of these chemicals. There would hence be no need to scan the field
directly by an external magnetic sensor.


This work is organized as follows.
After this introduction, the use of nonlinear master equations
is motivated in Sec.~\ref{sec2} in a general context.
Sec.~\ref{sec3} outlines the RPM focusing on the evolution of separated RPs and
their creation and recombination.
Sec.~\ref{sec10} introduces the model environment and applies it to derive a
general description of reactions caused by the RP-encounters. Exponential and
master-equation models are described as limiting cases.
The remainder of the paper limits to a simplified case that ignores dephasing
in the triplet subspace.
Sec.~\ref{sec4} reconsiders the master equations in order to put them in
relation to existing literature, and Sec.~\ref{sec5} discusses the statistics
and effect of multiple encounters.
After that, Sec.~\ref{sec6} investigates the RP-evolution prior to fluorescence
detection described by a nonlinear master equation.
Sec.~\ref{sec7} generalizes to the scenario of a RP-ensemble rather than a
single chemical system and derives a nonlinear master equation for the evolution
of an ensemble member given a fluorescence record.
Sec.~\ref{sec8} discusses types of encounters allowed by our model and derives
the singlet yield of the exponential model.
Finally, a summary and outlook on future work is provided in Sec.~\ref{sec9}.
Auxiliary information on master equations, jump operators, and non-Hermitian
coherent evolution is given in an appendix.

\section{
\label{sec2}
Nonlinear master equations}
Recently, nonlinear evolution equations have been suggested for the description
of the RPM \cite{Kom09,Kom11}, which initiated a controversy on the correct
reaction operator. Nonlinear master equations have in fact been discussed for
a long time to allow for the description of conditional state evolutions. In
what follows, we provide an introduction into their emergence and meaning, which
is not restricted to the context of the RPM.
\subsection{
\label{sec2.1}
Equivalence of trace-reducing and nonlinear master equations}
We consider a linear non-trace (but positivity) preserving evolution
\begin{eqnarray}
\label{lnte}
  \frac{\partial}{\partial{t}}\hat{\varrho}_{\mathrm{N}}
  &=&\mathcal{L}(t)\hat{\varrho}_{\mathrm{N}},
  \\
\label{lntes}
  \hat{\varrho}_{\mathrm{N}}(t)
  &=&\mathcal{T}_+\mathrm{e}^{\int_0^t\mathrm{d}\tau\mathcal{L}(\tau)}
  \hat{\varrho}(0),
\end{eqnarray}
where the index N denotes an improper density matrix, that must be normalized
according to
\begin{equation}
\label{rho}
  \hat{\varrho}
  =\frac{\hat{\varrho}_{\mathrm{N}}}{\mathrm{Tr}\hat{\varrho}_{\mathrm{N}}}.
\end{equation}
The formal solution of (\ref{lnte}) has been expressed in
(\ref{lntes}) using positive time ordering $\mathcal{T}_+$. Here we assume that 
$\mathrm{Tr}[\hat{\varrho}_{\mathrm{N}}(0)]$ $\!=$ $\!1$, so that we can write
$\hat{\varrho}(0)$ for short. Inserting (\ref{lnte}) into the time derivative of
(\ref{rho}) gives
\begin{equation}
\label{nlme}
  \frac{\partial}{\partial{t}}\hat{\varrho}
  =\left(\mathcal{L}-\langle\mathcal{L}\rangle\right)\hat{\varrho},
  \quad\langle\mathcal{L}\rangle=\mathrm{Tr}(\mathcal{L}\hat{\varrho}),
\end{equation}
whose solution is given by (\ref{rho}) together with (\ref{lntes}). In contrast
to (\ref{lnte}), (\ref{nlme}) contains a nonlinearity caused by the term
$\langle\mathcal{L}\rangle$. Note that this notation is consistent with the
fact that a state for which the time derivative in (\ref{nlme}) vanishes
constitutes an eigenstate of $\mathcal{L}$,  
$\mathcal{L}\hat{\varrho}$ $\!=$ $\!\langle\mathcal{L}\rangle\hat{\varrho}$.
We may also write
\begin{eqnarray}
\label{ppdot}
  p(t)&=&\mathrm{Tr}\left[\hat{\varrho}_{\mathrm{N}}(t)\right],
  \quad
  \dot{p}=\mathrm{Tr}(\mathcal{L}\hat{\varrho}_{\mathrm{N}}),
  \\
\label{lnpdot}
  \langle\mathcal{L}\rangle&=&\frac{\dot{p}}{p}
  =\frac{\mathrm{d}}{\mathrm{d}t}\ln{p}(t).
\end{eqnarray}
In summary, (\ref{lnte}) and (\ref{nlme}) can be transformed into each other
by the normalization (\ref{rho}). A nonlinear trace-preserving master equation
(\ref{nlme}) allows one to describe the state evolution conditioned on a given
history. The nonlinearity results from describing a continuous transformation
associated with a non-unit probability as a trace-preserving evolution. In this
case, $p(t)$ in (\ref{ppdot}) is the probability in the state transformation
(\ref{rho}), and $\langle\mathcal{L}\rangle$ represents a normalized probability
rate. Independent of this, a linear non-trace preserving master equation
(\ref{lnte}) is obtained if $\hat{\varrho}_{\mathrm{N}}$ is a projection of the
normalized system state $\hat{\varrho}$ onto a given subspace of interest, and
$p(t)$ is then the probability to find the state in this subspace. Such a case
occurs in the description of leakage phenomena \cite{WKB09}.

Eq.~(\ref{nlme}) may also contain an incomplete trace-preservation, e.g., the
term $\langle\mathcal{L}\rangle$ may be replaced with
$\langle\mathcal{L}_{\mathrm{P}}\rangle$, where $\mathcal{L}$ $\!=$
$\!\mathcal{L}_{\mathrm{P}}$ $\!+$ $\!\mathcal{L}_{\mathrm{Q}}$ is some
decomposition. The resulting equation is then neither linear nor trace
preserving. In Sec.~\ref{sec6}, we will in the context of the RPM derive an
evolution equation for the state of a chemical system which refers to a given
outcome of a continuous measurement (absence of a recombination fluorescence
signal). In a first step, this equation is then made trace-preserving as done in
(\ref{nlme}), which introduces a nonlinearity. In a second step, the equation is
projected onto a subspace of interest (the RP-subspace of the chemical system,
which we call R-subspace for short), after which it no longer preserves the
trace. Since a leakage out of the subspace of interest may not contradict the
outcome of the continuous measurement (a recombination of the RP may occur
undetected), the second step does \emph{not} recover linearity and  the equation
thus obtained is indeed neither linear nor trace preserving.
\subsection{
\label{sec2.2}
Jumpless continuous observation}
To give an example of a nonlinear trace preserving evolution, we first consider
a discrete input-output relation as generated by an ``instrument'' [a quantum
information (QI) term for an input-output device that performs a measurement].
We may start from a decomposition
$\mathcal{A}_{\mathrm{CPT}}$ $\!=$ $\!\sum_i\mathcal{A}_i$ of a completely
positive trace preserving (CPT) map $\mathcal{A}_{\mathrm{CPT}}$ into completely
positive but trace reducing maps $\mathcal{A}_i$. A state transformation can
then be written as
\begin{eqnarray}
\label{ishort}
  \mathcal{A}_{\mathrm{CPT}}\hat{\varrho}&=&\sum_i\hat{\varrho}_{\mathrm{N},i},
  \quad\hat{\varrho}_{\mathrm{N},i}=\mathcal{A}_i\hat{\varrho},
  \\
\label{instrument}
  &=&\sum_ip_i\hat{\varrho}_i,
  \quad\hat{\varrho}_i=\frac{1}{p_i}\mathcal{A}_i\hat{\varrho},
  \quad{p}_i=\langle\mathcal{A}_i\rangle.\quad
\end{eqnarray}
Eq.~(\ref{instrument}) describes the action of an instrument, which operates
on an input state $\hat{\varrho}$ and displays with probability $p_i$ a
classical output $i$ and as quantum output a transformed input state
$\hat{\varrho}_i$. The averaged output $\mathcal{A}_{\mathrm{CPT}}\hat{\varrho}$
is then obtained by ignoring the classical output. Alternatively,
(\ref{instrument}) can be written in short form (\ref{ishort}).
Physically, an instrument may be realized by performing a coupling $\hat{U}$ of
the system with an auxiliary system A (which in QI-jargon is usually referred to
as ``ancilla'') in a known state $\hat{\varrho}_{\mathrm{A}}$ after
which the ancilla is measured. Ascribing measurement outcome $i$ to an element
$\hat{\Pi}_i$ of a positive operator-valued measure
(POVM, i.e., $\hat{\Pi}_i$ $\!\ge$ $\!0$, $\sum_i\hat{\Pi}_i$ $\!=$ $\!1$)
acting on the ancilla, the transformation of the input state associated with the
classical outcome $i$ is
\begin{equation}
\label{Ai}
  \mathcal{A}_i\hat{\varrho}
  =\mathrm{Tr}_{\mathrm{A}}(\hat{U}\hat{\varrho}
  \otimes\hat{\varrho}_{\mathrm{A}}\hat{U}^\dagger\hat{\Pi}_i).
\end{equation}
A detection event associated with $\hat{\Pi}_i$ thus induces a transformation
$\mathcal{A}_i$. The $\hat{\Pi}_i$ (and $\hat{\varrho}_{\mathrm{A}}$) could be
replaced with projectors by replacing $\hat{U}$ with an enlarged operator, the
use of a POVM is however more convenient, since it is (as $\hat{U}$) associated
with the operation of a given physical device.

In order to make the transition to a continuous evolution equation, we assume
that between time $t$ and $t$ $\!+$ $\!\mathrm{d}t$, a transformation
$\mathcal{A}_0$ is carried out with probability $p$ $\!=$ $\!r\mathrm{d}t$,
where $r$ is some given probability density, e.g., due to an instrument which is
activated at random times with a rate $r$. Ignoring the normalization of the
state and probability of success, the change of state is
\begin{eqnarray}
  \hat{\varrho}_{\mathrm{N}}(t+\mathrm{d}t)
  &=&\hat{\varrho}_{\mathrm{N}}(t)
  +r\,\mathrm{d}t\,\Delta\hat{\varrho}_{\mathrm{N}},
  \\
  \Delta\hat{\varrho}_{\mathrm{N}}
  &=&\mathcal{A}_0\hat{\varrho}_{\mathrm{N}}
  -\hat{\varrho}_{\mathrm{N}},
\end{eqnarray}
which gives (\ref{lnte}) with
\begin{equation}
\label{LofA}
  \mathcal{L}=r(\mathcal{A}_0-1).
\end{equation}
If normalization is taken into account, the change of state now becomes
\begin{eqnarray}
  \hat{\varrho}(t+\mathrm{d}t)
  &=&\hat{\varrho}(t)+r\,\mathrm{d}t\,p_0\,\Delta\hat{\varrho},
  \\
  \Delta\hat{\varrho}
  &=&p_0^{-1}\mathcal{A}_0\hat{\varrho}-\hat{\varrho},
\end{eqnarray}
where $p_0$ $\!=$ $\!\langle\mathcal{A}_0\rangle$ is the probability that
the particular $\mathcal{A}_0$ is realized under the condition that the
instrument (\ref{instrument}) is activated. This can also be written as
(\ref{nlme}) with (\ref{LofA}). Note that (\ref{LofA}) vanishes for the identity
transformation, $\mathcal{A}_0$ $\!=$ $\!1$, as it should be. In case that
$\mathcal{A}_0$ $\!=$ $\!\mathcal{A}_{\mathrm{CPT}}$ preserves the trace, the
nonlinearity disappears in (\ref{nlme}),
$\langle\mathcal{L}\rangle$ $\!=$ $\!0$.
Other terms $\tilde{\mathcal{L}}$ may be added to (\ref{LofA}), to account for
additional (trace preserving) effects.

Eq.~(\ref{LofA}) implies that $\mathcal{A}_0$ describes the absence of any
jump-like events such as detector clicks observed at discrete instances of time.
This conditional dark evolution is a special case of a continuous measurement of
variables with a continuous spectrum $\bm{x}\in\mathbb{R}^n$, for which 
$\mathcal{A}_0$ is replaced with $\mathcal{A}_{\bm{x}}$, where $\bm{x}(t)$
describes a specific measurement record (history). We conclude with two examples
of jumpless conditional evolutions.
\subsubsection{Example of occasional projections}
Let us assume that
\begin{equation}
\label{Pex}
  \mathcal{A}_0=\mathcal{P},
\end{equation}
where $\mathcal{P}$ $\!=$ $\!\mathcal{P}^2$ is a projection, repeated at random
times with rate $r$. By spectral decomposition,
$\alpha^\mathcal{P}$ $\!=$ $\!1$ $\!+$ $\!(\alpha-1)\mathcal{P}$, so that
$\mathcal{L}$ $\!=$ $\!r(\mathcal{P}-1)$ and $\mathrm{e}^{\mathcal{L}t}$ $\!=$
$\!\mathrm{e}^{-rt}(\mathrm{e}^{rt})^{\mathcal{P}}$ $\!=$
$\!\mathrm{e}^{-rt}(1-\mathcal{P})+\mathcal{P}$, and the state evolution
(\ref{rho}) becomes
\begin{eqnarray}
  \hat{\varrho}(t)&=&p^{-1}\mathcal{G}(t,0)\hat{\varrho}(0),
  \\
  p(t)&=&\mathrm{Tr}[\mathcal{G}(t,0)\hat{\varrho}(0)],
  \\
  \mathcal{G}(t,0)&=&\mathrm{e}^{-rt}(1-\mathcal{P})+\mathcal{P}.
\end{eqnarray}
An initial state $\hat{\varrho}(0)$ is thus asymptotically projected onto
$\hat{\varrho}(\infty)$ $\!=$ $\!p^{-1}\mathcal{P}\hat{\varrho}(0)$, and
$p$ $\!=$ $\!\mathrm{Tr}[\mathcal{P}\hat{\varrho}(0)]$ is the corresponding
probability.

Care must be taken to ensure that $\mathcal{A}_0$ is physical in the sense that
it describes the outcome of a measurement according to
(\ref{ishort}) -- (\ref{Ai}). If instead of (\ref{Pex}) we used the
complementary projection $\mathcal{A}_0$ $\!=$ $\!1$ $\!-$ $\!\mathcal{P}$, and
inserted $\mathcal{P}(\bm{\cdot})$ $\!=$ $\!\hat{P}\bm{\cdot}\hat{P}$ with some
projector $\hat{P}$, then (\ref{lnte}) and (\ref{LofA}) would suggest
$\frac{\partial}{\partial{t}}\hat{\varrho}_{\mathrm{N}}$ $\!=$
$\!-r\hat{P}\hat{\varrho}_{\mathrm{N}}\hat{P}$, which is just the equation shown
in \cite{Hab76} to be unphysical. The correct form should read 
$\mathcal{A}_0$ $\!=$ $\!\mathcal{A}_{\mathrm{CPT}}$ $\!-$ $\!\mathcal{P}$,
where $\mathcal{A}_{\mathrm{CPT}}$ $\!=$ $\!\mathcal{P}$ $\!+$ $\!\mathcal{Q}$
with $\mathcal{Q}(\bm{\cdot})$ $\!=$ $\!\hat{Q}\bm{\cdot}\hat{Q}$, and
$\hat{Q}$ $\!=$ $\!1$ $\!-$ $\!\hat{P}$ is the orthogonal complement. That is,
the projectors add to the identity, whereas the corresponding projective maps
add to the trace-preserving map (here again a projective map), but not the
identity. With this we get
$\frac{\partial}{\partial{t}}\hat{\varrho}_{\mathrm{N}}$ $\!=$
$\!r(\hat{P}\hat{\varrho}_{\mathrm{N}}\hat{P}
-\{\hat{P},\hat{\varrho}_{\mathrm{N}}\})$, which is just the type of equation
considered in \cite{Jon10}.
\subsubsection{Example of one-atom maser
\label{sec:one-atom-maser}}
\begin{figure}[ht]
\includegraphics[width=7cm]{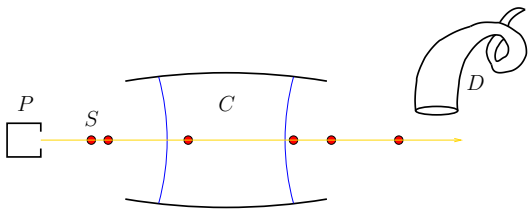}
\caption{\label{fig5}
(color online).
Scheme of a one-atom (or micro-) maser. A preparation device P prepares atoms S,
which are regarded as two-level systems, in a given initial state. The atoms
traverse a resonator at random time instances, where they interact with a cavity
field C for a fixed time period. Afterwards they pass a detector D which either
detects them in one out of two fixed basis states or fails to detect them.
}
\end{figure}

Another example is given in \cite{Bri94}, where at random times but with a
given rate $r$, diagnosis atoms enter a cavity
(i.e., $p$ $\!=$ $\!r\mathrm{d}t$ is the probability that an atom enters between
$t$ and $t$ $\!+$ $\!\mathrm{d}t$). The atoms traverse the cavity field, after
which they pass a detector as depicted in Fig.~\ref{fig5}. 
The atoms are regarded as two-level systems and the
possible detection events are
\begin{eqnarray}
  \hat{\Pi}_A&=&p_A|A\rangle\langle{A}|,
  \\
  \hat{\Pi}_B&=&p_B|B\rangle\langle{B}|,
  \\
  \hat{\Pi}_0&=&(1-p_A)|A\rangle\langle{A}|+(1-p_B)|B\rangle\langle{B}|,
\end{eqnarray}
corresponding to the detection of state $A$ ($B$) with classical detection
efficiencies $p_A$ ($p_B$), or a failed detection. The induced transformations
of the cavity field are 
\begin{eqnarray}
  \mathcal{A}_A&=&p_A\mathcal{A},
  \\
  \mathcal{A}_B&=&p_B\mathcal{B},
  \\
\label{A0}
  \mathcal{A}_0&=&(1-p_A)\mathcal{A}+(1-p_B)\mathcal{B}, 
\end{eqnarray}
whereas $\mathcal{A}_{\mathrm{CPT}}=\mathcal{A}+\mathcal{B}$ describes the
transformation of the cavity field if the detector is ignored.
The evolution of the cavity field between detector clicks is then described by
(\ref{nlme}) together with (\ref{LofA}), where (\ref{A0}) is used for
$\mathcal{A}_0$. While a broken detector, $p_A$ $\!=$ $\!p_B$ $\!=$ $\!0$,
reduces dark to unconditional evolution,
$\mathcal{A}_0$ $\!=$ $\!\mathcal{A}_{\mathrm{CPT}}$, a perfect detector,
$p_A$ $\!=$ $\!p_B$ $\!=$ $\!1$, renders failed counts impossible,
$\mathcal{A}_0$ $\!=$ $\!0$. An additional $\tilde{\mathcal{L}}$ with
$\langle\tilde{\mathcal{L}}\rangle$ $\!=$ $\!0$ may be added to
$\mathcal{L}$ to take into account additional effects such as a unitary
evolution of the cavity field during the passage of the atoms.
Here, $\langle\mathcal{L}\rangle$ $\!\equiv$ $\!\langle{r}(\mathcal{A}_0$ $\!-$
$\!1)\rangle$ $\!=$ $\!-rp_{\mathrm{click}}(t)$, where
$p_{\mathrm{click}}(t)$ $\!=$ $\!\langle1$ $\!-$ $\!\mathcal{A}_0\rangle$ $\!=$
$\!p_A\mathrm{Tr}(\mathcal{A}\hat{\varrho})$ $\!+$
$\!p_B\mathrm{Tr}(\mathcal{B}\hat{\varrho})$ is the probability that an atom
that entered the cavity at time $t$ is detected (in state A or B). Defining
$k$ $\!=$ $\!rp_{\mathrm{click}}$ and comparing with (\ref{ppdot}) and
(\ref{lnpdot}) gives as normalization factor the probability
$p(t)$ $\!=$ $\!\mathrm{Tr}\left[\hat{\varrho}_{\mathrm{N}}(t)\right]$ $\!=$
$\!\mathrm{e}^{-\int_0^t\mathrm{d}\tau{k}(\tau)}$ $\!=$
$\!\mathrm{e}^{-kt}$ of the state transformation, i.e, the probability that no
click occurs during the time interval $(0,t]$. In the last identity we have
assumed that $k$ remains constant. Hence
$\frac{\mathrm{d}}{\mathrm{d}t}[1-p(t)]$ $\!=$ $\!k\mathrm{e}^{-kt}$
is the waiting time distribution until the first click is observed.
\section{
\label{sec3}
Spin conversion within the exponential model}
After having motivated the use of non-trace preserving and nonlinear master
equations in the previous section, we will now discuss an application of
(\ref{lnte}) in the context of the RPM. We consider a chemical system consisting
of two electron spins which, upon their spatial separation, evolve under the
influence of a local nuclear spin environment and possible external magnetic
fields, while performing some mechanical motion. This individual evolution is
therefore occasionally interrupted by transient short-time re-encounters of the
electrons, where random direct electron interactions affect the two-electron
spin state. On the other hand, such encounters are a necessary pre-condition for
a possible recombination of the RP. Our model describes these effects as a
generalized measurement by means of an interaction with a minimal model
environment, that is switched on during the time of encounters. We thus
decompose the generator of the evolution (\ref{lnte}) according to
\begin{eqnarray}
\label{Lbe}
  \mathcal{L}&=&\mathcal{L}_{\mathrm{betw}}+\mathcal{L}_{\mathrm{enc}},
  \\
\label{Lbetw}
  \mathcal{L}_{\mathrm{betw}}
  &=&-\mathrm{i}[\hat{H},\bm{\cdot}]+\mathcal{L}_{\mathrm{diss}}.
\end{eqnarray}
The term $\mathcal{L}_{\mathrm{enc}}$ describes the effect of the encounters,
such as possible dephasing and recombination, and thus gives rise to the various
reaction operators suggested for the RPM. It is one of the main subjects of
this work, and in the next section we will derive expressions for it. The other
term $\mathcal{L}_{\mathrm{betw}}$ takes into account all effects which cannot
be attributed to the encounters, in particular interactions with local nuclear
spins and external magnetic fields, which can be described by means of a
Hamiltonian $\hat{H}$, and potential additional dissipative effects
$\mathcal{L}_{\mathrm{diss}}$ affecting the two separated electron spins.
Although in practice such additional decoherence due to interactions with the
environment of the radicals (spin relaxation) is likely, it may be
indistinguishable from similar effects originating from the encounters and
happens on longer timescales. In this work, we will hence ignore the latter
term, i.e., we set $\mathcal{L}_{\mathrm{diss}}$ $\!=$ $\!0$. In this section
we will however briefly outline the origin of $\hat{H}$ that gives rise to a
unitary spin evolution. Within an (idealized) exponential model, one single
encounter is sufficient to terminate the unitary spin evolution as it induces a
chemical transition to reaction products. This allows one to divide the total
RPM-reaction cycle into three steps and place it within a simple state
preparation-evolution-measurement scheme, which we outline in this section.
\subsection{RP-creation and initial state}
Upon absorption of a photon by the chemical system, one of two paired electrons
in a singlet state is excited from the highest occupied to the lowest unoccupied
molecular orbital (HOMO-LUMO transition) and subsequently transferred to a
separate location, leaving behind the other electron. Each of the two locations
now possesses an unpaired electron and thus together they form a RP.
If we assume that this formation process itself occurs fast on the time scale of
the subsequent unitary evolution, that the two electrons originate from a single
pair of covalently bound electrons in a singlet state, and that other effects
such as spin-orbit coupling can be neglected, the electron spin state of a newly
formed RP can be assumed to be the singlet state $|S\rangle$ $\!=$
$\!(|\!\uparrow\downarrow\rangle-|\!\downarrow\uparrow\rangle)/\sqrt{2}$.
Since the nuclear spin states are not affected by this formation process, the
initial state of the chemical system can be assumed to be
$\hat{\varrho}(0)$ $\!=$ $\!|S\rangle\langle{S}|\otimes
\frac{\hat{\mathbb{I}}_{\mathrm{nuc}}}
{\mathrm{Tr}(\hat{\mathbb{I}}_{\mathrm{nuc}})}$.

For a magnetic field strength of $B=$ 1T as used in ESR spectroscopy, the free
electron resonance
$\nu_{\mathrm{res}}/B$ $\!=$ $\!\mu_{\mathrm{B}}g/h$ $\!=$ 28.02 GHz/T lies in
the microwave range ($\lambda$ $\!=$ $\!c/\nu$ $\!\approx$ 1cm), and for the
geomagnetic field $B$ $\!=$ $\!47\mu{T}$ in the medium-wave radio-frequency
region ($\nu_{\mathrm{res}}$ $\!=$ 1.316 MHz, $\lambda$ $\!\approx$ 200m).
For comparison, at room temperature $T$ $\!=$ $\!298$K, thermal radiation
$\nu_{\mathrm{th}}$ $\!=$ $\!k_{\mathrm{B}}T/h$ $\!=$ $\!6.211\cdot10^{12}$Hz
$=$ $\!2.22\cdot10^2\nu_{\mathrm{res}}(1T)$ $\!=$
$\!4.72\cdot10^6\nu_{\mathrm{res}}(47\mu{T})$ is orders of magnitude higher in
energy and falls in the mid-infrared,
$\lambda_{\mathrm{th}}$ $\!\approx$ 50$\mu$m. We can therefore conclude, that
each of the nuclear spins by themselves are in the maximally mixed state
$\frac{1}{2}\hat{\mathbb{I}}_2$. The same holds for the RP-electrons due to
their initial state (see below), and remains so at later times, since the
electron state in thermal equilibrium is consistent with it.
\subsection{Unitary evolution}
In the subsequent evolution step, the chemical system remains in its new
conformation or in the form of two radicals diffusing in solution.
The hyperfine coupling of each of the electron spins to the nearby nuclear spins
under the influence of an external such as the geomagnetic field drives a spin
interconversion (intersystem crossing) of the two-electron spin state, that is,
the initial singlet state gradually acquires overlap with the triplet state
manifold during the time of separation. Typical nuclides, which reside on the
chemical system itself, are $^1$H of spin 1/2 and $^{14}$N of spin 1, whereas
$^{12}$C and $^{16}$O are of spin 0. This evolution would occasionally be
interrupted (in the multiple-encounter model discussed below) or terminated (in
the exponential model) by encounters of the radicals, during which direct
interactions between the electrons such as short-range dipolar and exchange
interactions dominate over the hyperfine couplings.

\begin{figure}[ht]
\includegraphics[width=6cm]{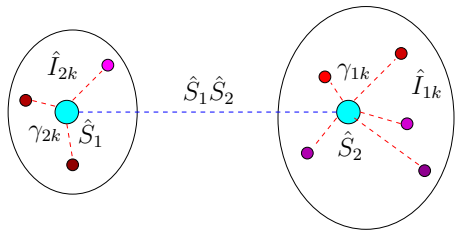}
\caption{\label{fig1}
(color online).
Simple model of a RP consisting of two electron spins $\hat{\bm{S}}_m$
interacting with surrounding nuclear spins $\hat{\bm{I}}_{mk}$ via hyperfine
couplings $\bm{\gamma}_{mk}$. For sufficiently separated spins, cross-couplings
such as $\hat{S}_1\hat{S}_2$ can be neglected. The radicals together with their
local spin environments then form two separate subsystems. 
}
\end{figure}

We thus assume that the state of the chemical system evolves unitarily during
such time intervals, with a Hamiltonian given by the Zeeman and hyperfine
interactions according to \cite{Tie12a},
\begin{equation}
\label{H}
  \hat{H}=\sum_{m=1}^2\hat{H}_m,
  \quad
  \hat{H}_m=\hat{\bm{S}}_m\cdot(\bm{B}+\hat{\bm{I}}_m).
\end{equation}
$\hat{\bm{S}}_m$ $\!=$ $\!\mu_{\mathrm{B}}\frac{g_m}{2}\hat{\bm{\sigma}}_m$ is
the operator of the electron spin at radical $m$, where $\hat{\bm{\sigma}}_m$ is
the respective vector of Pauli matrices,
$\mu_{\mathrm{B}}$ $\!=$ $\!\frac{e\hbar}{2m_{\mathrm{e}}}$ is the
Bohr magneton, and $g_m\approx{g}_{\mathrm{e}}\approx2$ is the respective
effective $g$-factor, which may vary slightly from the value ${g}_{\mathrm{e}}$
of a free electron due to the molecular environment. The $\hat{\bm{I}}_m$
$\!=$ $\!\sum_{k=1}^{N_m}\bm{\gamma}_{mk}\cdot\hat{\bm{I}}_{mk}$ are given by
the operators $\hat{\bm{I}}_{mk}$ of nuclear spin $k$ located at radical $m$ and
the respective hyperfine coupling tensors $\bm{\gamma}_{mk}$ with respect to
electron spin $\hat{\bm{S}}_m$. $\bm{B}$ is the classical external magnetic
field whose value can be regarded to be the same for both electron spins.
Eq.~(\ref{H}) disregards the dipolar [$\mathcal{O}(R^{-3})$] and exchange
interaction [$\mathcal{O}(\mathrm{e}^{-R})$] between spins belonging to
different radicals, assuming that radicals diffusing in solution are separated
by a sufficiently large distance $R$. If the radicals are located on the same
molecule, the two interactions may cancel each other \cite{Efi08}. Eq.~(\ref{H})
also neglects any interactions of the nuclear spins with the external magnetic
field and among themselves due to the mass difference between electrons and
nuclei, $m_{\mathrm{e}}$ $\!=$ $\!\mathcal{O}(10^{-3}m_{\mathrm{nuc}})$. This
model is illustrated in Fig.~\ref{fig1}.

For isotropic hyperfine couplings, the
$\bm{\gamma}_{mk}$ $\!=$ $\!\gamma_{mk}\mathbb{I}_3$
become for each $m$ and $k$ proportional to the $3\times3$ identity matrix
$\mathbb{I}_3$. For a chemical compass, anisotropic couplings are required to
obtain a dependence of the spin dynamics on the direction of the external
magnetic field $\bm{B}$. An alternative is to model the nuclear spins by means
of a classical inhomogeneous magnetic field, i.e. a modified effective $g$
tensor.

Because of (\ref{H}), the unitary evolution of the chemical system factorizes
into two terms, each acting only on one of the radicals. One may identify the
electron spins as the system of interest with free Hamiltonian
$\hat{\bm{S}}_m\cdot\bm{B}$ coupled by the interaction Hamiltonian
$\hat{\bm{S}}_m\cdot\hat{\bm{I}}_m$ to the nuclear spins forming local spin
environments. In contrast to the assumption of a large bath being weakly coupled
to the system, the number of relevant nuclear spins may be small and their
hyperfine couplings too large to justify a master equation approach, however.

As illustrated in \cite{Tie12a}, quantities defined on the electron spins such
as state purity or fidelity with the initial state exhibit recurrences in the
simplest case of a single nuclear spin coupled isotropically to one of the
electrons. If more nuclear spins are added, recurrence times grow and for about
5 neighboring nuclear spins, a number comparable to those in relevant
bio-molecules, recurrence times become long compared to the time scales of the
chemical reactions. 
\subsection{Re-encounter and RP-recombination}
The third step is initiated by the molecule returning to its original
conformation or a sufficiently close encounter of the diffusing radicals,
followed either by a direct relaxation to the original singlet ground state,
which is often accompanied by fluorescence, or via a different state such as a
metastable triplet state. The probability of these different reaction channels
is determined by the overlap of the electron spin state with the singlet or
triplet states, since direct relaxation conserves the spin state. This last
detection step can hence be understood as an imperfect measurement of the spin
character of the electrons at a generally random time that corresponds to an
encounter of the radicals.

\begin{figure}[ht]
\includegraphics[width=3cm]{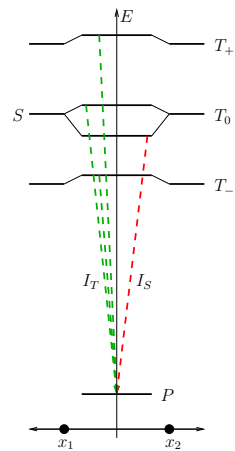}
\caption{\label{fig2}
(color online).
V-type three-manifold model of the chemical system sketching the levels of
energy $E$ as a function of the locations $x_1$ and $x_2$ of the radicals.
Transitions can occur between the stable ground state (P-subspace) and the
excited RP space (R-subspace), spanned by singlet (S) and triplet
($\mathrm{T}_j$) states. Transitions from the singlet ($I_S$) can be
distinguished from transitions from the triplet manifold ($I_T$) by their
fluorescence. For high magnetic fields, the $\mathrm{T}_\pm$-states are
separated from the $\mathrm{T}_0$-state. In practice, the transitions may take
place via transient intermediate steps, which are not shown here for simplicity.
}
\end{figure}

Let us now look at the state space of the two RP electron spins.
For a sufficiently high external magnetic field, the three triplet
(T$_0$, T$_{\pm}$) levels are energetically well separated. The structure of
the spin system can then be simplified by limiting to the subspace consisting
of the S, T(=T$_0$), and reaction product (P) state, making up effectively a
V-type three level system \cite{Tie12} as shown in Fig.~\ref{fig2}. 

The simplest way to include the spin environment into the model is to assume
just one single nuclear spin which is coupled to one of the RP spins,
and which is treated classically \cite{Cai11a}. The S and T states are then no
longer eigenstates of the total Hamiltonian, which causes coherent S-T
oscillations. In general, the effect of the environment can be visualized in
Fig.~\ref{fig2} as a broadening of the levels to energy bands.

With the distance-dependent exchange interaction, levels shift as shown in
Fig.~\ref{fig2}. For example, if the two RP electron spins are equal
(e.g., for a homogeneous magnetic field in $z$-direction), and an
(isotropic $x,y,z$ or symmetric $x,y$) exchange interaction with positive
coupling constant is added, the eigenstates of the RP spins remain eigenstates
of the total Hamiltonian, but the energy of the singlet state is lowered,
whereas the energy of the triplet states is raised. In a classical model, one
may interpret this as a repulsive force felt by radicals approaching in the
triplet states so that they are less likely to get close enough to recombine.
In a singlet state, they experience an attractive force so that the chance of
recombination is increased. The RP recombination itself is a chemical reaction
involving energies in the optical range and is therefore treated as an
interaction with a separate environment, typically a zero temperature bosonic
bath \cite{Iva10}.

In our model, we distinguish between time intervals of unitary evolution and
encounter events, which are assumed to be necessary for a recombination. By
coupling the chemical system to a minimal model environment, each encounter is
modeled as a general measurement process defined by phenomenological parameters
such as relaxation, dephasing, and detection constants. We assume that these
constants could in principle be determined from a microscopic model which takes
into account details such as the exchange interaction during an encounter, state
densities of the radiation field at optical frequencies, or excitations of
vibronic molecular levels. Measurement outcomes are distinct singlet or triplet
fluorescence signals, accompanied by a population of separate states of the
model environment. This allows us to reduce to one single ground state P of the
chemical system as shown in Fig.~\ref{fig2}.
Since for low external magnetic fields, one cannot rely on an energetic
separation of the triplet levels, our analysis involves all three triplets and
not just the $\mathrm{T}_0$-state.
\subsection{Readout}
We conclude with resuming the approach used recently in \cite{cai10} that
generalizes the evaluation of the singlet yield well established in the
literature (see, e.g. \cite{Ste89} for a review). There, a yield
\begin{equation}
\label{average}
  \Phi=\int_{0}^{\infty}\mathrm{d}t\;{p}(t)f(t)
\end{equation}
is computed by averaging a given function
$f(t)$ $\!=$ $\!f[\hat{\varrho}_{\mathrm{S}}(t)]$ of the RP state
$\hat{\varrho}_{\mathrm{S}}(t)$ at time $t$ over the probability distribution
$p(t)$ of the time of RP recombination, which is assumed to happen at
the first encounter. The latter is exponentially distributed,
\begin{equation}
\label{em}
  p(t)=r\mathrm{e}^{-rt},
\end{equation}
where $r$ is the encounter rate.
Quantities considered are the singlet state fidelity
\begin{equation}
  f(t)=\langle{S}|\hat{\varrho}_{\mathrm{S}}(t)|{S}\rangle,
\end{equation}
for which (\ref{average}) is called singlet yield $\Phi_{\mathrm{S}}$, or some
appropriate measure $f(t)$ $\!=$ $\!E(t)$ of entanglement such as the
concurrence, for which (\ref{average}) gives the entanglement yield
$\Phi_{\mathrm{E}}$, along with the lifetime of entanglement, defined as
$T_{\mathrm{E}}$ $\!=$ $\!\mathrm{max}\{t|E(t)>0\}$.
The respective magnetic field sensitivity of the yields is obtained by
replacing $f$ in (\ref{average}) with
\begin{equation}
  \Lambda=\partial{f}/\partial{B},
\end{equation}
since $p(t)$ does not depend on $B$.
\section{
\label{sec10}
Reaction operators}
After having explained the magnetic spin interactions included in 
$\mathcal{L}_{\mathrm{betw}}$ in (\ref{Lbe}), we will now focus on the
RP-reactions corresponding to the second term $\mathcal{L}_{\mathrm{enc}}$.
Since chemical reactions are preceded by encounters, we describe them as random
interactions with a simple environment that is required, for consistency, in a
quantum mechanical description. Physically speaking, it provides the energy
necessary for the creation of a RP and absorbs the energy released in
the recombination thereof. We assume that the encounters have no effect on the
state of the nuclear spins. In this section, we present a general detailed
derivation of the reaction operators, distinguishing between the singlet and all
three triplet states
$j$ $\!=$ $\!\{\mathrm{S},\mathrm{T}_0,\mathrm{T}_+,\mathrm{T}_-\}$. This allows
us to incorporate decoherence effects in the triplet subspace in case coherences
in the latter are either present in the initial state or are dynamically
generated by external RF magnetic fields. In the two sections following
thereafter, we will limit attention to simplified expressions, only
distinguishing between quantities of singlet and triplet character,
$j$ $\!=$ $\!\{\mathrm{S},\mathrm{T}\}$. A reader merely interested in a
comparison with the related literature may skip the present section completely.
\subsection{Initial state}
Physically, RP and reaction product constitute macroscopically distinguishable
classical states of the chemical system, and a coherent superposition of them
would in practice be converted by decoherence processes to a corresponding
mixture on an unresolvable short time scale, analogous to the fate of
``Schr\"odinger's cat''. About the initial state
$\hat{\varrho}(0)$ $\!=$ $\!\hat{\varrho}_0$ we thus assume that
\begin{equation}
\label{inicon}
  \hat{Q}_j\hat{\varrho}\hat{Q}_{\mathrm{P}}
  =\hat{Q}_{\mathrm{P}}\hat{\varrho}\hat{Q}_j=0
  \quad\Leftrightarrow\quad
  \hat{\varrho}=\hat{\varrho}_{\mathrm{R}}+\hat{\varrho}_{\mathrm{P}}
\end{equation}
for $j$ $\!=$ $\!{\mathrm{S}},{\mathrm{T}_i}$, where
$\hat{\varrho}_{\mathrm{R}}$ $\!=$ $\!\mathcal{Q}_{\mathrm{R}}\hat{\varrho}$
$\!\equiv$ $\!\hat{Q}_{\mathrm{R}}\hat{\varrho}\hat{Q}_{\mathrm{R}}$ and
$\hat{\varrho}_{\mathrm{P}}$ $\!=$ $\!\mathcal{Q}_{\mathrm{P}}\hat{\varrho}$
$\!\equiv$ $\!\hat{Q}_{\mathrm{P}}\hat{\varrho}\hat{Q}_{\mathrm{P}}$, with
$\hat{Q}_{\mathrm{R}}$ and $\hat{Q}_{\mathrm{P}}$ being the projectors onto the
R- and P-subspaces, cf. (\ref{STPdec}) and (\ref{projs}). Eq.~(\ref{inicon})
means that the state is a mixture of its RP and product state components, i.e.,
there is no coherence between them. This is a basic assumption in spin
chemistry, and our evolution equations are consistent with it in the sense that
no such coherence is generated over time.
[Because of assuming (\ref{inicon}), the above-mentioned additional decoherence
processes are not required in our model. Consequently, if the model is extended
to allow for a continuous excitation of new RPs, it must be ensured that the
excitation process is consistent with (\ref{inicon}), cf. the comments in
Sec.~\ref{sec9.2}.]
\subsection{Model environment}
We adopt a phenomenological approach assuming an interaction with a ``minimum''
environment by an effective model Hamiltonian
\begin{equation}
\label{HI}
  \hat{H}_{\mathrm{I}}=\sum_{j={\mathrm{S}},{\mathrm{T}_i}}\bigl(
  \pi_j\hat{L}_j\otimes|\pi_j\rangle\langle0|
  +\delta_j\hat{Q}_j\otimes|\delta_j\rangle\langle{0}|
  +h.c.\bigr),
\end{equation}
where the sum runs over all four two-electron spin states,
$j$ $\!=$ $\!\{\mathrm{S},\mathrm{T}_0,\mathrm{T}_+,\mathrm{T}_-\}$, and
$|0\rangle$, $|\pi_j\rangle$, $|\delta_j\rangle$ are orthonormal environment
states. The $\hat{L}_j$ are defined in (\ref{Lj}), and the 
$\hat{Q}_j$ $\!=$ $\!\hat{L}_j^\dagger\hat{L}_j$ are the aforementioned
projectors onto the two-electron spin states $j$. The coefficients $\pi_j$ and
$\delta_j$ describe decay and dephasing from the respective states $j$. Below we
distinguish between a master equation model, where $\hat{H}_{\mathrm{I}}$ acts
constantly over time, and an encounter model, where $\hat{H}_{\mathrm{I}}$ acts
only over short periods of time that correspond to encounters of the radicals.
Of particular interest in both models are the following special cases:

\emph{Triplet symmetry:}
In this case, we assume identical absolute values of the decay and dephasing
coefficients for all triplet states,
\begin{equation}
\label{symmcondmaps1}
  |\pi_{\mathrm{T}_i}|=|\pi_{\mathrm{T}}|,\quad
  |\delta_{\mathrm{T}_i}|=|\delta_{\mathrm{T}}|.
\end{equation}

\emph{Triplet symmetry without triplet dephasing:}
In this case, we assume in addition to (\ref{symmcondmaps1}), that there is no
dephasing within the triplet subspace,
\begin{equation}
\label{symmcondmaps1a}
  |\delta_{\mathrm{T}}|=0,
\end{equation}
i.e., that the triplet subspace is ``dephasing-free''.
\subsection{
\label{secMEM}
Master equation model}
\subsubsection{Equation and its solution in full space}
Extending the approach followed in \cite{Tie12}, we apply the situation
described in App.~\ref{secPES} to (\ref{HI}). In (\ref{Lk}), therefore
$\hat{c}_0$ $\!=$ $\!0$ and by writing the coupling strengths separately,
\begin{equation}
\label{rjdj}
  r_j=2t|\pi_j|^2,\quad d_j=2t|\delta_j|^2,\quad(j={\mathrm{S}},{\mathrm{T}_i}),
\end{equation}
Eq.~(\ref{TCLpe0}) can be written as
\begin{equation}
\label{30neu}
  \frac{\partial}{\partial{t}}\hat{\varrho}
  =\sum_{j={\mathrm{S}},{\mathrm{T}_i}}\bigl[r_j\mathcal{L}(\hat{L}_j)
  +d_j\mathcal{L}(\hat{Q}_j)\bigr]\hat{\varrho}.
\end{equation}
The solution of (\ref{30neu}) is given by
\begin{eqnarray}
  \hat{\varrho}(t)
  &=&\hat{\varrho}_{0\mathrm{P}}
  +\sum_{j,k={\mathrm{S}},{\mathrm{T}_i}}\mathrm{e}^{-{\eta}_{jk}t}
  \mathcal{Q}_{jk}\hat{\varrho}_0
  \nonumber\\
  &+&\!\!\!\sum_{j={\mathrm{S}},{\mathrm{T}_i}}
  \!\Bigl[(\mathrm{e}^{-r_jt}\!-\!\mathrm{e}^{-\eta_{jj}t})
  \mathcal{Q}_j\hat{\varrho}_0\!+\!(1\!-\!\mathrm{e}^{-r_jt})
  \hat{L}_j\hat{\varrho}_0\hat{L}_j^\dagger\Bigr]
  \nonumber\\
  &=&\hat{\varrho}_{0\mathrm{P}}
  +\sum_{j,k={\mathrm{S}},{\mathrm{T}_i}}^{(j\neq{k})}\mathrm{e}^{-{\eta}_{jk}t}
  \mathcal{Q}_{jk}\hat{\varrho}_0
  \nonumber\\
  &+&\sum_{j={\mathrm{S}},{\mathrm{T}_i}}
  \Bigl[\mathrm{e}^{-r_jt}\mathcal{Q}_j\hat{\varrho}_0+(1-\mathrm{e}^{-r_jt})
  \hat{L}_j\hat{\varrho}_0\hat{L}_j^\dagger\Bigr],
\label{gensol1}
\end{eqnarray}
where
\begin{equation}
\label{etajjp}
  \eta_{jk}=\frac{r_{j}+r_{k}+d_{j}+d_{k}}{2}.
\end{equation}
Obviously, $\hat{\varrho}(0)$ $\!=$ $\!\hat{\varrho}_{0\mathrm{P}}$ $\!+$
$\!\hat{\varrho}_{0\mathrm{R}}$ and (if all $r_j$ $\!>$ $\!0$) the asymptotic
state is
$\hat{\varrho}(\infty)$ $\!=$ $\!\hat{Q}_{\mathrm{P}}$.
Furthermore, (\ref{gensol1}) obeys (\ref{inicon}) at any time.

\emph{Triplet symmetry:}
If the decay and dephasing rates are the same for all triplet states, i.e.,
(\ref{symmcondmaps1}) holds, so that
\begin{equation}
\label{symmcond}
  r_{\mathrm{T}_i}=r_{\mathrm{T}},\quad
  d_{\mathrm{T}_i}=d_{\mathrm{T}},
\end{equation}
the coefficients (\ref{etajjp}) reduce to
$\eta_{jj}$ $\!=$ $\!r_j$ $\!+$ $\!d_j$ with $j={\mathrm{S}},{\mathrm{T}}$ and
$\eta_{\mathrm{ST}}$ $\!=$ $\!\eta_{\mathrm{TS}}$ $\!=$ $\!\eta$ defined below
in (\ref{coeff}). In this case, (\ref{30neu}) can be simplified to
\begin{eqnarray}
  \frac{\partial}{\partial{t}}\hat{\varrho}
  &=&\sum_{j={\mathrm{S}},{\mathrm{T}}}\Bigl[r_j
  \bigl(\langle\hat{Q}_j\rangle\hat{Q}_{\mathrm{P}}
  -\frac{1}{2}\{\hat{Q}_j,\hat{\varrho}\}\bigr)
  +d_j\mathcal{L}(\hat{Q}_j)\hat{\varrho}\Bigr]
  \nonumber\\
  &&
  +d_{\mathrm{T}}\mathcal{Q}_{\mathrm{coh}}\hat{\varrho},
\label{30neus}
\end{eqnarray}
where
\begin{eqnarray}
  \langle\hat{Q}_j\rangle&=&\mathrm{Tr}(\hat{\varrho}\hat{Q}_j)
  =\langle\mathcal{Q}_j\rangle
  =\mathrm{Tr}(\mathcal{Q}_j\hat{\varrho})
  \nonumber\\
  &=&\mathrm{Tr}(\mathcal{Q}_j\hat{\varrho}_{\mathrm{R}})
  =\mathrm{Tr}(\hat{Q}_j\hat{\varrho}_{\mathrm{R}}),
\label{avQ}
\end{eqnarray}
and
\begin{equation}
\label{Qcoh}
  \mathcal{Q}_{\mathrm{coh}}
  \!=\!\sum_{j={\mathrm{T}_i}}\mathcal{L}(\hat{Q}_j)
  \!-\!\mathcal{L}(\hat{Q}_{\mathrm{T}})
  \!=\!\sum_{j={\mathrm{T}_i}}\mathcal{Q}_j\!-\!\mathcal{Q}_{\mathrm{T}}
  \!=\!-\sum_{j,k={\mathrm{T}_i}}^{(j\neq{k})}\mathcal{Q}_{jk}
\end{equation}
removes the triplet coherences. (The first two sums here run over all three
triplet states $T_0$, $T_+$, $T_-$, and the double sum on the right excludes
the terms for which the triplet states $T_j$ and $T_k$ are the same.)
Its solution follows from (\ref{gensol1}),
\begin{eqnarray}
  \hat{\varrho}(t)&=&
  \hat{\varrho}_{0\mathrm{P}}+
  \mathrm{e}^{-\eta{t}}\hat{\varrho}_{0\mathrm{R}}
  \nonumber\\
  &+&\!\!\!\sum_{j={\mathrm{S}},{\mathrm{T}}}\!\Bigl[
  (\mathrm{e}^{-r_jt}\!-\!\mathrm{e}^{-\eta{t}})
  \mathcal{Q}_j\hat{\varrho}_0
  \!+\!(1\!-\!\mathrm{e}^{-r_jt})
  \hat{Q}_{\mathrm{P}}\mathrm{Tr}(\hat{\varrho}_0\hat{Q}_j)
  \Bigr]
  \nonumber\\
  &&+\Bigl[\mathrm{e}^{-r_{\mathrm{T}}t}
  -\mathrm{e}^{-(r_{\mathrm{T}}+d_{\mathrm{T}})t}\Bigr]
  \mathcal{Q}_{\mathrm{coh}}\hat{\varrho}_0.
\label{gensol2}
\end{eqnarray}

\emph{Triplet symmetry without triplet dephasing:}
If both (\ref{symmcondmaps1}) and (\ref{symmcondmaps1a}) hold, so that in
addition to (\ref{symmcond}), there is no triplet dephasing,
\begin{equation}
\label{symmcond0}
  d_{\mathrm{T}}=0,
\end{equation}
(\ref{30neus}) finally reduces to (\ref{genme}) below, and (\ref{gensol2})
reduces to (\ref{gensol}). We will discuss these simplified equations in the
next section. Here we just note that since (\ref{gensol}) obeys (\ref{inicon})
at any time, we can apply the first identity in (\ref{STid}) to interchange the
dephasing terms in (\ref{genme}), i.e., we can replace
$\sum_{j={\mathrm{S}},{\mathrm{T}}}d_j\mathcal{L}(\hat{Q}_j)$ $\!=$
$\!d\sum_{j={\mathrm{S}},{\mathrm{T}}}\mathcal{L}(\hat{Q}_j)$, and
(\ref{gensol}) depends only on the sum of the $d_j$ via $\eta$, cf.
(\ref{coeff}).
\subsubsection{Reduced equation and its solution in the R-subspace}
Limiting to the R-subspace, (\ref{TCLpe0R}) gives
\begin{equation}
\label{36neu}
  \frac{\partial}{\partial{t}}\hat{\varrho}_{\mathrm{R}}
  =\sum_{j={\mathrm{S}},{\mathrm{T}_i}}
  \Bigl[-\frac{r_j}{2}\{\hat{Q}_j,\hat{\varrho}_{\mathrm{R}}\}
  +d_j\mathcal{L}(\hat{Q}_j)\hat{\varrho}_{\mathrm{R}}\Bigr].
\end{equation}
Its solution is obtained directly as subspace projection of (\ref{gensol1}).

\emph{Triplet symmetry:}
Assuming again that (\ref{symmcondmaps1}) and with it (\ref{symmcond}) hold,
(\ref{36neu}) becomes
\begin{equation}
\label{36neus}
  \frac{\partial}{\partial{t}}\hat{\varrho}_{\mathrm{R}}
  =\sum_{j={\mathrm{S}},{\mathrm{T}}}
  \Bigl[-\frac{r_j}{2}\{\hat{Q}_j,\hat{\varrho}_{\mathrm{R}}\}
  +d_j\mathcal{L}(\hat{Q}_j)\hat{\varrho}_{\mathrm{R}}\Bigr]
  +d_{\mathrm{T}}\mathcal{Q}_{\mathrm{coh}}\hat{\varrho}_{\mathrm{R}},
\end{equation}
with its solution following as subspace projection of (\ref{gensol2}).

\emph{Triplet symmetry without triplet dephasing:}
If both (\ref{symmcondmaps1}) and (\ref{symmcondmaps1a}) hold, so that
in addition to (\ref{symmcond}), we have (\ref{symmcond0}), the last term in
(\ref{36neus}) vanishes, which thus reduces to (\ref{genmeR37}) discussed below.
The corresponding solution reduces to (\ref{gensolR}).
\subsection{
\label{secEM}
Encounter model}
Instead of applying a time-independent interaction with the effective 9-level
model environment according to (\ref{HI}), on which the derivation of the master
equation (\ref{30neu}) [and hence the simplified version (\ref{genme}) below] is
based, we now add a stochastic time-dependence by multiplying (\ref{HI}) with a
time-dependent function $f(\tau)$ that reflects the diffusion process. For
simplicity, we set it to a positive constant $\kappa/\mathrm{d}t$ during those
intervals $\mathrm{d}t$ within $(0,t]$ during which an encounter occurs and
$f(\tau)$ is set to zero for other times.
\subsubsection{Encounter maps for perfect detection efficiency}
The detector, with which the measurements on the model environment are
performed, is described by a POVM consisting of the projectors
\begin{eqnarray}
  \hat{\Pi}_j&=&|\pi_j\rangle\langle\pi_j|,\quad
  (j={\mathrm{S}},{\mathrm{T}_i}),
  \\
  \hat{\Pi}_0&=&|0\rangle\langle0|
  +\sum_{j={\mathrm{S}},{\mathrm{T}_i}}|\delta_j\rangle\langle\delta_j|,
\end{eqnarray}
which induce corresponding transformations (\ref{Ai}) of the chemical system
according to
\begin{eqnarray}
\label{Ajm}
  \mathcal{A}_j\hat{\varrho}
  &=&\langle\pi_j|\hat{U}|0\rangle\hat{\varrho}
  \langle0|\hat{U}^\dagger|\pi_j\rangle,
  \\
\label{A0m}
  \mathcal{A}_0\hat{\varrho}
  &=&\langle0|\hat{U}|0\rangle\hat{\varrho}\langle0|\hat{U}^\dagger|0\rangle
  \!+\!\!\!\!\sum_{j={\mathrm{S}},{\mathrm{T}_i}}\!\!
  \langle\delta_j|\hat{U}|0\rangle
  \hat{\varrho}\langle0|\hat{U}^\dagger|\delta_j\rangle.\quad\quad\quad
\end{eqnarray}
Here, $|0\rangle$ is our assumed initial environment state and
$\hat{U}$ $\!=$ $\!\mathrm{e}^{-\mathrm{i}\kappa\hat{H}_{\mathrm{I}}}$ is given
by (\ref{HI}), where in what follows, we omit the index I for simplicity.
The sum
\begin{equation}
  1=\hat{\Pi}_0+\sum_{j={\mathrm{S}},{\mathrm{T}_i}}\hat{\Pi}_j
\end{equation}
then corresponds to the CPT-map
\begin{equation}
\label{ACPT}
  \mathcal{A}_{\mathrm{CPT}}\hat{\varrho}
  =\mathcal{A}_0\hat{\varrho}
  +\sum_{j={\mathrm{S}},{\mathrm{T}_i}}\mathcal{A}_j\hat{\varrho}
  =\mathrm{Tr}_{\mathrm{E}}
  (\hat{U}|0\rangle\hat{\varrho}\langle0|\hat{U}^\dagger),
\end{equation}
that describes the effect of an unmeasured encounter. In order to give explicit
expressions for the $\mathcal{A}_j$ and $\mathcal{A}_0$, we make use of
(\ref{HI}) and write
\begin{eqnarray}
  \hat{H}^{2n}|0\rangle
  &=&\delta_{n0}\hat{Q}_{\mathrm{P}}|0\rangle
  +\sum_{j={\mathrm{S}},{\mathrm{T}_i}}c_j^n\hat{Q}_j|0\rangle,
  \\
  \hat{H}^{2n+1}|0\rangle&=&\sum_{j={\mathrm{S}},{\mathrm{T}_i}}c_j^n
  (\pi_j\hat{L}_j|\pi_j\rangle+\delta_j\hat{Q}_j|\delta_j\rangle),
\end{eqnarray}
where $n$ $\!=$ $\!0,1,2,\ldots$ and
\begin{eqnarray}
  c_j&=&|\pi_j|^2+|\delta_j|^2,
  \\
  \varphi_j&=&\kappa\sqrt{c_j}.
\end{eqnarray}
With this we obtain
\begin{eqnarray}
\label{U0}
  \hat{U}|0\rangle&=&\hat{Q}_{\mathrm{P}}|0\rangle
  +\sum_{j={\mathrm{S}},{\mathrm{T}_i}}
  \bigl[\cos(\varphi_j)\hat{Q}_j|0\rangle
  \nonumber\\
  &&-\mathrm{i}\kappa\,\mathrm{sinc}(\varphi_j)\bigl(\pi_j\hat{L}_j|\pi_j\rangle
  +\delta_j\hat{Q}_j|\delta_j\rangle\bigr)\bigr],
\end{eqnarray}
where $\mathrm{sinc}(x)$ $\!=$ $\!\sin(x)/x$. This gives
\begin{eqnarray}
  \langle0|\hat{U}|0\rangle
  &=&\hat{Q}_{\mathrm{P}}
  +\sum_{j={\mathrm{S}},{\mathrm{T}_i}}\cos(\varphi_j)\hat{Q}_j,
  \\
  \langle\delta_j|\hat{U}|0\rangle
  &=&-\mathrm{i}\kappa\delta_j\,\mathrm{sinc}(\varphi_j)\hat{Q}_j,
  \\
  \langle\pi_j|\hat{U}|0\rangle
  &=&-\mathrm{i}\kappa\pi_j\,\mathrm{sinc}(\varphi_j)\hat{L}_j,
\end{eqnarray}
and we get 
\begin{eqnarray}
\label{Amj1}
  \mathcal{A}_j
  &=&\tilde{r}_j\hat{L}_j\bm{\cdot}\hat{L}_j^\dagger,
  \quad(j={\mathrm{S}},{\mathrm{T}_i}),
  \\
  \mathcal{A}_0
  &=&\mathcal{Q}_{\mathrm{P}}
  +\sum_{j,k={\mathrm{S}},{\mathrm{T}_i}}{c}_{jk}\mathcal{Q}_{jk}
  +\sum_{j={\mathrm{S}},{\mathrm{T}_i}}\tilde{d}_j\mathcal{Q}_j
  \quad\quad
  \nonumber\\
  &=&\mathcal{Q}_{\mathrm{P}}
  +\sum_{j={\mathrm{S}},{\mathrm{T}_i}}(1-\tilde{r}_j)\mathcal{Q}_j
  +\sum_{j,k={\mathrm{S}},{\mathrm{T}_i}}^{(j\neq{k})}{c}_{jk}\mathcal{Q}_{jk}
  \quad\quad
  \nonumber\\
  &=&1-\sum_{j={\mathrm{S}},{\mathrm{T}_i}}\tilde{r}_j\mathcal{Q}_j
  -\sum_{j,k={\mathrm{S}},{\mathrm{T}_i}}^{(j\neq{k})}
  (1-{c}_{jk})\mathcal{Q}_{jk},
\label{Am01}
\end{eqnarray}
where
\begin{eqnarray}
\label{alpha1}
  \tilde{r}_j&=&\kappa^2|\pi_j|^2\,\mathrm{sinc}^2(\varphi_j)
  =\frac{|\pi_j|^2}{|\pi_j|^2+|\delta_j|^2}\sin^2(\varphi_j),\quad
  \\
  \tilde{d}_j&=&\frac{|\delta_j|^2\sin^2(\varphi_j)}{|\pi_j|^2+|\delta_j|^2}
  =\sin^2\varphi_j-\tilde{r}_j,
  \\
  {c}_{jk}&=&\cos\varphi_j\cos\varphi_k,
\label{cjk}
\end{eqnarray}
In (\ref{Am01}) we have used that (\ref{inicon}) and hence (\ref{substjk}) holds
for all times.

\emph{Triplet symmetry:}
If (\ref{symmcondmaps1}) holds, one obtains in analogy to (\ref{symmcond})
\begin{equation}
\label{symmcondmaps2}
  \tilde{r}_{\mathrm{T}_i}=\tilde{r}_{\mathrm{T}},\quad
  \tilde{d}_{\mathrm{T}_i}=\tilde{d}_{\mathrm{T}},\quad
  \varphi_{\mathrm{T}_i}=\varphi_{\mathrm{T}},
\end{equation}
and if we cannot distinguish from which particular triplet state a triplet
fluorescence came, so that
\begin{equation}
\label{PiT}
  \hat{\Pi}_{\mathrm{T}}=\sum_{j={\mathrm{T}_i}}\hat{\Pi}_j
  \quad\Rightarrow\quad
  \mathcal{A}_{\mathrm{T}}=\sum_{j={\mathrm{T}_i}}\mathcal{A}_j,
\end{equation}
we can use (\ref{subst}) to simplify (\ref{Amj1}) and (\ref{Am01}) to
\begin{eqnarray}
\label{Am21}
  \mathcal{A}_j
  &=&\tilde{r}_j\langle\mathcal{Q}_j\rangle
  \hat{Q}_{\mathrm{P}},\quad(j={\mathrm{S}},{\mathrm{T}}),
  \\
\label{Am22}
  \mathcal{A}_0
  &=&(1\!-\!\tilde{\eta})
  \!+\!\tilde{\eta}\Bigl[\mathcal{Q}_{\mathrm{P}}
  +\!\!\!\sum_{j={\mathrm{S}},{\mathrm{T}}}
  (1\!-\!\tilde{\eta}_j)\mathcal{Q}_j\Bigr]
  \!+\!\tilde{d}_{\mathrm{T}}\mathcal{Q}_{\mathrm{coh}},\quad\quad
\end{eqnarray}
where $\mathcal{Q}_{\mathrm{coh}}$ is defined in (\ref{Qcoh}) and
\begin{equation}
\label{beta1}
  \tilde{\eta}_j=\frac{\tilde{r}_j}{\tilde{\eta}},
  \quad
  \tilde{\eta}=1-\cos(\varphi_{\mathrm{S}})\cos(\varphi_{\mathrm{T}}).
  \\
\end{equation}

\emph{Triplet symmetry without triplet dephasing:}
If both (\ref{symmcondmaps1}) and (\ref{symmcondmaps1a}) hold, so that in
addition to (\ref{symmcondmaps2}), we also assume in analogy to
(\ref{symmcond0}) that
\begin{equation}
\label{symmcondmaps0}
  \tilde{d}_{\mathrm{T}}=0,
\end{equation}
then (\ref{Am21}) and (\ref{Am22}) reduce to (\ref{Ajrho}) and (\ref{A0rho})
below, respectively.
\subsubsection{
\label{sec3b}
Encounter maps for imperfect detection efficiency}
In this work, we model finite detection efficiencies according to
\begin{eqnarray}
\label{Pijeta}
  \hat{\Pi}_j^{(\mathrm{D})}
  &=&\eta_j^{(\mathrm{D})}\hat{\Pi}_j,
  \quad(j={\mathrm{S}},{\mathrm{T}_i}),
  \\
\label{Pi0eta}
  \hat{\Pi}_0^{(\mathrm{D})}
  &=&\sum_{j={\mathrm{S}},{\mathrm{T}_i}}
  \bigl(1-\eta_j^{(\mathrm{D})}\bigr)\hat{\Pi}_j
  +\hat{\Pi}_0,
\end{eqnarray}
and analogous for $\mathcal{A}_j^{(\mathrm{D})}$ and
$\mathcal{A}_0^{(\mathrm{D})}$, cf. (\ref{Ai}). This allows for missed
$j$-clicks but disregards the possibility of dark counts.
$\eta_j^{(\mathrm{D})}$ $\!\in$ $\![0,1]$ are the
$j$ $\!=$ $\!{\mathrm{S}},{\mathrm{T}_i}$ detection efficiencies.
Alternatively, we may only refer to singlet and triplet detection without
resolving the triplet states. To do so, we set in analogy to (\ref{PiT})
\begin{equation}
  \hat{\Pi}_{\mathrm{T}}^{(\mathrm{D})}
  =\sum_{j={\mathrm{T}_i}}\hat{\Pi}_j^{(\mathrm{D})},
  \quad
  \eta_{\mathrm{T}_i}^{(\mathrm{D})}=\eta_{\mathrm{T}}^{(\mathrm{D})},
\end{equation}
and in (\ref{Pijeta}) and (\ref{Pi0eta}) we replace $\mathrm{T}_i$ with
$\mathrm{T}$, so that the resulting expressions also refer to
$j$ $\!=$ $\!{\mathrm{S}},{\mathrm{T}}$ singlet and triplet detection
efficiencies.
\subsection{
\label{secCM}
Master equation model as a limiting case of the encounter model}
Identifying $2t$ $\!=$ $\!\kappa^2r$ in the limit $\kappa$ $\!\ll$ $\!1$ reduces
(\ref{alpha1}) and (\ref{cjk}) to (\ref{rjdj}) and (\ref{etajjp}), respectively,
i.e., $r\tilde{r}_j$ $\!\to$ $\!r_j$ and $r(1$ $\!-$ $\!{c}_{jk})$ $\!\to$
$\!\eta_{jk}$. In this case, (\ref{Amj1}) and (\ref{Am01}) yield the weak
encounter limits of the maps as
\begin{eqnarray}
\label{Ajwl}
  r\mathcal{A}_j&\to&r_j\hat{L}_j\bm{\cdot}\hat{L}_j^\dagger,
  \quad(j={\mathrm{S}},{\mathrm{T}_i}),
  \\
\label{A0wl}
  r\mathcal{A}_0&\to&r
  +\sum_{j={\mathrm{S}},{\mathrm{T}_i}}d_j\mathcal{Q}_j
  -\sum_{j,k={\mathrm{S}},{\mathrm{T}_i}}\eta_{jk}\mathcal{Q}_{jk}.
\end{eqnarray}
Applying further $\{\hat{Q}_j,\bm{\cdot}\}$ $\!=$
$\!\sum_{k={\mathrm{S}},{\mathrm{T}_i}}(\mathcal{Q}_{jk}$ $\!+$
$\!\mathcal{Q}_{kj})$, where $j$ $\!=$ $\!{\mathrm{S}},{\mathrm{T}_i}$, we
reproduce (\ref{30neu}) in the sense of
\begin{eqnarray}
  r(\mathcal{A}_{\mathrm{CPT}}-1)&=&r\Bigl(
  \sum_{j={\mathrm{S}},{\mathrm{T}_i}}\mathcal{A}_j+\mathcal{A}_0-1\Bigr)
  \\
  &\to&
\label{lim1}
  \sum_{j={\mathrm{S}},{\mathrm{T}_i}}\bigl[r_j\mathcal{L}(\hat{L}_j)
  +d_j\mathcal{L}(\hat{Q}_j)\bigr].\quad
\end{eqnarray}
The more general case of an imperfect detection is described by
$\mathcal{A}_j^{(\mathrm{D})}$ $\!=$ $\!\eta_j^{(\mathrm{D})}\mathcal{A}_j$,
where $j$ $\!=$ $\!{\mathrm{S}},{\mathrm{T}_i}$, and
$\mathcal{A}_0^{(\mathrm{D})}$ $\!=$ $\!\mathcal{A}_0$ $\!+$
$\!\sum_{j={\mathrm{S}},{\mathrm{T}_i}}
\bigl(1-\eta_j^{(\mathrm{D})}\bigr)\mathcal{A}_j$
$\!=$ $\!\mathcal{A}_{\mathrm{CPT}}$ $\!-$
$\!\sum_{j={\mathrm{S}},{\mathrm{T}_i}}\eta_j^{(\mathrm{D})}\mathcal{A}_j$,
cf. (\ref{Pijeta}), (\ref{Pi0eta}), which gives as generator of an imperfect
dark evolution
\begin{eqnarray}
  \mathcal{L}_{\mathrm{enc}}&=&
  r\bigl(\mathcal{A}_0^{(\mathrm{D})}-1\bigr)
  \\
  &=&r(\mathcal{A}_{\mathrm{CPT}}-1)
  -\sum_{j={\mathrm{S}},{\mathrm{T}_i}}\eta_j^{(\mathrm{D})}r\mathcal{A}_j
  \\
  &\to&
  \!\!\sum_{j={\mathrm{S}},{\mathrm{T}_i}}\Bigl\{r_j\bigl[\mathcal{L}(\hat{L}_j)
  \!-\!\eta_j^{(\mathrm{D})}\hat{L}_j\bm{\cdot}\hat{L}_j^\dagger\bigr]
  \!+\!d_j\mathcal{L}(\hat{Q}_j)\!\Bigr\}
  \quad\quad
  \\
  &=&
  \sum_{j={\mathrm{S}},{\mathrm{T}_i}}r_j
  \Bigl[\bigl(1-\eta_j^{(\mathrm{D})}\bigr)
  \hat{L}_j\bm{\cdot}\hat{L}_j^\dagger-\frac{1}{2}\{\hat{Q}_j,\bm{\cdot}\}
  \Bigr]
  \nonumber\\
  &&+\sum_{j={\mathrm{S}},{\mathrm{T}_i}}d_j\mathcal{L}(\hat{Q}_j).
\label{Lencmel}
\end{eqnarray}
With this, the non-trace preserving but linear equation for the imperfect dark
evolution reads $\frac{\partial}{\partial{t}}\hat{\varrho}_{\mathrm{N}}$ $\!=$
$\!(\mathcal{L}_{\mathrm{betw}}$ $\!+$ $\!\mathcal{L}_{\mathrm{enc}})
\hat{\varrho}_{\mathrm{N}}$. For $\eta_j^{(\mathrm{D})}$ $\!=$ $\!0$ it reduces
to the linear and trace-preserving master equation. The projection of this
equation to the R-subspace,
$\frac{\partial}{\partial{t}}\hat{\varrho}_{\mathrm{N,R}}$ $\!=$
$\!(\mathcal{L}_{\mathrm{betw,R}}$ $\!+$ $\!\mathcal{L}_{\mathrm{enc,R}})
\hat{\varrho}_{\mathrm{N,R}}$ does not depend on $\eta_j^{(\mathrm{D})}$.

\emph{Triplet symmetry:}
If (\ref{symmcondmaps1}) holds, we again introduce (\ref{PiT}), which simplifies
(\ref{Ajwl}) and (\ref{A0wl}) to
\begin{eqnarray}
\label{Ajwls}
  r\mathcal{A}_j&\to&r_j\langle\mathcal{Q}_j\rangle\hat{Q}_{\mathrm{P}},\quad
  (j={\mathrm{S}},{\mathrm{T}}),
  \\
\label{A0wls}
  r\mathcal{A}_0
  &\to&r
  \!+\!\!\!\sum_{j={\mathrm{S}},{\mathrm{T}}}\Bigl[
  d_j\mathcal{L}(\hat{Q}_j)-\frac{r_j}{2}\{\hat{Q}_j,\bm{\cdot}\}\Bigr]
  \!+\!d_{\mathrm{T}}\mathcal{Q}_{\mathrm{coh}}.\quad\quad
\end{eqnarray}

\emph{Triplet symmetry without triplet dephasing:}
If both (\ref{symmcondmaps1}) and (\ref{symmcondmaps1a}) hold, the last term in
(\ref{A0wls}) vanishes, and we obtain
\begin{eqnarray}
  \mathcal{L}_{\mathrm{enc}}&=&
  r\bigl(\mathcal{A}_0^{(\mathrm{D})}-1\bigr)
    \\
  &=&
  r(\mathcal{A}_{\mathrm{CPT}}-1)
  -\sum_{j={\mathrm{S}},{\mathrm{T}}}\eta_j^{(\mathrm{D})}r\mathcal{A}_j
  \\
  &=&r\Bigl[
  \sum_{j={\mathrm{S}},{\mathrm{T}}}\bigl(1-\eta_j^{(\mathrm{D})}\bigr)
  \mathcal{A}_j+\mathcal{A}_0-1\Bigr]
  \\
  &\to&
\label{lim2}
  \sum_{j={\mathrm{S}},{\mathrm{T}}}\Bigl\{r_j
  \bigl[\bigl(1-\eta_j^{(\mathrm{D})}\bigr)
  \langle\hat{Q}_j\rangle\hat{Q}_{\mathrm{P}}
  -\frac{1}{2}\{\hat{Q}_j,\bm{\cdot}\}\bigr]
  \nonumber\\
  &&+d_j\mathcal{L}(\hat{Q}_j)\Bigr\}
\end{eqnarray}
[keeping in mind that $d_{\mathrm{T}}$ $\!=$ $\!0$ and
$\mathcal{L}(\hat{Q}_{\mathrm{S}})$ $\!=$ $\!\mathcal{L}(\hat{Q}_{\mathrm{T}})$,
cf. (\ref{STid})].
While for $\eta_j^{(\mathrm{D})}$ $\!=$ $\!0$ this reduces to (\ref{genme})
below, the projection of (\ref{lim2}) to the R-subspace and restriction to
$d_j$ $\!=$ $\!0$ gives the Haberkorn equation, independent of
$\eta_j^{(\mathrm{D})}$.
\subsection{
Exponential model as a limiting case of the encounter model}
We define the exponential model by restricting to perfect encounters defined by
$\delta_j$ $\!=$ $\!0$ and $\varphi_j$ $\!=$ $\!(k_j+\frac{1}{2})\pi$  with
$j$ $\!=$ $\!{\mathrm{S}},{\mathrm{T}_i}$, for which $\tilde{r}_j$ $\!=$ $\!1$
and $\tilde{d}_j$ $\!=$ $\!{c}_{jk}$ $\!=$ $\!0$, so that
$\mathcal{A}_j$ $\!=$ $\!\hat{L}_j\bm{\cdot}\hat{L}_j^\dagger$ and
$\mathcal{A}_0$ $\!=$ $\!\mathcal{Q}_{\mathrm{P}}$ (cf. Sec.~\ref{sec8.1} below
for a discussion within the simplified case). Since such an encounter leads
with certainty to a RP-recombination, there are no multiple encounters and
consequently no encounter-induced dephasing of the RPs.
An imperfect dark evolution is determined by $\mathcal{A}_0^{(\mathrm{D})}$
$\!=$ $\!\mathcal{Q}_{\mathrm{P}}$ $\!+$
$\!\sum_{j={\mathrm{S}},{\mathrm{T}_i}}\bigl(1-\eta_j^{(\mathrm{D})}\bigr)
\hat{L}_j\bm{\cdot}\hat{L}_j^\dagger$, which gives
\begin{eqnarray}
  \mathcal{L}_{\mathrm{enc}}&=&
  r\bigl(\mathcal{A}_0^{(\mathrm{D})}-1\bigr)
  \\
  &=&r\sum_{j={\mathrm{S}},{\mathrm{T}_i}}
  \Bigl[\bigl(1-\eta_j^{(\mathrm{D})}\bigr)
  \hat{L}_j\bm{\cdot}\hat{L}_j^\dagger-\frac{1}{2}\{\hat{Q}_j,\bm{\cdot}\}
  \Bigr].
\label{Lenceml}
\end{eqnarray}
This is a special case of the master equation limit (\ref{Lencmel}) with
$d_j$ $\!=$ $\!0$ and for which $r_j$ $\!=$ $\!r$ equals the encounter rate.
If we project (\ref{Lencmel}) onto the R-space we obtain
(for $\mathcal{L}_{\mathrm{diss}}$ $\!=$ $\!0$ and $d_j$ $\!=$ $\!0$) an
equation of the Haberkorn type,
\begin{equation}
  \frac{\partial}{\partial{t}}\hat{\varrho}_{\mathrm{R}}
  =-\mathrm{i}[\hat{H},\hat{\varrho}_{\mathrm{R}}]
  -\frac{1}{2}\sum_{j={\mathrm{S}},{\mathrm{T}_i}}r_j
  \{\hat{Q}_j,\hat{\varrho}_{\mathrm{R}}\}.
\end{equation}
If the R-component of the state is pure, then this purity is preserved by the
equation,
$\hat{\varrho}_{\mathrm{R}}$ $\!=$ $\!|\Psi\rangle_{\mathrm{R}}\langle\Psi|$,
with $|\Psi(t)\rangle_{\mathrm{R}}$ $\!=$
$\!\hat{U}_{\mathrm{eff}}(t)|\Psi_0\rangle_{\mathrm{R}}$, where
$\hat{U}_{\mathrm{eff}}$ $\!=$ $\!\mathcal{T}_+\mathrm{e}^{-\mathrm{i}
\int_0^t\mathrm{d}\tau\hat{H}_{\mathrm{eff}}(\tau)}$ is non-unitary and
constructed with $\hat{H}_{\mathrm{eff}}$ $\!=\hat{H}$
$\!-\frac{\mathrm{i}}{2}\sum_{j={\mathrm{S}},{\mathrm{T}_i}}r_j\hat{Q}_j$.
For a pure initial state
$\hat{\varrho}(0)$ $\!=$ $\!|\Psi_0\rangle\langle\Psi_0|$ and a perfect dark
evolution (for which $\eta_j^{(\mathrm{D})}$ $\!=$ $\!1$), the normalized
physical state $|\Psi(t)\rangle$ in fact coincides with the stochastic wave
function \cite{Ple98,*Bru00,*bookBreuer}, which is in the general case only
a mathematical construction. Perfect observation is required to determine the
moment of the R$\to$P transition (jump). Moreover, since in the exponential
model limit $r_j$ $\!=$ $\!r$ holds, the normalized state evolves here solely
due to $\hat{H}$, until a fluorescence click eventually detects the jump
(assuming that $r$ $\!>$ $\!0$). The waiting time distribution of the jump is
obtained from the state norm just as with the stochastic wave function method.
\subsection{
Number of independent parameters}
The model interaction (\ref{HI}) to our 9-level environment is determined by the
complex coefficients $\pi_j$ and $\delta_j$, as well as the interaction time $t$
in case of the master equation model, or the coefficient $\kappa$ in case of the
encounter model. There are however only 8 real free parameters entering as
$2t|\pi_j|^2$ and $2t|\delta_j|^2$ the master equation model in (\ref{rjdj}), or
entering as $\kappa^2|\pi_j|^2$ and $\kappa^2|\delta_j|^2$ the encounter model
in (\ref{alpha1}) -- (\ref{cjk}), where in both cases
$j$ $\!=$ $\!\mathrm{S},\mathrm{T}_0\mathrm{T}_+,\mathrm{T}_-$.

\emph{Triplet symmetry without triplet dephasing:}
If (\ref{symmcondmaps1}) and (\ref{symmcondmaps1a}) hold, both models depend on
the three real parameters $|\pi_{\mathrm{S}}|^2$, $|\pi_{\mathrm{T}}|^2$, and
$|\delta_{\mathrm{S}}|^2$ (scaled with $2t$ or $\kappa^2$, respectively). The
model environment is in this case reduced to the $9$ $\!-$ $\!3$ $\!=$ $\!6$
states $\{|0\rangle,|\pi_j\rangle,|\delta_{\mathrm{S}}\rangle\}$, where
$j$ $\!=$ $\!\mathrm{S},\mathrm{T}_i$.

In the master equation model, the free parameters determine in turn
$r_{\mathrm{S}}$, $r_{\mathrm{T}}$, and $d_{\mathrm{S}}$ given by
\begin{equation}
\label{p1}
  r_j=2t|\pi_j|^2,\quad d_j=2t|\delta_j|^2,\quad(j={\mathrm{S}},{\mathrm{T}}).
\end{equation}
In the encounter model, they determine $\tilde{r}_{\mathrm{S}}$,
$\tilde{r}_{\mathrm{T}}$, and $\tilde{\eta}$ given by (\ref{alpha1}) and
(\ref{beta1}), i.e.,
\begin{eqnarray}
\label{p2}
  \tilde{r}_j&=&
  \kappa^2|\pi_j|^2\mathrm{sinc}^2(\varphi_j),\quad
  (j={\mathrm{S}},{\mathrm{T}}),
  \\
\label{p3}
  \tilde{\eta}&=&1-\cos(\varphi_{\mathrm{S}})\cos(\varphi_{\mathrm{T}}),
\end{eqnarray}
where $\varphi_j$ $\!=$ $\!\kappa\sqrt{|\pi_j|^2+|\delta_j|^2}$. 
Remember that $\delta_{\mathrm{T}}$ $\!=$ $\!0$ in both models.
(Note that setting $\delta_{\mathrm{T}}$ $\!\neq$ $\!0$ in
(\ref{p1}) -- (\ref{p3}) is legitimate in the description of an effective
S -- T$_0$ -- qubit, which corresponds to the case of a sufficiently high
external magnetic field such that the T$_\pm$ states remain unpopulated. If we
want to restrict to this case right from the start, we can limit to a model
environment consisting of the 5 states
$\{|0\rangle,|\pi_j\rangle,|\delta_j\rangle\}$, where
$j$ $\!=$ $\!\mathrm{S},\mathrm{T}_0$.)
\section{
\label{sec4}
Master equation model reconsidered}
We now focus on the reaction operators as obtained with the master equation
model in the simplified case of triplet symmetry without triplet dephasing.
They generate transitions from the RP to the reaction product subspace of
the chemical system. We express the equations both with and without including
the reaction product to facilitate comparison with the literature. A solution of
the equations is straightforward if any unitary evolution (due to, e.g.,
coupling with local spin baths) is ignored.
\subsection{
\label{sec4.1}
Decay and dephasing combined}
\subsubsection{Equation and its solution in full space}
Our starting point is a Lindblad-type equation
\begin{equation}
\label{genme}
  \frac{\partial}{\partial{t}}\hat{\varrho}
  =\sum_{j={\mathrm{S}},{\mathrm{T}}}\Bigl[r_j
  \bigl(\langle\hat{Q}_j\rangle\hat{Q}_{\mathrm{P}}
  -\frac{1}{2}\{\hat{Q}_j,\hat{\varrho}\}\bigr)
  +d_j\mathcal{L}(\hat{Q}_j)\hat{\varrho}\Bigr],
\end{equation}
which is obtained from (\ref{30neu}) under the additional assumptions
(\ref{symmcond}) and (\ref{symmcond0}) as described in the previous section.
The $r_j$ and $d_j$ are arbitrary positive constants, that describe decay and
dephasing, respectively, and are related to the coupling coefficients with our
model environment via (\ref{p1}). $\langle\hat{Q}_j\rangle$ is defined in
(\ref{avQ}). The $\hat{Q}_{\mathrm{P}}$ and $\hat{Q}_j$ are projectors onto the
reaction product and singlet/triplet ($j$ $\!=$ $\!{\mathrm{S}},{\mathrm{T}}$)
RP spaces, and the Lindblad operator $\mathcal{L}(\hat{c})$ is defined
in (\ref{Lc}). Remember the interchangeability of the dephasing terms as
mentioned in the comments following (\ref{symmcond0}).

The terms $\langle\hat{Q}_j\rangle\hat{Q}_{\mathrm{P}}$ describe transitions
(``jumps'') to the reaction product (P) space. While singlet transitions
$\langle\hat{Q}_{\mathrm{S}}\rangle\hat{Q}_{\mathrm{P}}$ $\!=$
$\!\hat{L}_{\mathrm{S}}\hat{\varrho}\hat{L}_{\mathrm{S}}^\dagger$ can be
described in terms of a corresponding jump operator $\hat{L}_{\mathrm{S}}$, cf.
(\ref{Lj}), an analogous substitution for $j$ $\!=$ $\!\mathrm{T}$ requires
limitation to a given triplet state, e.g. $|\mathrm{T}_0\rangle$. (Note that a
symmetrized construction $\hat{L}_{\mathrm{T}}$ $\!=$
$\!\sum_{j=\mathrm{T}_i}\hat{L}_j$ would only describe transitions from a single
given state $|\mathrm{T}\rangle$ $\!=$
$\!\frac{1}{\sqrt{3}}\sum_{j=\mathrm{T}_i}|\mathrm{T}_j\rangle$.)

Keeping in mind (\ref{inicon}), the solution of (\ref{genme}) can be written as
\begin{eqnarray}
\label{gensol}
  &&\hat{\varrho}(t)=
  \hat{\varrho}_{0\mathrm{P}}+
  \mathrm{e}^{-\eta{t}}\hat{\varrho}_{0\mathrm{R}}
  \\
  &&+\!\!\!\sum_{j={\mathrm{S}},{\mathrm{T}}}\!\Bigl[
  (\mathrm{e}^{-r_jt}\!-\!\mathrm{e}^{-\eta{t}})
  \hat{Q}_j\hat{\varrho}_0\hat{Q}_j
  \!+\!(1\!-\!\mathrm{e}^{-r_jt})
  \mathrm{Tr}(\hat{\varrho}_0\hat{Q}_j)\hat{Q}_{\mathrm{P}}\Bigr],
  \nonumber
\end{eqnarray}
where 
\begin{equation}
\label{coeff}
  \eta=r+d,\quad
  r=(r_{\mathrm{S}}+r_{\mathrm{T}})/2,\quad
  d=(d_{\mathrm{S}}+d_{\mathrm{T}})/2.
\end{equation}
Since (\ref{genme}) is a simplified version of (\ref{30neu}) [and
(\ref{30neus})], Eq.~(\ref{gensol}) is a simplified version of (\ref{gensol1})
[and (\ref{gensol2})]. In particular, for a singlet or triplet initial state,
$\hat{\varrho}_0$ $\!=$ $\!\hat{Q}_j$, (\ref{gensol}) becomes independent of the
dephasing terms,
\begin{equation}
  \hat{\varrho}(t)=
  \mathrm{e}^{-r_jt}\hat{Q}_j
  +(1-\mathrm{e}^{-r_jt})\hat{Q}_{\mathrm{P}}.
\end{equation}

For $r_j$ $\!\neq$ $\!0$, (\ref{gensol}) evolves into the steady state
$\hat{\varrho}(\infty)$ $\!=$ $\!\hat{Q}_{\mathrm{P}}$. In particular, for 
equal reaction rates, $r_j$ $\!=$ $\!{r}$,
(\ref{gensol}) becomes
\begin{eqnarray}
  \hat{\varrho}(t)&=&
  \mathrm{e}^{-{r}t}
  \bigl[\hat{\varrho}_{0\mathrm{P}}
  +\mathrm{e}^{-{d}t}\hat{\varrho}_{0\mathrm{R}}
  \nonumber\\
  &&+(1-\mathrm{e}^{-{d}t})
  \!\!\!\sum_{j={\mathrm{S}},{\mathrm{T}}}
  \!\hat{Q}_j\hat{\varrho}_0\hat{Q}_j\bigr]
  +(1-\mathrm{e}^{-{r}t})\hat{Q}_{\mathrm{P}}.\quad\quad
\label{symsol}
\end{eqnarray}
\subsubsection{Reduced equation and its solution in the R-subspace}
Before we consider special cases of (\ref{genme}), we mention that
we may project (\ref{genme}) onto the R-subspace, which gives an equation for
the RP component $\hat{\varrho}_{\mathrm{R}}$ of the state
$\hat{\varrho}$ of the chemical system,
\begin{eqnarray}
\label{genmeR37}
  \frac{\partial}{\partial{t}}\hat{\varrho}_{\mathrm{R}}
  &=&\sum_{j={\mathrm{S}},{\mathrm{T}}}
  \Bigl[-\frac{r_j}{2}\{\hat{Q}_j,\hat{\varrho}_{\mathrm{R}}\}
  +d_j\mathcal{L}(\hat{Q}_j)\hat{\varrho}_{\mathrm{R}}\Bigr]
  \\
\label{genmeR}
  &=&\!\sum_{j={\mathrm{S}},{\mathrm{T}}}
  \gamma_j\bigl(p_j\hat{Q}_j\hat{\varrho}_{\mathrm{R}}\hat{Q}_j
  -\frac{1}{2}\{\hat{Q}_j,\hat{\varrho}_{\mathrm{R}}\}\bigr)
  \\
\label{genmeRan}
  \!&=&\!\eta\bigl[(1\!-\!\eta_{\mathrm{S}})
  \hat{Q}_{\mathrm{S}}\hat{\varrho}_{\mathrm{R}}\hat{Q}_{\mathrm{S}}
  \!+\!(1\!-\!\eta_{\mathrm{T}})
  \hat{Q}_{\mathrm{T}}\hat{\varrho}_{\mathrm{R}}\hat{Q}_{\mathrm{T}}
  \!-\!\!\!\hat{\varrho}_{\mathrm{R}}\bigr],\quad\quad
\end{eqnarray}
where
\begin{equation}
  \gamma_j=r_j+d_j,\quad
  p_j=d_j/\gamma_j,\quad
  \eta_j=r_j/\eta,
\end{equation}
so that $0$ $\!\le$ $\!p_j$ $\!\le$ $\!1$ in agreement with \cite{Hab76}.
Its solution follows immediately from (\ref{gensol}),
\begin{equation}
\label{gensolR}
  \hat{\varrho}_{\mathrm{R}}(t)=
  \mathrm{e}^{-\eta{t}}\hat{\varrho}_{0\mathrm{R}}
  \!+\!\!\!\sum_{j={\mathrm{S}},{\mathrm{T}}}
  [\mathrm{e}^{-(1-p_j)\gamma_jt}\!-\!\mathrm{e}^{-\eta{t}}]
  \hat{Q}_j\hat{\varrho}_{0\mathrm{R}}\hat{Q}_j.
\end{equation}
\subsection{
\label{sec4.2}
Special cases of interest}
In the general case, an encounter causes a reaction, which can in turn result in
dephasing and/or recombination (``decay'' and ``relaxation'' are used as
synonyms for the latter).
Our assumption of a simultaneous action of relaxation and dephasing terms
\cite{Shu10,Tie12} is in agreement with the special cases of pure decay
(Haberkorn model \cite{Hab76}), pure dephasing (which coincides with the
non-reaction term considered by Kominis \cite{Kom09,Kom11}), and balanced decay
and dephasing (Jones-Hore model \cite{Jon10,Jon11a}). In our notation, these
cases can be recovered as follows:
\subsubsection{Pure decay}
The first and second term on the right hand side of (\ref{genme}) describe decay
and dephasing, respectively. Their ratio determines the
coefficients $p_j$ in (\ref{genmeR}) in the R-subspace. The limiting
case of pure decay, i.e., absence of dephasing, $d_j$ $\!=$ $\!0$, leads to
$p_j$ $\!=$ $\!0$, so that (\ref{genmeR}) reduces to
\begin{equation}
  \frac{\partial}{\partial{t}}\hat{\varrho}_{\mathrm{R}}
  =-\frac{1}{2}\sum_{j={\mathrm{S}},{\mathrm{T}}}
  r_j\{\hat{Q}_j,\hat{\varrho}_{\mathrm{R}}\},
\end{equation}
and (\ref{gensolR}) can be written as
\begin{eqnarray}
\label{gensolh}
  \hat{\varrho}_{\mathrm{R}}(t)
  &=&\hat{U}_{\mathrm{eff}}(t)\hat{\varrho}_{0\mathrm{R}}
  \hat{U}_{\mathrm{eff}}^\dagger(t),
  \\
  \hat{U}_{\mathrm{eff}}(t)
  &=&\mathrm{e}^{-\frac{r_{\mathrm{S}}}{2}t}\hat{Q}_{\mathrm{S}}
  +\mathrm{e}^{-\frac{r_{\mathrm{T}}}{2}t}\hat{Q}_{\mathrm{T}}.
\end{eqnarray}
An initial state whose projection onto the R-subspace leads to a pure
non-normalized state $|\Psi_0\rangle_{\mathrm{R}}$ will hence keep its purity
during evolution.
We can then alternatively write
$\frac{\partial}{\partial{t}}|\Psi\rangle_{\mathrm{R}}$ $\!=$
$\!-\mathrm{i}\hat{H}_{\mathrm{eff}}|\Psi\rangle_{\mathrm{R}}$, with
$\hat{H}_{\mathrm{eff}}$ $\!=$
$\!-\frac{\mathrm{i}}{2}\sum_{j={\mathrm{S}},{\mathrm{T}}}r_j\hat{Q}_j$
and $|\Psi(t)\rangle_{\mathrm{R}}$ $\!=$
$\!\hat{U}_{\mathrm{eff}}(t)|\Psi_0\rangle_{\mathrm{R}}$.
Note that since $\hat{H}_{\mathrm{eff}}$ is non-Hermitian,
$\hat{U}_{\mathrm{eff}}$ $\!=$
$\!\mathrm{e}^{-\mathrm{i}\hat{H}_{\mathrm{eff}}t}$ is non-unitary, cf.
App.~\ref{secNHC}.

As in (\ref{symsol}), we can consider the symmetric case $r_j$ $\!=$ $\!{r}$,
for which (\ref{genme}) and (\ref{symsol}) reduce to
\begin{eqnarray}
  \frac{\partial}{\partial{t}}\hat{\varrho}
  &=&-{r}\bigl(\hat{\varrho}-\hat{Q}_{\mathrm{P}}\bigr),
  \\
  \hat{\varrho}(t)&=&
  \mathrm{e}^{-{r}t}\hat{\varrho}_0+(1-\mathrm{e}^{-{r}t})
  \hat{Q}_{\mathrm{P}},
\end{eqnarray}
and hence
\begin{eqnarray}
  \frac{\partial}{\partial{t}}\hat{\varrho}_{\mathrm{R}}
  &=&-{r}\hat{\varrho}_{\mathrm{R}},
  \\
  \hat{\varrho}_{\mathrm{R}}(t)&=&
  \mathrm{e}^{-{r}t}\hat{\varrho}_{0\mathrm{R}},
\end{eqnarray}
or $|\Psi(t)\rangle_{\mathrm{R}}$ $\!=$
$\!\mathrm{e}^{-\frac{r}{2}t}|\Psi_0\rangle_{\mathrm{R}}$ for a pure state.
Obviously, this evolution commutes with any additional unitary evolution in the
R-subspace.
\subsubsection{Pure dephasing}
In the absence of recombination, $r_j$ $\!=$ $\!0$, there is only a dephasing
caused by the encounters of the RPs. The trace of $\hat{\varrho}_{\mathrm{R}}$
thus remains preserved, while the state evolves into a mixture of singlet and
triplet states. Therefore
\begin{eqnarray}
\label{49}
 \frac{\partial}{\partial{t}}\hat{\varrho}
  &=&2d\bigl(\hat{Q}_j\hat{\varrho}\hat{Q}_j
  -\frac{1}{2}\{\hat{Q}_j,\hat{\varrho}\}\bigr),
  \\
  \hat{\varrho}(t)&=&
  \!\hat{\varrho}_{0\mathrm{P}}\!+\!
  \mathrm{e}^{-{d}t}\hat{\varrho}_{0\mathrm{R}}
  \!+\!(1\!-\!\mathrm{e}^{-{d}t})
  \!\!\!\sum_{j={\mathrm{S}},{\mathrm{T}}}\!\hat{Q}_j\hat{\varrho}_0\hat{Q}_j,
  \quad
\label{sold}
\end{eqnarray}
where $\hat{Q}_j$ in (\ref{49}) can either stand for $\hat{Q}_{\mathrm{S}}$ or
$\hat{Q}_{\mathrm{T}}$, cf. (\ref{STid}), and $d$ is given in (\ref{coeff}).
The preservation of trace of $\hat{\varrho}_{\mathrm{R}}$ can also be seen from
(\ref{genmeR}) by considering the limiting case $p_j$ $\!=$ $\!1$.
\subsubsection{Balanced decay and dephasing}
A third special case of interest refers to identical rates of dephasing and
decay (or reaction and recombination, respectively),
$d_j$ $\!=$ $\!r_j$. Therefore, (\ref{genme}) can equivalently be written as
[using (\ref{inicon})]
\begin{eqnarray}
  \frac{\partial}{\partial{t}}\hat{\varrho}
  &=&-r_{\mathrm{S}}(\hat{\varrho}
  -\hat{Q}_{\mathrm{T}}\hat{\varrho}\hat{Q}_{\mathrm{T}}
  -\hat{Q}_{\mathrm{P}}\hat{\varrho}\hat{Q}_{\mathrm{P}}
  -\langle\hat{Q}_{\mathrm{S}}\rangle\hat{Q}_{\mathrm{P}})
  \nonumber\\
  &&-r_{\mathrm{T}}(\hat{\varrho}
  -\hat{Q}_{\mathrm{S}}\hat{\varrho}\hat{Q}_{\mathrm{S}}
  -\hat{Q}_{\mathrm{P}}\hat{\varrho}\hat{Q}_{\mathrm{P}}
  -\langle\hat{Q}_{\mathrm{T}}\rangle\hat{Q}_{\mathrm{P}}),\quad
\end{eqnarray}
so that
\begin{eqnarray}
  \frac{\partial}{\partial{t}}\hat{\varrho}_{\mathrm{R}}
  &=&-r_{\mathrm{S}}(\hat{\varrho}_{\mathrm{R}}
  -\hat{Q}_{\mathrm{T}}\hat{\varrho}_{\mathrm{R}}\hat{Q}_{\mathrm{T}})
  -r_{\mathrm{T}}(\hat{\varrho}_{\mathrm{R}}
  -\hat{Q}_{\mathrm{S}}\hat{\varrho}_{\mathrm{R}}\hat{Q}_{\mathrm{S}})
  \quad\quad
  \\
  &=&-\sum_{j={\mathrm{S}},{\mathrm{T}}}
  r_j(\{\hat{Q}_j,\hat{\varrho}_{\mathrm{R}}\}
  -\hat{Q}_j\hat{\varrho}_{\mathrm{R}}\hat{Q}_j).
\end{eqnarray}
The right hand side is hence the sum of the respective terms describing
pure dephasing and pure decay.
\subsection{
\label{sec4.3}
Comparison with the model of Kominis}
In what follows, we compare our approach including the models of Haberkorn,
Jones, and Hore, with that of Kominis. We will thereby reaffirm that the latter
approach differs from the others (which are essentially extensions of the
original model of Haberkorn that include additional decoherence), and explain
why we cannot support the latter.

We first summarize our results. All theories are of the form
\begin{equation}
\label{genf}
  \frac{\partial}{\partial{t}}\hat{\varrho}_{\mathrm{R}}
  =-\mathrm{i}[\hat{H},\hat{\varrho}_{\mathrm{R}}]
  +\mathcal{L}\hat{\varrho}_{\mathrm{R}},
\end{equation}
where $\hat{\varrho}_{\mathrm{R}}$ is the state of the RPs and nuclear
spins ($\mathrm{Tr}\hat{\varrho}_{\mathrm{R}}\le1$) and $\hat{H}$ includes the
coupling between the former and the latter. Our model is described by
(\ref{genmeR37}), which suggests that the reaction operator
$\mathcal{L}$ is the sum of a relaxation term $\mathcal{L}_{\mathrm{rel}}$ that
describes recombination of the RPs, and a dephasing term
$\mathcal{L}_{\mathrm{dep}}$ that generates a transition of the RP
state from a coherent superposition to an incoherent mixture of singlet and
triplet components,
\begin{eqnarray}
\label{cg}
  \mathcal{L}
  &=&\mathcal{L}_{\mathrm{rel}}(r_j)+\mathcal{L}_{\mathrm{dep}}(d_j),
  \\
\label{cr}
  \mathcal{L}_{\mathrm{rel}}(r_j)\hat{\varrho}_{\mathrm{R}}
  &=&-\frac{1}{2}\sum_{j={\mathrm{S}},{\mathrm{T}}}r_j
  \{\hat{Q}_j,\hat{\varrho}_{\mathrm{R}}\},
  \\
\label{cd}
  \mathcal{L}_{\mathrm{dep}}(d_j)\hat{\varrho}_{\mathrm{R}}
  &=&\sum_{j={\mathrm{S}},{\mathrm{T}}}d_j
  \bigl(\hat{Q}_j\hat{\varrho}_{\mathrm{R}}\hat{Q}_j
  -\frac{1}{2}\{\hat{Q}_j,\hat{\varrho}_{\mathrm{R}}\}\bigr).\quad
\end{eqnarray}
Here, $r_{\mathrm{S}}$, $r_{\mathrm{T}}$ and $d_{\mathrm{S}}$, $d_{\mathrm{T}}$
are the corresponding singlet and triplet relaxation and dephasing rates.
For $d_j$ $\!=$ $\!0$ we obtain pure relaxation (Haberkorn model),
$r_j$ $\!=$ $\!d_j$ gives balanced relaxation and dephasing (Jones-Hore model),
while for $r_j$ $\!=$ $\!0$ we obtain pure dephasing. The mentioned models are
therefore consistent with each other. If we assume that the exact values of the
rates depend on the experimental details, in general no precise realization of
any of these specific cases can be expected. Note that (\ref{cd}) is a Lindblad
operator and (\ref{cr}) is a remnant of a Lindblad operator left by leaving out
the reaction products, cf. (\ref{genme}).

An alternative model has been put forward by Kominis in \cite{Kom09,Kom11}.
There, instead of (\ref{cg}), the following expressions have been proposed,
\begin{eqnarray}
\label{kg}
  \mathcal{L}_{\mathrm{Kom}}
  &=&\mathcal{L}_{\mathrm{nr}}(k_j)+\mathcal{L}_{\mathrm{r}}(k_j),
  \\
\label{knr}
  \mathcal{L}_{\mathrm{nr}}&=&\mathcal{L}_{\mathrm{dep}}(k_j),
  \\
\label{kr}
  \mathcal{L}_{\mathrm{r}}&=&(1-p_{\mathrm{coh}})
  \mathcal{L}_{\mathrm{inc}}+p_{\mathrm{coh}}\mathcal{L}_{\mathrm{coh}},
\end{eqnarray}
with \emph{ad hoc} terms
\begin{eqnarray}
  \mathcal{L}_{\mathrm{inc}}(k_j)\hat{\varrho}_{\mathrm{R}}
  &=&-\sum_{j={\mathrm{S}},{\mathrm{T}}}
  k_j\hat{Q}_j\hat{\varrho}_{\mathrm{R}}\hat{Q}_j
  \nonumber\\
\label{kinc}
  &\equiv&-\sum_{j={\mathrm{S}},{\mathrm{T}}}k_j
  \mathcal{Q}_j\hat{\varrho}_{\mathrm{R}},
  \\
  \mathcal{L}_{\mathrm{coh}}(k_j)\hat{\varrho}_{\mathrm{R}}
  &=&-\sum_{j={\mathrm{S}},{\mathrm{T}}}k_j\mathrm{Tr}\Bigl[\hat{Q}_j
  \frac{\hat{\varrho}_{\mathrm{R}}}{\mathrm{Tr}(\hat{\varrho}_{\mathrm{R}})}
  \Bigr]\hat{\varrho}_{\mathrm{R}}
  \nonumber\\
\label{kcoh}
  &\equiv&-\sum_{j={\mathrm{S}},{\mathrm{T}}}k_j
  \langle\mathcal{Q}_j\rangle_{\scriptscriptstyle{\hat{\tilde{\varrho}}}}
  \hat{\varrho}_{\mathrm{R}},
  \\
\label{pcoh}
  p_{\mathrm{coh}}
  &=&\frac{\mathrm{Tr}[(\hat{Q}_{\mathrm{S}}\hat{\varrho}_{\mathrm{R}})
  (\hat{Q}_{\mathrm{T}}\hat{\varrho}_{\mathrm{R}})]}
  {\mathrm{Tr}(\hat{Q}_{\mathrm{S}}\hat{\varrho}_{\mathrm{R}})
  \mathrm{Tr}(\hat{Q}_{\mathrm{T}}\hat{\varrho}_{\mathrm{R}})},
\end{eqnarray}
where in (\ref{kinc}) and (\ref{kcoh}) we have simplified the notation, defining
$\langle\mathcal{Q}_j\rangle_{\scriptscriptstyle{\hat{\tilde{\varrho}}}}$
$\!\equiv$ $\!\mathrm{Tr}(\hat{Q}_j\hat{\tilde{\varrho}})$ with
$\hat{\tilde{\varrho}}$ $\!=$
$\!\frac{\hat{\varrho}_{\mathrm{R}}}{\mathrm{Tr}(\hat{\varrho}_{\mathrm{R}})}$
for convenience.

The controversy on the proper reaction operators can be summarized as follows.
In \cite{Kom11c} it was claimed that the Jones-Hore theory \cite{Jon10} (and
with it (\ref{cg})) rests on assumptions built into the theory ``by hand'' and
leads to ``ambiguous conclusions'' for the state of the unrecombined RPs
(e.g. Eqs.~(2) and (3) in \cite{Kom11c}). Subsequently, it had been shown in
\cite{Jon11} that this asserted ambiguity is due to a wrong interpretation of
the improper RP-density matrix in \cite{Kom11c} which omits the possibility that
the RP is converted to a reaction product, and that Eq.~(2) but not
Eq.~(3) in \cite{Kom11c} is correct. In a later response \cite{Kom11m}, it
was claimed that the master equation for the state of unrecombined RPs
(Eq.~(2) in \cite{Kom11c}) cannot be applied to a single RP and that it is
``problematic'' because it is nonlinear.

In light of this discussion, one may apply the above claims also to (\ref{kg}):
it rests on the factor (\ref{pcoh}) built into the theory by hand without a
derivation, and this leads to ambiguous results, because one could equally well
use, e.g., $p_{\mathrm{coh}}^2$ with analogous properties (see appendix of
\cite{Kom11}) instead of $p_{\mathrm{coh}}$. We thus suggest to omit
$p_{\mathrm{coh}}$ completely and replace (\ref{kr}) with
\begin{eqnarray}
  \mathcal{L}_{\mathrm{r}}&=&
  \mathcal{L}_{\mathrm{inc}}+\mathcal{L}_{\mathrm{coh}},
  \\
  \mathcal{L}_{\mathrm{inc}}(k_j)\hat{\varrho}_{\mathrm{R}}
  &=&-\sum_{j={\mathrm{S}},{\mathrm{T}}}k_j
  \hat{Q}_j\hat{\varrho}_{\mathrm{R}}\hat{Q}_j,
  \\
  \mathcal{L}_{\mathrm{coh}}(k_j)\hat{\varrho}_{\mathrm{R}}
  &=&-\frac{k_{\mathrm{S}}+k_{\mathrm{T}}}{2}
  (\hat{Q}_{\mathrm{S}}\hat{\varrho}_{\mathrm{R}}\hat{Q}_{\mathrm{T}}
  +\hat{Q}_{\mathrm{T}}\hat{\varrho}_{\mathrm{R}}\hat{Q}_{\mathrm{S}}),\quad
\end{eqnarray}
which still carries the spirit of (\ref{kr}), but is linear and coincides with
$\mathcal{L}_{\mathrm{rel}}(k_j)$ defined in (\ref{cr}).
Why is linearity important? Linearity is a requirement of a master equation
which does not refer to any condition (such as absence of recombination). The
theories of Haberkorn and Jones-Hore, and consequently (\ref{cg}) are linear,
and since (\ref{kg}) is meant to describe the same physical problem, it should
be linear too.

There is an alternative way of comparing (\ref{cg}) with (\ref{kg}). To begin
with, there are many different ways to decompose a given $\mathcal{L}$. Let us
rewrite (\ref{kg}) in order to separate the term with $p_{\mathrm{coh}}$ and
compare it with $\mathcal{L}_{\mathrm{dep}}$ in (\ref{cg}). We thus write
\begin{eqnarray}
\label{c1}
  \mathcal{L}&=&\mathcal{L}_{\mathrm{rel}}(r_j)
  +\sum_{j={\mathrm{S}},{\mathrm{T}}}d_j
  \bigl(\mathcal{Q}_j-\frac{1}{2}\{\hat{Q}_j,\bm{\cdot}\}\bigr),
  \\
\label{c2}
  \mathcal{L}_{\mathrm{Kom}}&=&\mathcal{L}_{\mathrm{rel}}(k_j)
  +p_{\mathrm{coh}}\Bigl[\sum_{j={\mathrm{S}},{\mathrm{T}}}k_j
  \bigl(\mathcal{Q}_j
  -\langle\mathcal{Q}_j\rangle_{\scriptscriptstyle{\hat{\tilde{\varrho}}}}\bigr)
  \Bigr],\quad
\end{eqnarray}
where the dot ($\bm{\cdot}$) in (\ref{c1}) marks the location where
$\hat{\varrho}_{\mathrm{R}}$ must be inserted. We can observe the following
facts: $\mathcal{L}_{\mathrm{nr}}(k_j)$ by itself in (\ref{kg}) corresponds to
$\mathcal{L}_{\mathrm{dep}}(d_j)$ in (\ref{cg}) if we identify
$d_j$ $\!=$ $\!k_j$. 
On the other hand, if we compare (\ref{c1}) with (\ref{c2}), i.e., with Kominis'
reaction and non-reaction terms combined, we see that it is now the relaxation
terms which coincide if we identify $r_j$ $\!=$ $\!k_j$.

The remaining terms in (\ref{c1}) and (\ref{c2}) are both of the form
$a(\mathcal{Q}_j+b)$, but they do not coincide, even if we set
$d_j$ $\!=$ $\!k_j$ (which would correspond to the Jones-Hore case).
The term in (\ref{c1}) is the Lindbladian $\mathcal{L}_{\mathrm{dep}}(d_j)$
representing the dephasing component in (\ref{cg}). It is
\emph{linear, trace-free, and vanishes if applied to an incoherent mixture} of
singlet and triplet components (i.e., to the already dephased steady state
$\hat{\varrho}_{\mathrm{R}}$ of pure dephasing,
$\mathcal{L}_{\mathrm{dep}}\hat{\varrho}_{\mathrm{R}}$ $\!=$ $\!0$).
If recombination is accounted for by the first term $\mathcal{L}_{\mathrm{rel}}$
in (\ref{c2}), then the second term in (\ref{c2}) must emulate the above
properties. It indeed subtracts
$\langle\mathcal{Q}_j\rangle_{\scriptscriptstyle{\hat{\tilde{\varrho}}}}$ to
make it trace-free [this fixes $b$ in $a(\mathcal{Q}_j+b)$], and multiplies it
with a coherence factor $p_{\mathrm{coh}}$ to make it disappear if applied to an
incoherent mixture (for which $p_{\mathrm{coh}}$ $\!=$ $\!0$)
[this fixes $a$ in $a(\mathcal{Q}_j+b)$].
However, this comes at the price of introducing two nonlinear terms, namely the
visibility factor $p_{\mathrm{coh}}$ that has been suggested heuristically,
invoking a comparison to the double slit interference, and the term
$\langle\mathcal{Q}_j\rangle_{\scriptscriptstyle{\hat{\tilde{\varrho}}}}$.

Here, we repeat, based on general considerations as described in
Sec.~\ref{sec2}, that, unless an equation refers to a specific record of
measurement results, the master equation should be linear. This condition is met
by the Lindblad term in (\ref{c1}), which also complies with the most general
form of a generator of a linear, trace and positivity preserving state
evolution. It is not met by (\ref{c2}). On these grounds, we can again rule out
the second term in (\ref{c2}) in agreement with \cite{Jon11}.
In the next section we demonstrate the correctness of $\mathcal{L}$ given in
(\ref{cg}) from a completely different consideration.

Let us now return to the description of the evolution of an \emph{unrecombined}
RP, which is conditional and consequently described by an equation obtained by
formulating (\ref{genf}) in a trace-preserving non-linear form,
\begin{eqnarray}
\label{genfn}
  \frac{\partial}{\partial{t}}\hat{\tilde{\varrho}}
  &=&-\mathrm{i}[\hat{H},\hat{\tilde{\varrho}}]
  +(\mathcal{L}-\langle\mathcal{L}\rangle)\hat{\tilde{\varrho}},
  \\
  \langle\mathcal{L}\rangle&=&\mathrm{Tr}(\mathcal{L}\hat{\tilde{\varrho}})
  =-\sum_{j={\mathrm{S}},{\mathrm{T}}}r_j
  \langle\mathcal{Q}_j\rangle_{\scriptscriptstyle{\hat{\tilde{\varrho}}}},
\end{eqnarray}
where $\hat{\tilde{\varrho}}$ $\!=$
$\!\frac{\hat{\varrho}_{\mathrm{R}}}{\mathrm{Tr}(\hat{\varrho}_{\mathrm{R}})}$
is normalized but acts on the R-subspace and describes the state of a RP before
it has recombined. It is here the condition `before' (and the accompanied forced
state normalization) that results in the nonlinearity introduced via
$\langle\mathcal{L}\rangle$. The evolution without any conditions should be
linear, however, as is (\ref{genf}), which also preserves the trace if the
reaction products are included. 
In our more general picture (cf. Figs.~\ref{fig6b} and \ref{fig6a}),
Eq.~(\ref{genfn}) can be recovered as the R-subspace projection of the evolution
of the chemical system for a perfect dark evolution in the master
equation limit and the case of triplet symmetry without triplet dephasing.
Conversely, Eq.~(\ref{genfn}) generalizes Eq.~(2) in \cite{Kom11c} and
\cite{Jon11}. Note that (\ref{genfn}) refers to a \emph{single} RP. In the more
general case of a fluorescing ensemble of independent identical RPs, the dark
evolution equation (\ref{genfn}) must be replaced with a more general
stochastic master equation (\ref{sME}).
\section{
\label{sec5}
Encounter model reconsidered}
\subsection{
\label{sec5.1}
External evolution of the chemical system: statistics of the encounters}
Consider a chemical system in the form of a RP diffusing in a solvent.
We assume that the radicals evolve unitarily unless they meet, which results in
some non-unitary effect. Regarding each individual encounter event as
instantaneous on the timescale of the unitary evolution between the encounters,
an encounter is described by some map $\mathcal{A}$, whereas the unitary
evolution between consecutive encounters at times $t_j$ and
$t_k$ $\!\ge$ $\!t_j$ is described by a map $\mathcal{U}(t_k,t_j)$. The quantity
sought is the state $\hat{\varrho}(t)$ of the chemical system that has evolved
from a given initial state $\hat{\varrho}(0)$. We regard the diffusion process
itself as classical and independent of the state $\hat{\varrho}$, and average
over the (unknown) number of encounters $k$ that occur within the interval
$(0,t]$,
\begin{equation}
\label{state}
  \hat{\varrho}(t)=\sum_{k=0}^\infty\hat{\varrho}|_k(t)p_k.
\end{equation}
Here, $\hat{\varrho}|_k(t)$ is the state under the condition that $k$ encounters
have occurred (at unknown times, so it is averaged over the possible motions
leading to $k$ encounters before time $t$) and $p_k$ is the probability of
having $k$ encounters. We now divide the interval $(0,t]$ into $n$ equal
intervals $\mathrm{d}t$ $\!=$ $\!t/n$ and assume that at most one encounter
occurs within a given $\mathrm{d}t$, which happens with probability
$p$ $\!=$ $\!r\mathrm{d}t$. This gives a binomial distribution
\begin{equation}
\label{binomdist}
  p_k=\mathrm{b}_{kn}(p)\equiv\binom{n}{k}p^k(1-p)^{n-k}
\end{equation}
for the total number of encounters within $(0,t]$, that can for $n\to\infty$ be
approximated by a Poisson distribution
$p_k$ $\!=$ $\!\frac{(rt)^k}{k!}\mathrm{e}^{-rt}$. 
In particular, the probability that no encounter occurs during $(0,t]$ can also
be obtained directly as the limit
$(1-\frac{rt}{n})^n$ $\!\to$ $\!\mathrm{e}^{-rt}$ for $n$ $\!\to$ $\!\infty$.
Thus $1-\mathrm{e}^{-rt}$ is the probability that the first encounter occurs
before time $t$, i.e., the cumulative distribution function, from which we
obtain the probability density of the time of the first encounter
$r\mathrm{e}^{-rt}$ as time derivative, cf. the analogous argumentation
following (\ref{A0}). This exponential distribution hence describes the waiting
time distribution between encounters that occur at rate $r$ and has e.g. been
used in \cite{cai10} to compute average values of state dependent quantities of
interest.

If these $k$ encounters occur at times $t_k,\ldots,t_1$, the state reads
\begin{equation}
\label{stateknowntimes}
  \hat{\varrho}|_{t_k,\ldots,t_1}(t)=\mathcal{U}(t,t_k)\cdots
  \mathcal{A}\,\mathcal{U}(t_2,t_1)\mathcal{A}\,\mathcal{U}(t_1,0)
  \hat{\varrho}(0).
\end{equation}
Here we have assumed $\mathcal{A}$ to be completely positive trace-preserving,
i.e., no measurements are read out during $(0,t]$, cf. (\ref{ACPT}). In the
Heisenberg picture with respect to the unitary evolution between the encounters
we can rewrite (\ref{stateknowntimes}) as
\begin{equation}
\label{stateHknowntimes}
  \hat{\varrho}_{\mathrm{H}}|_{t_k,\ldots,t_1}(t)
  =\mathcal{A}_{\mathrm{H}}(t_k)\cdots
  \mathcal{A}_{\mathrm{H}}(t_2)\mathcal{A}_{\mathrm{H}}(t_1)\hat{\varrho}(0),
\end{equation}
with $\hat{\varrho}_{\mathrm{H}}(t)$ $\!=$
$\!\mathcal{U}^\dagger(t,0)\hat{\varrho}(t)$ and 
$\mathcal{A}_{\mathrm{H}}(t)$ $\!=$
$\!\mathcal{U}^\dagger(t,0)\mathcal{A}(t)\mathcal{U}(t,0)$. While the time
argument in $\mathcal{A}(t)$ indicates that $\mathcal{A}$ could in principle
carry an explicit time dependence, here we assume that
$\mathcal{A}(t)$ $\!=$ $\!\mathcal{A}$ is constant.
The time of each encounter occurring within $(0,t]$ is uniformly distributed
over this interval (in the above discrete model, $p$ does not depend on
the location of the interval $\mathrm{d}t$ within $(0,t]$). When averaging over
the times of the encounters we must however take into account that
(\ref{stateHknowntimes}) implies ${t}\ge{t}_k\ge\cdots\ge{t}_2\ge{t}_1\ge0$. We
do so by a formal time ordering operator $\mathcal{T}_+$ in the time averaging,
that is, replacing $\frac{1}{t^k}\int_0^{t}\mathrm{d}t_k
\cdots\int_0^{t}\mathrm{d}t_2\int_0^{t}\mathrm{d}t_1$ with
\begin{equation}
  \mathcal{T}_+\frac{1}{t^k}\!\!\int_0^{t}\!\!\mathrm{d}t_k
  \cdots\!\!\int_0^{t}\!\!\mathrm{d}t_2\!\!\int_0^{t}\!\!\mathrm{d}t_1
  \equiv\frac{k!}{t^k}\!\!\int_0^{t}\!\!\mathrm{d}t_k
  \cdots\!\!\int_0^{t_3}\!\!\mathrm{d}t_2\!\!\int_0^{t_2}\!\!\mathrm{d}t_1,
\end{equation}
and the averaging over the times of the $k$ encounters is carried out as
\begin{equation}
  \hat{\varrho}_{\mathrm{H}}|_k(t)
  =\mathcal{T}_+\frac{1}{t^k}\!\!\int_0^{t}\!\!\mathrm{d}t_k
  \cdots\!\!\int_0^{t}\!\!\mathrm{d}t_2\!\!\int_0^{t}\!\!\mathrm{d}t_1
  \hat{\varrho}_{\mathrm{H}}|_{t_k,\ldots,t_1}(t).
\end{equation}
Note that this assumes that there is only one type of encounter $\mathcal{A}$.
If there were $k$ different encounters
$\mathcal{A}^{[1]},\ldots,\mathcal{A}^{[k]}$ instead, then the $k$-encounter
average would be
\begin{eqnarray}
  \hat{\varrho}_{\mathrm{H}}|_k(t)
  &=&\frac{1}{t^k}\!\!\int_0^{t}\!\!\mathrm{d}t_k
  \cdots\!\!\int_0^{t_3}\!\!\mathrm{d}t_2\!\!\int_0^{t_2}\!\!\mathrm{d}t_1
  \sum_{j=1}^{k!}
  \nonumber\\
  &\times&
  \mathcal{A}_{\mathrm{H}}^{[\pi_j(k)]}(t_k)\cdots
  \mathcal{A}_{\mathrm{H}}^{[\pi_j(2)]}(t_2)
  \mathcal{A}_{\mathrm{H}}^{[\pi_j(1)]}(t_1)
  \hat{\varrho}(0),\quad\quad
\end{eqnarray}
where $\pi_1,\ldots,\pi_{k!}$ are the permutations of the labels $1,\ldots,k$ of
the encounters. For (\ref{state}) we thus obtain
\begin{eqnarray}
\label{Heisexp}
  \hat{\varrho}_{\mathrm{H}}(t)&=&\mathcal{T}_+\sum_{k=0}^\infty
  \frac{r^k}{k!}\mathrm{e}^{-rt}\!\!\int_0^{t}\!\!\mathrm{d}t_k
  \cdots\!\!\int_0^{t}\!\!\mathrm{d}t_2\!\!\int_0^{t}\!\!\mathrm{d}t_1
  \nonumber\\
  &\times&\mathcal{A}_{\mathrm{H}}(t_k)\cdots
  \mathcal{A}_{\mathrm{H}}(t_2)\mathcal{A}_{\mathrm{H}}(t_1)\hat{\varrho}(0).
\end{eqnarray}
We can hence write
\begin{equation}
  \hat{\varrho}_{\mathrm{H}}(t)=\mathcal{T}_+
  \mathrm{e}^{\int_0^{t}\!\!\mathrm{d}\tau\mathcal{L}_{\mathrm{H}}(\tau)}
  \hat{\varrho}(0),\quad
  \frac{\partial}{\partial{t}}\hat{\varrho}_{\mathrm{H}}(t)
  =\mathcal{L}_{\mathrm{H}}(t)\hat{\varrho}_{\mathrm{H}}(t),
\end{equation}
with
\begin{equation}
  \mathcal{L}_{\mathrm{H}}(t)=r[\mathcal{A}_{\mathrm{H}}(t)-1].
\end{equation}
Returning to the Schr\"odinger picture
$\hat{\varrho}(t)$ $\!=$ $\!\mathcal{U}(t,0)\hat{\varrho}_{\mathrm{H}}(t)$ and
$\mathcal{A}(t)$ $\!=$
$\!\mathcal{U}(t,0)\mathcal{A}_{\mathrm{H}}(t)\mathcal{U}^\dagger(t,0)$, we
obtain
\begin{eqnarray}
\label{statefinal}
  \frac{\partial}{\partial{t}}\hat{\varrho}(t)&=&
  \bigl[\mathcal{L}_{\mathrm{betw}}(t)+\mathcal{L}_{\mathrm{enc}}(t)\bigr]
  \hat{\varrho}(t),
  \\
\label{encounterL}
  \mathcal{L}_{\mathrm{betw}}(t)\hat{\varrho}(t)&=&
  -\mathrm{i}[\hat{H}(t),\hat{\varrho}(t)]
  +\mathcal{L}_{\mathrm{diss}}(t)\hat{\varrho}(t),
  \\
\label{encounterA}
  \mathcal{L}_{\mathrm{enc}}(t)&=&r(t)[\mathcal{A}(t)-1],
\end{eqnarray}
where we have used that 
$\mathcal{U}(t,0)\hat{\varrho}_{\mathrm{H}}(t)$ $\!=$
$\!\hat{U}(t)\hat{\varrho}_{\mathrm{H}}(t)\hat{U}^\dagger(t)$,
$\frac{\partial}{\partial{t}}\hat{U}(t)$ $\!=$
$\!-\mathrm{i}\hat{H}(t)\hat{U}(t)$. The two terms 
$\mathcal{L}_{\mathrm{betw}}$ and $\mathcal{L}_{\mathrm{enc}}$ describe the
change of state between and due to the encounters, respectively, cf.
(\ref{Lbe}) and (\ref{Lbetw}). In (\ref{encounterA}) we allowed for a
time-dependent encounter rate $r(t)$. To justify this, we now briefly consider
some possibilities of such time-dependence. In particular, we outline two
situations which suggest a rate declining with time. These situations serve as
complementary examples of an immobile and a mobile RP.

In the first scenario, the RP is bounded within some molecular
structure and its creation (e.g., accompanied by a conformational change of this
structure) releases some energy which is deposited in vibrational degrees of
freedom. An encounter thus requires overcoming an energetic barrier (such as
returning to the old conformation acting as an activation energy of the
back-reaction), the amount of which is of the order of
the vibrational energy released. The assumption that the initial molecular
vibrations are damped exponentially by dissipative processes with the
environment would suggest an exponentially declining encounter rate,
\begin{equation}
\label{ex1}
  r(t)=r\mathrm{e}^{-at},
\end{equation}
where $r$ $\!=$ $\!r(0)$, and $a$ is a constant that has been included for
dimensional reasons, being roughly the inverse of the lifetime of the
vibrational states.

In the second scenario, one radical's diffusion follows a three-dimensional
Brownian motion (e.g., an electron released at $t$ $\!=$ $\!0$), whereas its
radical partner (the remainder of the donor molecule) stays at rest. The
probability density $\phi$ of the location $\bm{r}$ of the mobile radical is
then a Gaussian $\phi(\bm{r},t)$ $\!=$
$\!(2\pi\sigma^2)^{-\frac{3}{2}}\mathrm{e}^{-\frac{\bm{r}^2}{2\sigma^2}}$,
whose variance $\sigma^2$ $\!=$ $\!2Dt$ depends via the diffusion constant
$D$ $\!=$ $\!k_{\mathrm{B}}T/(6\pi\eta{R})$ on the radical's radius $R$,
viscosity $\eta$, and temperature $T$. Since we neither know nor want to depend
on such details, we assume that initially, the RP is confined in an
appropriately sized cage. At $t$ $\!=$ $\!0$, when the encounter rate has
assumed a constant equilibrium value $r$, the walls of the cage are removed,
which would suggest an encounter rate declining as
\begin{equation}
\label{ex2}
  r(t)=\Bigl(r^{-\frac{1}{\mu}}+at\Bigr)^{-\mu},
\end{equation}
where $a$ is again a constant and $\mu$ $\!=$ $\!\frac{3}{2}$.

It is clear that these two schemes cannot capture the diversity of possible
realistic scenarios and here only serve as illustrations of an exponential and
an algebraic decline. Moreover, it may be possible to modify the encounter
statistics, e.g., by optically switching between different molecule
conformations \cite{Gue12}. In (\ref{encounterA}), we therefore leave $r(t)$
unspecified, assuming that it could attain an arbitrary time dependence in
principle.

Here, we focus on the simplest case of a constant rate $r$. A physical
justification of this is that the state change due to the encounters
$\mathcal{L}_{\mathrm{enc}}$ is competing with processes occurring between the
encounters $\mathcal{L}_{\mathrm{betw}}$. The dissipative component
$\mathcal{L}_{\mathrm{diss}}$ of these
processes sets a time frame over which the interesting dynamics takes place.
In the above examples (\ref{ex1}) and (\ref{ex2}), the long time tails for which
$r(t)/r(0)$ $\!\ll$ $\!1$ correspond to an equilibrated system, for which the
dynamics of interest (and the validity range of the model assumptions) have long
disappeared. We therefore replace the formal limit $t$ $\!\to$ $\!\infty$ with
some appropriate cutoff time $t$ $\!=$ $\!t_\infty$ which reflects the time
scale over which the chemical reaction takes place and assume a constant rate
$r$ $\!=$ $\!r(0)$ up to $t$. The number of encounters occurring up to $t$ then
follows the mentioned Poisson distribution
$p_k$ $\!=$ $\!\frac{(rt)^k}{k!}\mathrm{e}^{-rt}$, hence
$p_{k+1}$ $\!=$ $\!\frac{rt}{k+1}p_k$. For example, if $rt$ $\!=$ $\!10^{-1}$,
then up to $t_\infty$, 90\% of the RPs have no encounter, 9\% have a
single encounter, 0.5\% have two encounters and only 0.02\% have more than two
encounters. In this case, it would then be justified to restrict analysis to a
single encounter and compare the results with corrections due to a low number of
additional encounters.

If $\mathcal{A}$ does not preserve the trace because it is associated with a
probability, we must normalize the state which leads to the nonlinear master
equation. In this way, one can in the multiple encounter case also re-derive the
solution of the nonlinear master equation. Since the master equation describes
encounters occurring with a rate, its integration naturally yields multiple
encounters.
\subsection{
\label{sec5.2}
Internal evolution of the chemical system: effect of the encounters}
So far we have not specified $\mathcal{A}$. To facilitate comparison with the
previous section,  here and in the remaining sections we limit our encounter
model developed in Sec.~\ref{sec10} to the simplified case of triplet symmetry
without triplet dephasing. In particular, this means that the encounters have
the same effect on all triplet states, i.e., they do not ``resolve'' them, and
it is sufficient to use the triplet projector $\hat{Q}_{\mathrm{T}}$. Analogous
to (\ref{ACPT}), we decompose $\mathcal{A}$ $\!=$ $\!\mathcal{A}_{\mathrm{CPT}}$
$\!=$ $\!\mathcal{A}_0+\sum_{j={\mathrm{S}},{\mathrm{T}}}\mathcal{A}_j$, where
$\mathcal{A}_{\mathrm{S}}$ ($\mathcal{A}_{\mathrm{T}}$), and $\mathcal{A}_0$
describe the detection of a singlet (triplet) or no fluorescence signal,
respectively [cf. (\ref{PiT})]. The corresponding transformations of the
two-electron spin state are
\begin{eqnarray}
\label{Ajrho}
  \mathcal{A}_j&=&\tilde{r}_j
  \langle\mathcal{Q}_j\rangle\hat{Q}_{\mathrm{P}},\quad
  (j={\mathrm{S}},{\mathrm{T}}),
  \\
\label{A0rho}
  \mathcal{A}_0
  &=&(1-\tilde{\eta})
  +\tilde{\eta}\Bigl[\mathcal{Q}_{\mathrm{P}}
  +\sum_{j={\mathrm{S}},{\mathrm{T}}}
  (1-\tilde{\eta}_j)\mathcal{Q}_j\Bigr],
\end{eqnarray}
which are obtained from (\ref{Amj1}) and (\ref{Am01}) under the additional
assumptions (\ref{symmcondmaps2}), (\ref{PiT}), and (\ref{symmcondmaps0}) as
described in Sec.~\ref{sec10}. Here, $\hat{Q}_{\mathrm{P}}$ is the projector
onto the product subspace, cf. (\ref{STPdec}), the different $\mathcal{Q}$ are
projective maps, cf. (\ref{projs}), and for $\langle\mathcal{Q}_j\rangle$, cf.
(\ref{avQ}). The free parameters $\tilde{r}_j$ and $\tilde{\eta}$ are related to
the coupling coefficients with our model environment via (\ref{p2}) and
(\ref{p3}), respectively, and $\tilde{\eta}_j$ is defined as in (\ref{beta1}).
From their definition it follows directly that $\tilde{r}_j$ $\!\in$ $\![0,1]$, 
$\sum_{j={\mathrm{S}},{\mathrm{T}}}\tilde{r}_j$ $\!\in$ $\![0,2]$,
and $\tilde{\eta}$ $\!\in$ $\![0,2]$, and a numerical analysis reveals that
$\tilde{\eta}_j$ $\in$ $\![0,2]$ and
$\sum_{j={\mathrm{S}},{\mathrm{T}}}\tilde{\eta}_j$ $\in$ $\![0,2]$.
This suggests that in practice, encounters of different strength may occur,
rather than one single well-defined type of encounter.
In fact it is not guaranteed that during every encounter the radicals reach
the same proximity with the same orientation or that they stay in contact for
a fixed duration, and this would lead to an explicit random time dependence of
$\mathcal{A}$ in (\ref{stateHknowntimes}), which we disregard as mentioned in
the comments following (\ref{stateHknowntimes}). A consideration of specific
types of encounters corresponding to special values of the parameters is carried
out in Sec.~\ref{sec8.1}.
\section{
\label{sec6}
Dark evolution}
Let us consider the evolution of the chemical system due to the encounters.
Since the master equation approach follows from the encounter model as a
limiting case, we refer to the latter. Furthermore, since for a single RP, a
fluorescence signal marks the moment of detection of its recombination, we are
interested in the state evolution \emph{before} a fluorescence click, i.e., the
dark evolution of the RP. A \emph{perfect} dark evolution is characterized by
perfect detectors, $\eta_j^{(\mathrm{D})}$ $\!=$ $\!1$, that do not click, and
an initial absence of a reaction product,
$\hat{\varrho}_{\mathrm{P}}(0)$ $\!=$ $\!0$. An interesting behaviour is shown
by a \emph{near perfect} dark evolution, which only approximates a perfect dark
evolution due to small imperfections in the RP-preparation and/or fluorescence
detection, $\langle\mathcal{Q}_{\mathrm{R}}\rangle_0$ $\!=$
$\!1$ $\!-$ $\!\varepsilon_{\mathrm{R}}$ and
$\eta_j^{(\mathrm{D})}$ $\!=$ $\!1$ $\!-$ $\!\varepsilon_{j\mathrm{D}}$,
where $\varepsilon_{\mathrm{R}},\varepsilon_{j\mathrm{D}}$ $\!\ll$ $\!1$. 
For simplicity we assume that only one type
of encounter with given parameters $\tilde{\eta}$ and $\tilde{r}_j$ occurs. This
is justified because we may always restrict to an average encounter discussed in
Sec.~\ref{sec8.1} below. In analogy to Sec.~\ref{sec4.1} we consider the master
equations with and without inclusion of the reaction products. In order to allow
combined treatment of unconditional and conditional evolution, we introduce
singlet and triplet detection efficiencies
$\eta_j^{(\mathrm{D})}$ $\!\in$ $\![0,1]$ with
$j$ $\!=$ $\!{\mathrm{S}},{\mathrm{T}}$, with which an encounter accompanied
with singlet, triplet, or no fluorescence detection is generalized from
(\ref{Ajrho}) and (\ref{A0rho}) to
\begin{eqnarray}
\label{AjrhoD}
  \mathcal{A}_j^{(\mathrm{D})}&=&\eta_j^{(\mathrm{D})}\tilde{r}_j
  \langle\mathcal{Q}_j\rangle\hat{Q}_{\mathrm{P}},
  \quad(j={\mathrm{S}},{\mathrm{T}}),
  \\
  \mathcal{A}_0^{(\mathrm{D})}
  &=&(1-\tilde{\eta})
  +\tilde{\eta}\Bigl[\mathcal{Q}_{\mathrm{P}}
  +\sum_{j={\mathrm{S}},{\mathrm{T}}}
  (1-\tilde{\eta}_j)\mathcal{Q}_j\Bigr]
  \nonumber\\
  &&+\sum_{j={\mathrm{S}},{\mathrm{T}}}\bigl(1-\eta_j^{(\mathrm{D})}\bigr)
  \tilde{r}_j\langle\mathcal{Q}_j\rangle\hat{Q}_{\mathrm{P}}
\label{A0rhoD}
  \\
  &=&(1-\tilde{\eta})+\tilde{\eta}\mathcal{Q}
  +\sum_{j={\mathrm{S}},{\mathrm{T}}}\tilde{r}_j
  [\bigl(1-\eta_j^{(\mathrm{D})}\bigr)
  \langle\mathcal{Q}_j\rangle\hat{Q}_{\mathrm{P}}-\mathcal{Q}_j],
  \nonumber
\end{eqnarray}
cf. Sec.~\ref{sec3b}. Here we have defined a S-T-P dephasing operator
\begin{equation}
\label{STPQ}
  \mathcal{Q}
  =\mathcal{Q}_{\mathrm{P}}+\sum_{j={\mathrm{S}},{\mathrm{T}}}\mathcal{Q}_j,
\end{equation}
and (\ref{Ajrho}) and (\ref{A0rho}) are recovered for
$\eta_j^{(\mathrm{D})}$ $\!=$ $\!1$. As before, the CPT-map
$\mathcal{A}_{\mathrm{CPT}}$ $\!=$ $\!\mathcal{A}_0^{(\mathrm{D})}$ $\!+$
$\!\sum_{j={\mathrm{S}},{\mathrm{T}}}\mathcal{A}_j^{(\mathrm{D})}$ of an
unmeasured encounter is
\begin{equation}
  \mathcal{A}_{\mathrm{CPT}}
  =(1-\tilde{\eta})+\tilde{\eta}\mathcal{Q}+\sum_{j={\mathrm{S}},{\mathrm{T}}}
  \tilde{r}_j(\langle\mathcal{Q}_j\rangle\hat{Q}_{\mathrm{P}}-\mathcal{Q}_j).
\end{equation}
Non-unit $\eta_j^{(\mathrm{D})}$ may result from inefficiencies in the detection
itself or due to characteristics of the recombination process. For example, if
fluorescence is only observed in case of a singlet recombination, whereas a
triplet recombination occurs radiationless, we may set
$\eta_{\mathrm{S}}^{(\mathrm{D})}$ $\!=$ $\!1$ but
$\eta_{\mathrm{T}}^{(\mathrm{D})}$ $\!=$ $\!0$. In what follows, we consider
the state evolution of a single chemical system \emph{before} the occurrence of
a fluorescence signal which marks \emph{the detection of} its recombination,
i.e., the detection of the transition to the reaction product space. This
conditional evolution is given by (\ref{nlme}) together with (\ref{LofA}), where
for $\mathcal{A}_0$ we use (\ref{A0rhoD}). The special case
$\eta_j^{(\mathrm{D})}$ $\!=$ $\!0$ corresponds to ignoring any fluorescence
signal and the master equation recovers the unconditional, hence linear and
trace-preserving evolution,
$\mathcal{A}_0^{(\mathrm{D})}\bigl(\eta_j^{(\mathrm{D})}\!=\!0\bigr)$ $\!=$
$\!\mathcal{A}_{\mathrm{CPT}}$.
\subsection{Evolution in full space}
We write (\ref{nlme}) together with (\ref{LofA}) as
\begin{eqnarray}
\label{cMESTP}
  \frac{\partial}{\partial{t}}\hat{\varrho}
  &=&\left(\mathcal{L}-\langle\mathcal{L}\rangle\right)\hat{\varrho},
  \\
  \mathcal{L}&=&\mathcal{L}_{\mathrm{betw}}+\mathcal{L}_{\mathrm{enc}}.
\end{eqnarray}
$\mathcal{L}_{\mathrm{betw}}$ takes into account other effects between the
encounters, cf. (\ref{statefinal}) and (\ref{encounterL}). The effect of the
encounters themselves is described by
\begin{eqnarray}
  \mathcal{L}_{\mathrm{enc}}&=&
  r\bigl(\mathcal{A}_0^{(\mathrm{D})}-1\bigr)
  =\tilde{r}(\tilde{\mathcal{A}}_0^{(\mathrm{D})}-1),
  \\
\label{rtilde}
  \tilde{r}&=&\tilde{\eta}r,
  \\
  \tilde{\mathcal{A}}_0^{(\mathrm{D})}&=&\mathcal{Q}
  +\sum_{j={\mathrm{S}},{\mathrm{T}}}\tilde{\eta}_j
  [\bigl(1-\eta_j^{(\mathrm{D})}\bigr)
  \langle\mathcal{Q}_j\rangle\hat{Q}_{\mathrm{P}}-\mathcal{Q}_j],\quad
\end{eqnarray}
where we have applied (\ref{A0rhoD}).
Since $\langle\mathcal{L}_{\mathrm{betw}}\rangle$ $\!=$ $\!0$, we obtain
\begin{equation}
  \langle\mathcal{L}\rangle
  =-r\sum_{j={\mathrm{S}},{\mathrm{T}}}
  \langle\mathcal{A}_j^{(\mathrm{D})}\rangle
  =-r\sum_{j={\mathrm{S}},{\mathrm{T}}}
  \tilde{r}_j\eta_j^{(\mathrm{D})}\langle\mathcal{Q}_j\rangle.
\end{equation}
As mentioned above, (\ref{cMESTP}) becomes linear for
$\eta_j^{(\mathrm{D})}$ $\!=$ $\!0$, which reduces it to (\ref{statefinal}).

When no evolution takes place between the encounters as described by imposing
$\mathcal{L}_{\mathrm{betw}}$ $\!=$ $\!0$, the solution of (\ref{cMESTP}) is
straightforward. We first solve the linear part
$\frac{\partial}{\partial{t}}\hat{\varrho}_{\mathrm{N}}$ $\!=$
$\!\mathcal{L}_{\mathrm{enc}}\hat{\varrho}_{\mathrm{N}}$ without the
normalization $\langle\mathcal{L}\rangle$, which gives
$\hat{\varrho}_{\mathrm{N}}$, and then normalize it according to (\ref{rho}).
The solution of (\ref{cMESTP}) thus becomes
\begin{eqnarray}
  \hat{\varrho}_{\mathrm{N}}(t)&=&\hat{\varrho}_{0\mathrm{P}}
  +\mathrm{e}^{-\tilde{r}t}\hat{\varrho}_{0\mathrm{R}}
  +\!\!\!\sum_{j={\mathrm{S}},{\mathrm{T}}}\!\Bigl[
  \mathrm{e}^{-\tilde{r}t}(\mathrm{e}^{\tilde{r}(1-\tilde{\eta}_j)t}\!-\!1)
  \mathcal{Q}_j\hat{\varrho}_0
  \nonumber\\
\label{RhoN}
  &&-\bigl(1-\eta_j^{(\mathrm{D})}\bigr)
  (\mathrm{e}^{-\tilde{r}_jrt}-1)
  \langle\mathcal{Q}_j\rangle_0\hat{Q}_{\mathrm{P}}\Bigr],\quad
  \\
\label{TrRhoN}
  \mathrm{Tr}\hat{\varrho}_{\mathrm{N}}(t)&=&
  1-\sum_{j={\mathrm{S}},{\mathrm{T}}}\eta_j^{(\mathrm{D})}
  (1-\mathrm{e}^{-\tilde{r}_jrt})\langle\mathcal{Q}_j\rangle_0.
\end{eqnarray}
Eq.~(\ref{TrRhoN}) is the probability that no fluorescence click is obtained up
to time $t$. Assuming $\tilde{r}_j$ $\!\neq$ $\!0$, it reaches the asymptotic
value
\begin{equation}
  \mathrm{Tr}\hat{\varrho}_{\mathrm{N}}(\infty)
  =1-\sum_{j={\mathrm{S}},{\mathrm{T}}}\eta_j^{(\mathrm{D})}
  \langle\mathcal{Q}_j\rangle_0.
\end{equation}
The special case $\mathrm{Tr}\hat{\varrho}_{\mathrm{N}}(\infty)$ $\!=$ $\!0$ is
observed only if $\eta_j^{(\mathrm{D})}$ $\!=$ $\!1$ and
$\hat{\varrho}_{0\mathrm{P}}$ $\!=$ $\!0$. This corresponds to a perfect dark
evolution. Eq.~(\ref{TrRhoN}) then simplifies to
$\mathrm{Tr}\hat{\varrho}_{\mathrm{N}}(t)$ $\!=$
$\!\sum_{j={\mathrm{S}},{\mathrm{T}}}\mathrm{e}^{-\tilde{r}_jrt}
\langle\mathcal{Q}_j\rangle_0$, i.e., an exponential decline. This reflects the
fact that a preparation of longlived RPs by means of continuous measurement and
postselection is inefficient in the number of trials.
\subsection{Evolution in the R-subspace}
The reaction product has been included in (\ref{cMESTP}) to ensure that
the state of the chemical system has unit-trace. We may limit to the R-subspace
spanned by $\hat{Q}_{\mathrm{R}}$, i.e., derive from
(\ref{cMESTP}) an equation for the projection $\hat{\varrho}_{\mathrm{R}}$
$\!=$ $\!\mathcal{Q}_{\mathrm{R}}\hat{\varrho}$
$\!=$ $\!\hat{Q}_{\mathrm{R}}\hat{\varrho}\hat{Q}_{\mathrm{R}}$, cf.
(\ref{STPdec}) and (\ref{projs}), which is given by the projection of
$\mathcal{A}_{\mathrm{CPT}}$ (or $\mathcal{A}_0^{(\mathrm{D})}$) on the
R-subspace,
\begin{eqnarray}
  \mathcal{A}_{\mathrm{R}}
  &=&\mathcal{Q}_{\mathrm{R}}\mathcal{A}_{\mathrm{CPT}}\mathcal{Q}_{\mathrm{R}}
  =\mathcal{A}_{0\mathrm{R}}
  \nonumber\\
  &=&(1-\tilde{\eta})\mathcal{Q}_{\mathrm{R}}
  +\tilde{\eta}\sum_{j={\mathrm{S}},{\mathrm{T}}}
  (1-\tilde{\eta}_j)\mathcal{Q}_j.
\end{eqnarray}
This can be seen by applying
$(\mathcal{Q}_{\mathrm{R}}$ $\!+$ $\!\mathcal{Q}_{\mathrm{P}})$ to
(\ref{cMESTP}) and using (\ref{inicon}), 
which gives
$\frac{\partial}{\partial{t}}(\mathcal{Q}_{\mathrm{R}}$ $\!+$
$\!\mathcal{Q}_{\mathrm{P}})\hat{\varrho}$ $\!=$
$\!\bigl[(\mathcal{Q}_{\mathrm{R}}$ $\!+$
$\!\mathcal{Q}_{\mathrm{P}})\mathcal{L}(\mathcal{Q}_{\mathrm{R}}$ $\!+$
$\!\mathcal{Q}_{\mathrm{P}})$ $\!-$
$\!\langle\mathcal{L}\rangle\bigr](\mathcal{Q}_{\mathrm{R}}$ $\!+$
$\!\mathcal{Q}_{\mathrm{P}})\hat{\varrho}$.
Inserting (\ref{A0rhoD}) yields for the RP component
\begin{eqnarray}
\label{cMEST}
  \frac{\partial}{\partial{t}}\hat{\varrho}_{\mathrm{R}}
  &=&\left(\mathcal{L}_{\mathrm{R}}-\langle\mathcal{L}\rangle\right)
  \hat{\varrho}_{\mathrm{R}},
  \\
  \mathcal{L}_{\mathrm{R}}
  &=&\mathcal{L}_{\mathrm{betw,R}}+\mathcal{L}_{\mathrm{enc,R}},
  \\
\label{117}
  \mathcal{L}_{\mathrm{enc,R}}&=&
  r(\mathcal{A}_{0\mathrm{R}}-\mathcal{Q}_{\mathrm{R}})
  =\tilde{r}(\tilde{\mathcal{A}}_{0\mathrm{R}}-\mathcal{Q}_{\mathrm{R}}),
  \\
\label{genmeRanME}
  \tilde{\mathcal{A}}_{0\mathrm{R}}
  &=&\sum_{j={\mathrm{S}},{\mathrm{T}}}(1-\tilde{\eta}_j)\mathcal{Q}_j,
 \end{eqnarray}
where we have assumed that $\mathcal{L}_{\mathrm{betw}}\mathcal{Q}_{\mathrm{P}}$
$\!=$ $\!0$.
Equivalently, this can be written as
\begin{eqnarray}
  \frac{\partial}{\partial{t}}\hat{\varrho}_{\mathrm{R}}
  &=&\mathcal{L}_{\mathrm{betw,R}}\hat{\varrho}_{\mathrm{R}}
  +\tilde{r}[\mathcal{B}(\hat{\varrho}_{\mathrm{R}})-1]
  \hat{\varrho}_{\mathrm{R}},
  \\
  \mathcal{B}&=&\sum_{j={\mathrm{S}},{\mathrm{T}}}
  \Bigl[(1-\tilde{\eta}_j)\mathcal{Q}_j
  +\tilde{\eta}_j\eta_j^{(\mathrm{D})}\langle\mathcal{Q}_j\rangle\Bigr],
\end{eqnarray}
cf. (\ref{avQ}).

Limiting again to $\mathcal{L}_{\mathrm{betw}}$ $\!=$ $\!0$, the solution of
(\ref{cMEST}) is obtained from (\ref{RhoN}) and (\ref{TrRhoN}) as the part of
the conditional state $\hat{\varrho}$
$\!=$ $\!\hat{\varrho}_{\mathrm{N}}/\mathrm{Tr}\hat{\varrho}_{\mathrm{N}}$ lying
in the R-subspace. This gives
\begin{eqnarray}
  \hat{\varrho}_{\mathrm{R}}(t)
  &=&\frac{\mathrm{e}^{-\tilde{r}t}}{\mathrm{Tr}\hat{\varrho}_{\mathrm{N}}}
  \Bigl[1+\sum_{j={\mathrm{S}},{\mathrm{T}}}
  (\mathrm{e}^{\tilde{r}(1-\tilde{\eta}_j)t}-1)
  \mathcal{Q}_j\Bigr]\hat{\varrho}_{0\mathrm{R}},\quad\quad
  \\
\label{TrRhoR}
  \mathrm{Tr}\hat{\varrho}_{\mathrm{R}}(t)
  &=&\frac{\sum_{j={\mathrm{S}},{\mathrm{T}}}\mathrm{e}^{-\tilde{r}_jrt}
  \langle\mathcal{Q}_j\rangle_0}{1-\sum_{j={\mathrm{S}},{\mathrm{T}}}
  \eta_j^{(\mathrm{D})}(1-\mathrm{e}^{-\tilde{r}_jrt})
  \langle\mathcal{Q}_j\rangle_0}.
\end{eqnarray}
Eq.~(\ref{TrRhoR}) is the probability that the RP has not recombined up to time
$t$ under the condition that no fluorescence has been detected up to this time.
Assuming $\tilde{r}_j$ $\!\neq$ $\!0$, it reaches the asymptotic value
$\mathrm{Tr}\hat{\varrho}_{\mathrm{R}}(\infty)$ $\!=$ $\!0$
for $\mathrm{Tr}\hat{\varrho}_{\mathrm{N}}(\infty)$ $\!>$ $\!0$.

\begin{figure}[ht]
\includegraphics[width=4.2cm]{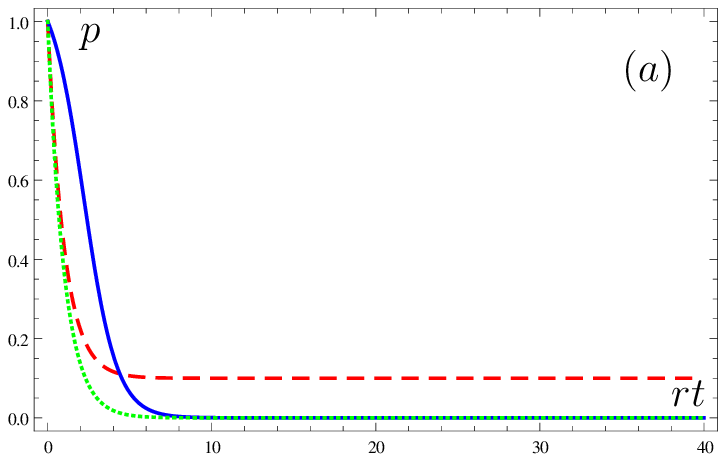}
\includegraphics[width=4.2cm]{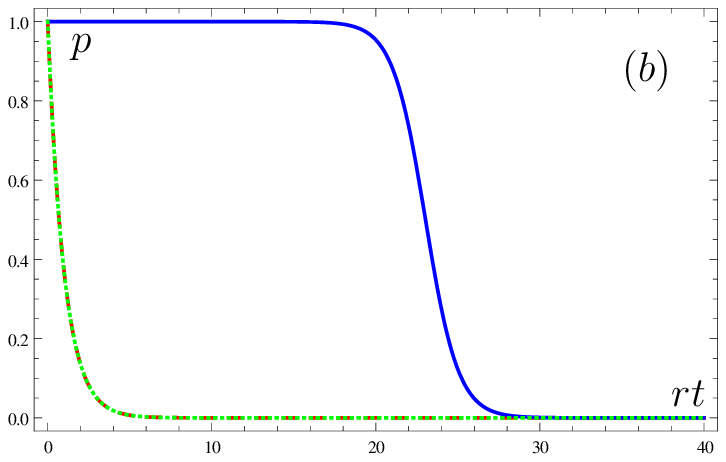}
\caption{\label{fig3}
(color online).
Temporal evolution of probabilities $p(rt)$ obtained with (\ref{TrRhoN}) and
(\ref{TrRhoR}).
Dashed red: probability $p(D)$ $\!:=$ $\!\mathrm{Tr}\hat{\varrho}_{\mathrm{N}}$
of detecting no fluorescence up to time $t$;
dotted green: probability $p(R)$ $\!=$ $\!p(R,D)$ $\!:=$ $\!\mathrm{Tr}
\hat{\varrho}_{\mathrm{R}}\mathrm{Tr}\hat{\varrho}_{\mathrm{N}}$
of having no recombination (and detecting no fluorescence) up to $t$;
solid blue: probability
$p(R|D)$ $\!:=$ $\!\mathrm{Tr}\hat{\varrho}_{\mathrm{R}}$
of having no recombination up to $t$ under the condition that no fluorescence
has been detected up to this time;
$\tilde{r}_j=1$, 
$\langle\mathcal{Q}_{\mathrm{S}}\rangle_0=1-10^{-10}$,
$\langle\mathcal{Q}_{\mathrm{T}}\rangle_0$ $\!=$
$\!1-\langle\mathcal{Q}_{\mathrm{S}}\rangle_0$,
$\eta_{\mathrm{T}}^{(\mathrm{D})}=0$;
(a) $\eta_{\mathrm{S}}^{(\mathrm{D})}=0.9$,
(b) $\eta_{\mathrm{S}}^{(\mathrm{D})}=1$.
}
\end{figure}

While in the special case of a perfect dark evolution,
$\mathrm{Tr}\hat{\varrho}_{\mathrm{N}}(\infty)$ $\!=$ $\!0$, the survival of the
RP follows from the absence of fluorescence,
$\mathrm{Tr}\hat{\varrho}_{\mathrm{R}}(t)$ $\!\equiv$ $\!1$, this does not hold
in the general case due to the possibility of missing the fluorescence photon or
the possibility that the RP had never been created to begin with. This
general behavior is shown in Fig.~\ref{fig3} for $\tilde{r}_j$ $\!=$ $\!1$,
i.e., the case of von Neumann encounters discussed below in (\ref{vNe}).
Since $r$ is the encounter rate, $rt$ is the time in units of the average
waiting time between encounters, i.e., the average number of encounters. The
solid blue curve shows the conditional (dark) survival probability of the
RP given by (\ref{TrRhoR}) as a function of the number of encounters. As
demonstrated in Fig.~\ref{fig3}(a), for significant inefficiencies in the
preparation of the RP
(such that $\sum_{j={\mathrm{S}},{\mathrm{T}}}\langle\mathcal{Q}_j\rangle_0$
$\!=$ $\!\langle\mathcal{Q}_{\mathrm{R}}\rangle_0$ $\!<$ $\!1$) or fluorescence
detection ($\eta_{\mathrm{S}}^{(\mathrm{D})}\eta_{\mathrm{T}}^{(\mathrm{D})}$
$\!<$ $\!1$), the conditional survival probability is not qualitatively
different from the unconditional one describing unobserved decay, shown as
dotted green line. For a near perfect dark evolution as shown in
Fig.~\ref{fig3}(b), the conditional survival probability of the RP undergoes an
abrupt collapse. This is in agreement with the intuitive reasoning of an
observer waiting for a fluorescence without avail. Initially, one would simply
assume that the RP has not yet recombined, but eventually conclude that the
preservation of the RP has failed, namely that either a fluorescence has been
missed or the RP-creation itself has failed. [Note that in (\ref{RhoN}), the
initial state is a mixture of its RP and reaction product component,
$\hat{\varrho}_{\mathrm{N}}(0)$ $\!=$ $\!\hat{\varrho}_{0\mathrm{P}}$ $\!+$
$\!\hat{\varrho}_{0\mathrm{R}}$.] As Fig.~\ref{fig3}(b) illustrates, this change
of reasoning happens at some critical transition time, which we may call the
dark survival time $t_{\mathrm{max}}$ of the RP, and which is determined by the
detection efficiency.

\begin{figure*}[ht]
\includegraphics[width=16cm]{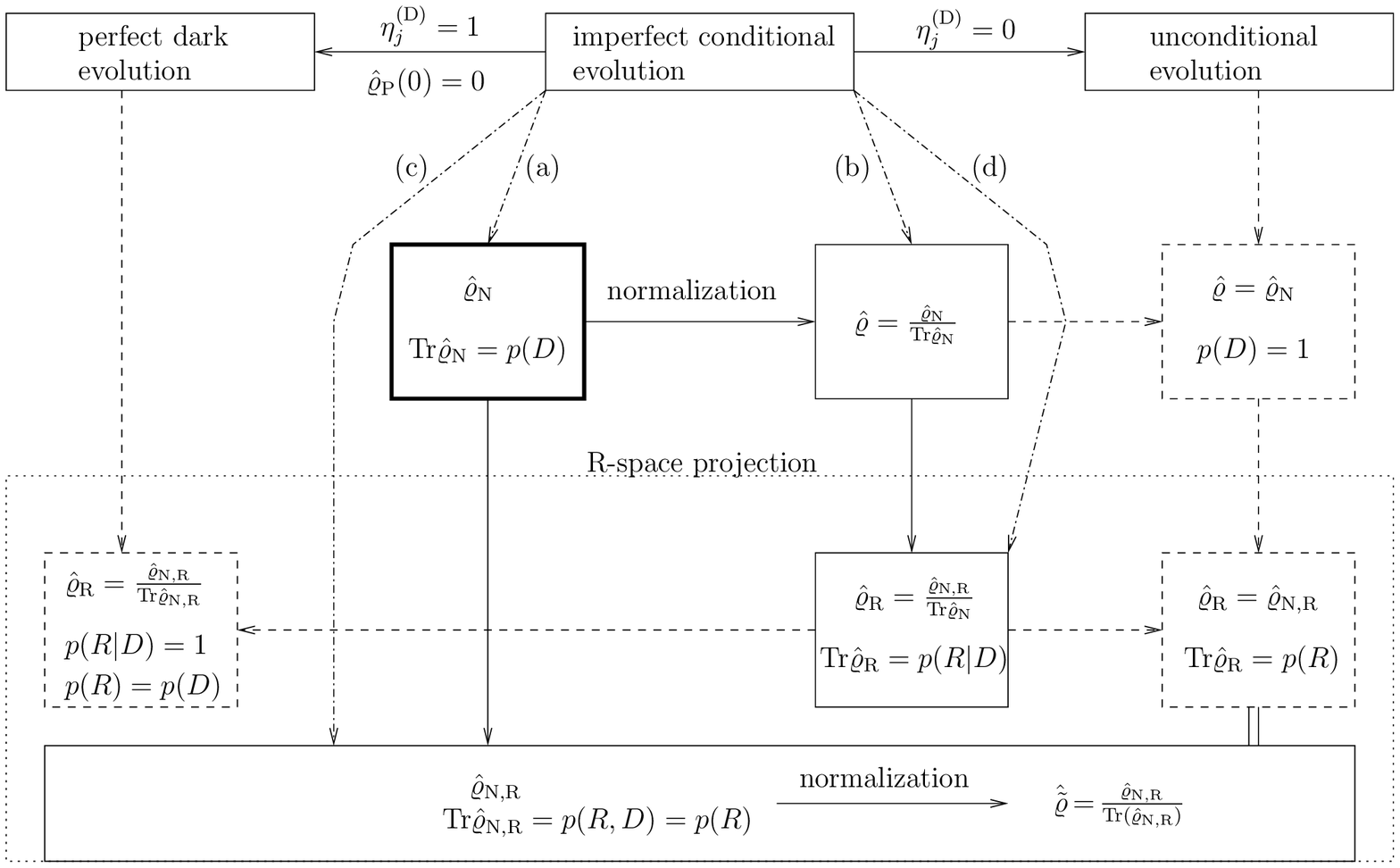}
\caption{\label{fig6b}
Summary of the different states distinguished in the text.
$\hat{\varrho}_{\mathrm{N}}$ is the non-normalized state of the chemical system,
including the reaction products corresponding to an imperfect dark evolution.
It is the solution of the master equation (a),
$\frac{\partial}{\partial{t}}\hat{\varrho}_{\mathrm{N}}$ $\!=$
$\!\mathcal{L}\hat{\varrho}_{\mathrm{N}}$, where the different possibilities
for $\mathcal{L}$ are summarized in Fig.~\ref{fig6a} below. Its normalized
version $\hat{\varrho}$ is the solution of the nonlinear master equation (b),
$\frac{\partial}{\partial{t}}\hat{\varrho}$ $\!=$
$\!(\mathcal{L}-\langle\mathcal{L}\rangle)\hat{\varrho}$. Both states can be
projected onto the radical pair (R)-subspace by omitting the reaction product
part. The projection $\hat{\varrho}_{\mathrm{N,R}}$ of
$\hat{\varrho}_{\mathrm{N}}$ solves the projected equation (c),
$\frac{\partial}{\partial{t}}\hat{\varrho}_{\mathrm{N,R}}$ $\!=$
$\!\mathcal{L}_{\mathrm{R}}\hat{\varrho}_{\mathrm{N,R}}$.
The projection $\hat{\varrho}_{\mathrm{R}}$ of
$\hat{\varrho}$ solves the projected nonlinear master equation (d),
$\frac{\partial}{\partial{t}}\hat{\varrho}_{\mathrm{R}}$ $\!=$
$\!(\mathcal{L}_{\mathrm{R}}-\langle\mathcal{L}\rangle)
\hat{\varrho}_{\mathrm{R}}$. The respective traces are the probabilities of
absence of fluorescence $p(D)$, absence of recombination $p(R)$ (i.e., the
expectation value of the projector onto the R-space), absence of both
fluorescence and recombination $p(R,D)$ and absence of recombination conditioned
on the absence of fluorescence $p(R|D)$, all up to time $t$, respectively.
Perfect detectors and an initial absence of a reaction product,
$\hat{\varrho}_{\mathrm{P}}(0)$ $\!=$ $\!0$, describe a perfect dark evolution,
whereas the absence of any measurement describes an unconditional evolution.
Both normalization and R-projection of the conditional states remove information
contained in their traces. In particular, the projected equation (c) is
independent of the detection efficiencies $\eta_j^{(\mathrm{D})}$, so that
$\hat{\varrho}_{\mathrm{N,R}}$ is unique including the limiting cases of
perfect dark and unconditional evolution. Since for an unconditional evolution
$\hat{\varrho}_{\mathrm{N,R}}$ $\!=$ $\!\hat{\varrho}_{\mathrm{R}}$ holds, we
omit the index N in these cases.
}
\end{figure*}

To give a simple estimate, we define $t_{\mathrm{max}}$ by the condition
$\mathrm{Tr}\hat{\varrho}_{\mathrm{R}}(t_{\mathrm{max}})=\frac{1}{2}$ and assume
equal relaxation rates, $\tilde{r}_j$ $\!=:$ $\!R$, and detection efficiencies,
$\eta_j^{(\mathrm{D})}$ $\!=:$ $\!\eta^{(\mathrm{D})}$, which gives
\begin{eqnarray}
\label{tmax}
  rt_{\mathrm{max}}&=&\frac{1}{R}\ln\frac{
  (2-\eta^{(\mathrm{D})})\langle\mathcal{Q}_{\mathrm{R}}\rangle_0}
  {1-\eta^{(\mathrm{D})}\langle\mathcal{Q}_{\mathrm{R}}\rangle_0}
  \\
\label{tmaxapp}
  &\approx&
  \frac{1}{R}\ln\frac{1}{\varepsilon_{\mathrm{R}}+\varepsilon_{\mathrm{D}}}.
\end{eqnarray}
In (\ref{tmaxapp}) we have assumed small errors (inefficiencies) of preparation,
$\langle\mathcal{Q}_{\mathrm{R}}\rangle_0$ $\!=$
$\!1$ $\!-$ $\!\varepsilon_{\mathrm{R}}$,
$\varepsilon_{\mathrm{R}}$ $\!\ll$ $\!1$, 
and detection,
$\eta^{(\mathrm{D})}$ $\!=$ $\!1$ $\!-$ $\!\varepsilon_{\mathrm{D}}$,
$\varepsilon_{\mathrm{D}}$ $\!\ll$ $\!1$. Eq.~(\ref{tmaxapp}) shows that the
survival time scales logarithmically with the errors, i.e., preparation of
longlived RPs by continuous measurement and postselection is not
robust with respect to experimental errors. A summary of the different types of
states discussed is given in Fig.~\ref{fig6b}.
\section{
\label{sec7}
Ensemble fluorescence}
The model so far assumes a single chemical system, and the master equation
describes the statistics obtained by repeating the experiment in independent
trials with this single system. In reality, if RPs are excited by a
light pulse, a number of them will be created at once. Similarly, a
photodetector may not be able to resolve from which location a given
fluorescence signal came from. Consider therefore a homogeneous ensemble
(a ``cloud'') of $n$ (sufficiently separated to neglect non-geminate encounters)
chemical systems diffusing in a solvent.
\subsection{Single effect $\mathcal{A}$ has occurred somewhere in the cloud}
As a first step, we consider the transformation of an individual system state if
we know that some transformation $\mathcal{A}$ has occurred within the cloud,
but we don't know where. We may think of $\mathcal{A}$ as describing an
encounter leading to a given (out of a set of possible) state transformations,
hence we do not demand $\mathcal{A}$ to be trace-preserving.
For the transformation of the total state, let us
therefore assume a symmetrized map
\begin{equation}
\label{Atots}
  \mathcal{A}_{\mathrm{tot}}=\frac{1}{n}\sum_{j=1}^n\mathcal{A}^{(j)},
\end{equation}
that acts on the total state
$\hat{\varrho}_{\mathrm{tot}}$ $\!=$ $\!\hat{\varrho}^{\otimes{n}}$ of the
cloud, i.e., the $n$ systems are initially prepared in the same state
$\hat{\varrho}$. We consider the resulting transformation of the state of a
single system $s$ from $\hat{\varrho}$ $\!=$ $\!\hat{\varrho}^{(s)}$ $\!=$
$\!\mathrm{Tr}_{\neq{s}}(\hat{\varrho}_{\mathrm{tot}})$ to
$\hat{\varrho}|_{\mathcal{A}}$ $\!=$
$\!\mathrm{Tr}_{\neq{s}}(\hat{\varrho}_{\mathrm{tot}}|_{\mathcal{A}})$,
i.e., we disregard any correlations between the systems created by
(\ref{Atots}). $\mathrm{Tr}_{\neq{s}}$ $\!=$ $\!\mathrm{Tr}_{1,\ldots,n\neq{s}}$
denotes the trace over all except system $s$. Omitting the system label $s$,
this gives for the transformed state 
\begin{eqnarray}
  &&\hat{\varrho}_{\mathrm{tot}}|_{\mathcal{A}}=\frac{1}{p_{\mathrm{tot}}}
  \mathcal{A}_{\mathrm{tot}}\hat{\varrho}_{\mathrm{tot}},
  \\
  &&\hat{\varrho}|_{\mathcal{A}}
  =\left(\frac{n-1}{n}
  +\frac{1}{n}\frac{\mathcal{A}}{\langle\mathcal{A}\rangle}\right)
  \hat{\varrho},
\end{eqnarray}
where the probability of the total transformation (\ref{Atots}) is
$p_{\mathrm{tot}}$ $\!=$
$\!\mathrm{Tr}(\mathcal{A}_{\mathrm{tot}}\hat{\varrho}_{\mathrm{tot}})$ $\!=$
$\!\langle\mathcal{A}\rangle$ $\!\equiv$
$\!\mathrm{Tr}(\mathcal{A}\hat{\varrho})$. We see that despite $\mathcal{A}$
having a finite effect on $\hat{\varrho}$, the change of state
\begin{equation}
\label{deltarho}
  \hat{\varrho}|_{\mathcal{A}}-\hat{\varrho}=\frac{1}{n}\left(
  \frac{\mathcal{A}}{\langle\mathcal{A}\rangle}-1\right)
  \hat{\varrho}
\end{equation}
becomes infinitesimal with increasing $n$, because the action of $\mathcal{A}$
is uniformly distributed over the whole cloud. [Note that this conclusion holds
for a given event $\mathcal{A}$. If all possible events are considered, then
(\ref{deltarho}) can still become arbitrarily large for given $n$ if a highly
unlikely event occurs.]
\subsection{Random occurrence of effect $\mathcal{A}$ in cloud}
As a second step, we consider the case that the event $\mathcal{A}$ occurs at
any location at random with a rate $r$. The analysis can be carried out in
close analogy to the case of multiple encounters. We consider a small time
interval $\mathrm{d}t$, such that the probability that a given system
experiences an encounter during $\mathrm{d}t$ is $p$ $\!=$ $\!r\mathrm{d}t$, and
we neglect the possibility of having more than one encounter during
$\mathrm{d}t$. The probability of having $k$ encounters during $\mathrm{d}t$
within the whole cloud is then again given by a binomial distribution
(\ref{binomdist}). If we neither know how many nor where encounters have
occurred, the overall transformation is a mixture of symmetrized maps,
\begin{eqnarray}
\label{Atot}
  \mathcal{A}_{\mathrm{tot}}
  &=&\sum_{k=0}^n{p}_k\mathcal{A}^{\otimes{k}}_{\mathrm{sym}},
  \\
\label{Atotk}
  \mathcal{A}^{\otimes{k}}_{\mathrm{sym}}
  &=&\binom{n}{k}^{-1}\sideset{}{'}\sum_{i_1,\ldots,i_k=1}^{n}
  \mathcal{A}^{(i_k)}\otimes\cdots\otimes\mathcal{A}^{(i_1)},
\end{eqnarray}
where $\sum^{'}$ runs over all $\binom{n}{k}$ possible selections of $k$ out of
$n$ systems experiencing an encounter and the factor $\binom{n}{k}^{-1}$ takes
into account that ${p}_k$ already contains the sum over all selections. Assume
that each of the $n$ individual systems is prepared in a state $\hat{\varrho}$.
The total map (\ref{Atot}) hence acts on the total state
$\hat{\varrho}_{\mathrm{tot}}$ $\!=$ $\!\hat{\varrho}^{\otimes{n}}$.
The state of the total system evolved by $\mathrm{d}t$ thus reads
\begin{eqnarray}
\label{sctot}
  &&\hat{\varrho}_{\mathrm{tot}}|_{\mathcal{A}}=\frac{1}{p_{\mathrm{tot}}}
  \mathcal{A}_{\mathrm{tot}}\hat{\varrho}_{\mathrm{tot}},
  \\
\label{ptot}
  &&p_{\mathrm{tot}}
  =\mathrm{Tr}(\mathcal{A}_{\mathrm{tot}}\hat{\varrho}_{\mathrm{tot}})
  =\bigl[1+p(\langle\mathcal{A}\rangle-1)\bigr]^n.
\end{eqnarray}
Performing again the trace $\hat{\varrho}^{(s)}|_{\mathcal{A}}$ $\!=$
$\!\mathrm{Tr}_{\neq{s}}(\hat{\varrho}_{\mathrm{tot}}|_{\mathcal{A}})$
and taking into account that the outcome does not depend on the location $s$ of
the individual system, we can omit the system index $s$ and write
\begin{equation}
\label{sc}
  \hat{\varrho}|_{\mathcal{A}}
  =\frac{1+p(\mathcal{A}-1)}{1+p(\langle\mathcal{A}\rangle-1)}\hat{\varrho}.
\end{equation}
While (\ref{sc}) does not require $p$ to be small, the derivation of a master
equation requires an infinitesimal change of state from
$\hat{\varrho}(t)$ $\!=$ $\!\hat{\varrho}$ to
$\hat{\varrho}(t+\mathrm{d}t)$ $\!=$ $\!\hat{\varrho}|_{\mathcal{A}}$. We 
perform in (\ref{sc}) a Taylor expansion to first order in $p$, which gives
\begin{equation}
\label{sct}
  \hat{\varrho}(t+\mathrm{d}t)\approx
  \bigl[1+p(\mathcal{A}-\langle\mathcal{A}\rangle)\bigr]
  \hat{\varrho}(t),
\end{equation}
leading to the nonlinear master equation (\ref{nlme}) together with
(\ref{LofA}), which we have derived in the beginning,
\begin{equation}
\label{me4cloud}
  \frac{\partial}{\partial{t}}\hat{\varrho}
  =r(\mathcal{A}-\langle\mathcal{A}\rangle)\hat{\varrho}.
\end{equation}
If $\mathcal{A}$ preserves the trace, i.e., it is a CPT-map,
$\langle\mathcal{A}\rangle$ $\!=$ $\!1$, the nonlinearity disappears. This is
just the case if $\mathcal{A}$ describes an unmeasured encounter.

The result (\ref{me4cloud}) is not surprising since we have expanded the system
into an ensemble and then again traced out the expansion made. It is clear that
during each time step, the total map (\ref{Atot}) builds up correlations between
the systems. Although they may be of interest for the description of the total
state of the RP ensemble, here, we are not interested in them, since we limit to
the state of one single RP picked out at random from the ensemble.
Therefore we additionally perform a projection onto the factorized state,
\begin{equation}
  \mathcal{P}\hat{\varrho}_{\mathrm{tot}}=
  \mathrm{Tr}_{\neq1}(\hat{\varrho}_{\mathrm{tot}})
  \otimes
  \mathrm{Tr}_{\neq2}(\hat{\varrho}_{\mathrm{tot}})
  \otimes\cdots\otimes
  \mathrm{Tr}_{\neq{n}}(\hat{\varrho}_{\mathrm{tot}}),
\end{equation}
similar to the projection onto the relevant part in the derivation of the
time-convolutionless and Nakajima-Zwanzig master equations in App.~\ref{secPES}.
Note that a (partial) trace describes ignorance of information rather than a
physical process.
\subsection{Given number $l$ of fluorescence signals from cloud}
We now consider the case that a photodetector has detected $l$ fluorescence
signals from a cloud of $n$ chemical systems. For simplicity of notation, we
denote by
$\mathcal{B}$ $\!=$ $\!\sum_{j={\mathrm{S}},{\mathrm{T}}}\mathcal{A}_j$ the
transformation associated with a fluorescence click and by $\mathcal{A}$
$\!=$ $\!\mathcal{A}_{\mathrm{CPT}}$ $\!-$ $\!\mathcal{B}$ the complementary
transformation associated with no fluorescence click. Note that the special case
$\eta_{\mathrm{S}}^{(\mathrm{D})}$ $\!=$ $\!1$ and 
$\eta_{\mathrm{T}}^{(\mathrm{D})}$ $\!=$ $\!0$ describes perfect singlet
fluorescence detection.
$\mathcal{A}_{\mathrm{CPT}}$ describes an encounter as such without performing
any measurements, hence
$\langle\mathcal{A}\rangle$ $\!+$ $\!\langle\mathcal{B}\rangle=1$. We proceed as
above but replace
(\ref{Atotk}) with 
\begin{eqnarray}
  \mathcal{A}^{\otimes{k,l}}_{\mathrm{sym}}
  &=&\binom{n}{k}^{-1}\sideset{}{'}\sum_{i_1,\ldots,i_k=1}^{n}\quad
  \sideset{}{'}\sum_{j_1,\ldots,j_l=1}^{k}
  \nonumber
  \\
\label{Akl}
  &\times&
  \prod_{s=1}^k
  [\delta_{\{j_l\},s}\mathcal{B}^{(i_s)}
  +(1-\delta_{\{j_l\},s})\mathcal{A}^{(i_s)}],
\end{eqnarray}
where $\sideset{}{'}\sum_{j_1,\ldots,j_l=1}^{k}$ runs over all $\binom{k}{l}$
possible selections of $l$ out of the $k$ sites $i_1,\ldots,i_k$, and
$\delta_{\{j_l\},s}$ $\!\equiv$ $\!\sum_{r=1}^l\delta_{j_r,s}$ ensures that
$\mathcal{B}^{(i_s)}$ acts if $i_s$ belongs to these sites selected, but
$\mathcal{A}^{(i_s)}$ acts if $i_s$ does not belong to these sites selected.
In other words, in each term in (\ref{Atotk}), i.e., among the $k$ sites which
may have experienced an encounter, we sum up the distributions of the $l$ sites
which may have been the origin of the $l$ clicks detected.

Instead of (\ref{Atotk}), we now use (\ref{Akl}) in (\ref{Atot}), which replaces
(\ref{sctot}) with
\begin{eqnarray}
  &&\hat{\varrho}_{\mathrm{tot}}|_l=\frac{1}{p(l)}
  \mathcal{A}_{\mathrm{tot}}|_l\hat{\varrho}_{\mathrm{tot}},
  \\
  &&\mathcal{A}_{\mathrm{tot}}|_l
  =\sum_{k=l}^n{p}_k\mathcal{A}^{\otimes{k,l}}_{\mathrm{sym}},
  \quad
  p(l)
  =\mathrm{Tr}(\mathcal{A}_{\mathrm{tot}}|_l\hat{\varrho}_{\mathrm{tot}}).
\end{eqnarray}
After some elementary algebra, we obtain analogous to (\ref{sc}) the
transformation of the individual system state $\hat{\varrho}|_l$ $\!=$
$\!\mathrm{Tr}_{\neq{s}}(\hat{\varrho}_{\mathrm{tot}}|_l)$ $\!=$
$\!\frac{1}{p(l)}\mathrm{Tr}_{\neq{s}}
(\mathcal{A}_{\mathrm{tot}}|_l\hat{\varrho}_{\mathrm{tot}})$ with
$\mathrm{Tr}_{\neq{s}}
(\mathcal{A}_{\mathrm{tot}}|_l\hat{\varrho}_{\mathrm{tot}})$ $\!=:$
$\!p(l)\mathcal{M}_l\hat{\varrho}$, so that $p(l)$ cancels out and
\begin{eqnarray}
  &&\hat{\varrho}|_l
  =\mathcal{M}_l\hat{\varrho},
  \\
\label{Ml}
  &&\mathcal{M}_l=(1-x)\frac{
  (1-p)+p\mathcal{A}}{1-p\langle\mathcal{B}\rangle}
  +x\frac{p\mathcal{B}}{p\langle\mathcal{B}\rangle},
\end{eqnarray}
where $x$ $\!=$ $\!\frac{l}{n}$ $\!\in$ $\![0,1]$ is the ratio of clicks
with respect to the number of chemical systems, i.e., the instantaneous
fluorescence yield. The probability of obtaining $l$ clicks is given by the
binomial distribution (\ref{binomdist}),
\begin{equation}
\label{sstat}
  p(l)=\mathrm{b}_{ln}\left(p\langle\mathcal{B}\rangle\right).
\end{equation}
$\mathcal{M}_l$ is by definition trace-preserving but nonlinear. Only if $l$
happens to coincide with the expectation value of (\ref{sstat}),
$x$ $\!=$ $\!p\langle\mathcal{B}\rangle$, the nonlinearity disappears and
(\ref{Ml}) reduces to
\begin{equation}
  \mathcal{M}_{np\langle\mathcal{B}\rangle}
  =(1-p)+p\mathcal{A}_{\mathrm{CPT}}
  =\sum_lp(l)\mathcal{M}_l,
\end{equation}
which corresponds to (\ref{me4cloud}) with $\mathcal{A}_{\mathrm{CPT}}$ plugged
in. In particular, $\mathcal{M}_0$ reproduces (\ref{sc}) with $\mathcal{A}$
describing an encounter without a fluorescence. The corresponding
probability is in agreement with (\ref{ptot}). The other extreme case is that
all systems in the cloud contribute a fluorescence click,
$\mathcal{M}_n$ $\!=$ $\!(p\langle\mathcal{B}\rangle)^{-1}p\mathcal{B}$, which
occurs with probability $p(n)$ $\!=$ $\!(p\langle\mathcal{B}\rangle)^n$.
If the cloud consists just of a single system, $n$ $\!=$ $\!1$, then
$\mathcal{M}_0$ and $\mathcal{M}_n$ are the only events possible.

The variance of (\ref{sstat}) is given by
$np\langle\mathcal{B}\rangle(1-p\langle\mathcal{B}\rangle)$.
To derive the evolution equation corresponding to (\ref{Ml}), we define
$z$ $\!=$ $\!(x-p\langle\mathcal{B}\rangle)/
[p\langle\mathcal{B}\rangle(1-p\langle\mathcal{B}\rangle)]$, use it to
substitute $x$, and further substitute
$\mathcal{A}$ $\!=$ $\!\mathcal{A}_{\mathrm{CPT}}-\mathcal{B}$. Analogous to
(\ref{sc}), performing in (\ref{Ml}) a Taylor expansion to first order in $p$
and considering $(\hat{\varrho}|_l-\hat{\varrho})/\mathrm{d}t$ gives the master
equation
\begin{eqnarray}
\label{sME}
  \frac{\partial}{\partial{t}}\hat{\varrho}
  &=&r\bigl[(\mathcal{A}_{\mathrm{CPT}}-1)
  +z(\mathcal{B}-\langle\mathcal{B}\rangle)\bigr]\hat{\varrho},
  \\
\label{x2z}
  z&=&\frac{\dot{x}-r\langle\mathcal{B}\rangle}{r\langle\mathcal{B}\rangle}.
\end{eqnarray}
Here, $\dot{x}$ is the number $l$ of fluorescence clicks obtained between $t$
and $t$ $\!+$ $\!\mathrm{d}t$, divided by $n$, hence $z(t)$ describes the
fluctuations of the fluorescence signal around its mean value and so does the
nonlinearity of (\ref{sME}), which increases with $|z|$. Typically, these
fluctuations are small and the same holds for the nonlinearity. In the limit of
large $n$, (\ref{sstat}) can be approximated by a Poisson or, in the continuous
limit, by a Gaussian distribution. Eq.~(\ref{sME}) remains valid in the latter
case, with $z(t)$ representing a Wiener process (with a variance not normalized
to one since we have chosen the denominator in (\ref{x2z}) such that (\ref{sME})
attains a simple form). 
Eq.~(\ref{sME}) opens the possibility to describe future experiments, where the
additional information gained from variation of the fluorescence signal of the
ensemble can be used to refine the model and with it the predictions of the
RP-state evolution as opposed to an approach based on the singlet yield alone.
This may also allow to compare different theories \cite{Del11}.
\subsection{Outlook on quantum control}
A possibility to modify the evolution of the RP state is to implement a quantum
control scheme, which in a general sense means that we introduce a time
dependence on demand into the evolution operator. One particular kind of control
suggested by (\ref{sME}) is a feedback loop, in which case this time dependence
is determined by the measurement record (\ref{x2z}). It must be stressed that in
contrast to a single RP which can generate at most one single fluorescence click
within its lifetime, (\ref{sME}) refers to a RP-ensemble, and hence a continuous
stream of clicks whose statistics can \emph{change} over time and be modified.
This feedback of $z$ (or $\dot{x}$) could act on an external magnetic field
which would modify the free Hamiltonian, $\hat{H}$ $\!=$ $\!\hat{H}[z(t)]$
(``magnetic control''), or it could act on auxiliary laser fields modifying the
encounter rate $r$ $\!=$ $\!r[z(t)]$, applying the optical ``switch''
investigated in \cite{Gue12} (``mechanical control''). Alternatively, one could
extend the model to allow for a continuous (e.g. optical) creation of new RPs,
and control their creation rate depending on $z(t)$ (``optical control''). 
\section{
\label{sec8}
Special cases of the simplified encounter model and their relationship to
existing work}
\subsection{
\label{sec8.1}
Types of encounters}
\subsubsection{Bright encounters}
Bright encounters are characterized by the maximum weight of recombination,
$\tilde{r}_j$ $\!=$ $\!1$, which is obtained if in the case of pure relaxation,
$\delta_j$ $\!=$ $\!0$, we set $\varphi_j$ $\!=$ $\!(k_j+\frac{1}{2})\pi$,
$k_j$ $\!\in$ $\!\mathbb{Z}$. As a consequence, $\tilde{\eta}$ $\!=$ $\!1$ and
\begin{equation}
\label{vNe}
  \mathcal{A}_j=\langle\mathcal{Q}_j\rangle\hat{Q}_{\mathrm{P}},
  \quad
  \mathcal{A}_0=\mathcal{Q}_{\mathrm{P}}.
\end{equation}
In this special case, which we will call a ``von Neumann encounter'', a
projective von Neumann measurement onto the S, T, P subspaces is performed,
followed by a transition to P. Just as an idealized photodetector performs a
photon number measurement by absorbing the radiation, this type of encounter
performs an S-T-measurement by recombining the RP. Eq.~(\ref{U0})
leads in this case to a maximum correlation between chemical system and
environment, $\hat{U}|0\rangle$ $\!=$ $\!\hat{Q}_{\mathrm{P}}|0\rangle$ $\!-$
$\!\mathrm{i}\sum_{j={\mathrm{S}},{\mathrm{T}_i}}\frac{\pi_j}{|\pi_j|}(-1)^{k_j}
\hat{L}_j|\pi_j\rangle$. If all encounters were of this type, the initial state
remained unchanged before the first fluorescence signal, 
$\mathcal{A}_{0\mathrm{R}}$ $\!=$ $\!0$, $\tilde{\eta}_j$ $\!=$ $\!1$.
Setting $\delta_j$ $\!=$ $\!0$ alone is not sufficient, since (\ref{A0rho}) then
still contains pure dephasing terms in general.
\subsubsection{Dark encounters}
Dark encounters are characterized by $\tilde{r}_j$ $\!=$ $\!0$, and hence the
probability of recombination becomes zero, $\mathcal{A}_j$ $\!=$ $\!0$. This is
the case either for (a) pure dephasing $\pi_j$ $\!=$ $\!0$, in which case
(\ref{A0rhoD}) reduces to
$\mathcal{A}_0$ $\!=$ $\!(1-\tilde{\eta})$ $\!+$ $\!\tilde{\eta}\mathcal{Q}$
with $\tilde{\eta}$ $\!\in$ $\![0,2]$ and $\mathcal{Q}$ defined in (\ref{STPQ}).
In particular, $\tilde{\eta}$ $\!=$ $\!1$ can be described as perfect dephasing,
$\mathcal{A}_0$ $\!=$ $\!\mathcal{A}$ $\!=$ $\!\mathcal{Q}$.
In case (b), $\varphi_j$ $\!=$ $\!k_j\pi$, $k_j$ $\!\in$ $\!\mathbb{Z}$, for
which
\begin{eqnarray}
\label{darkeven}
  \tilde{\eta}&=&0:\quad\mathcal{A}_0=1\hspace{1.25cm}
  (k_{\mathrm{S}}\!+\!k_{\mathrm{T}}\hspace{0.2cm}\rm{even}),
  \\
\label{darkodd}
  \tilde{\eta}&=&2:\quad\mathcal{A}_0=2\mathcal{Q}-1\quad
  (k_{\mathrm{S}}\!+\!k_{\mathrm{T}}\hspace{0.2cm}\rm{odd}).
\end{eqnarray}
$\mathcal{A}_0$ here describes a unitary transformation, which is  in
(\ref{darkeven}) the identity, and in (\ref{darkodd}) a reflection (i.e., Grover
diffusion map \cite{Gro96} or Householder transformation)
). This is not in contradiction to the open
system dynamics, since (\ref{U0}) shows that the environment ends in its initial
state $|0\rangle$, $\hat{U}|0\rangle$ $\!=$ $\![\hat{Q}_{\mathrm{P}}$ $\!+$
$\!\sum_{j={\mathrm{S}},{\mathrm{T}}}(-1)^{k_j}\hat{Q}_j]|0\rangle$.
Eq.~(\ref{darkeven}) includes the trivial encounter $\varphi_j$ $\!=$ $\!0$ as
a special case.
\subsubsection{Average encounters}
The free parameters allow for a continuous manifold of possible encounters. If
the encounters can be controlled, e.g., by means of the mechanical motion of the
radicals, their parameters can be chosen as desired. In practice, the motion of
the radicals will be uncontrolled, however, and the parameters of the encounters
are unknown. The unitary transformation
$\mathcal{U}$ $\!=$ $\!\hat{U}\bm{\cdot}\hat{U}^\dagger$ in (\ref{Ai}), and
consequently in (\ref{Ajm}) and (\ref{A0m}) must then be replaced with an
average according to
\begin{equation}
\label{twirling}
  \mathcal{U}
  \to\sum_\lambda{p}_\lambda\mathcal{U}^{(\lambda)}
  =\sum_\lambda{p}_\lambda\hat{U}^{(\lambda)}\bm{\cdot}
  \hat{U}^{(\lambda)\dagger},
\end{equation}
and hence
$\mathcal{A}_j$ $\!\to$ $\!\sum_\lambda{p}_\lambda\mathcal{A}_j^{(\lambda)}$
and
$\mathcal{A}_0$ $\!\to$ $\!\sum_\lambda{p}_\lambda\mathcal{A}_0^{(\lambda)}$,
where the ${p}_\lambda$ form some (discrete or continuous) probability
distribution. In general, the ${p}_\lambda$ depend on the experimental details
such as the statistics of the motion of the radicals. In the special case of
triplet symmetry without triplet dephasing, for which we assume that both
(\ref{symmcondmaps1}) and (\ref{symmcondmaps1a}) hold in all
$\hat{U}^{(\lambda)}$, we replace (\ref{Ajrho}) and (\ref{A0rho}) with
\begin{eqnarray}
\label{Ajrhom}
  \mathcal{A}_j&=&\overline{r}_j
  \langle\mathcal{Q}_j\rangle\hat{Q}_{\mathrm{P}},\quad
  (j={\mathrm{S}},{\mathrm{T}}),
  \\
\label{A0rhom}
  \mathcal{A}_0
  &=&(1-\overline{\eta})
  +\overline{\eta}\Bigl[\mathcal{Q}_{\mathrm{P}}
  +\sum_{j={\mathrm{S}},{\mathrm{T}}}
  (1-{\textstyle\frac{\overline{r}_j}{\overline{\eta}}})\mathcal{Q}_j\Bigr].
\end{eqnarray}
The form of the encounter maps thus remains the same, but their coefficients are
replaced with the averages
\begin{equation}
  \overline{r}_j=\sum_\lambda{p}_\lambda\tilde{r}_j^{(\lambda)},
  \quad
  \overline{\eta}=\sum_\lambda{p}_\lambda
  \tilde{\eta}^{(\lambda)},
\end{equation}
where the $\tilde{r}_j^{(\lambda)}$ and $\tilde{\eta}^{(\lambda)}$
correspond to the maps constructed with $\hat{U}^{(\lambda)}$.
The maximum uncertainty case then corresponds to a uniform distribution of all
possible $\hat{U}^{(\lambda)}$, for which (\ref{twirling}) describes the
(unbiased) twirling operation. Here, we just mention that since
$\tilde{\eta}^{(\lambda)}\in[0,2]$, there exist distributions that yield
$\overline{\eta}$ $\!=$ $\!1$ for which (\ref{A0rhom}) simplifies to
$\mathcal{A}_0$ $\!=$ $\!\mathcal{Q}_{\mathrm{P}}$ $\!+$
$\!\sum_{j={\mathrm{S}},{\mathrm{T}}}(1-\overline{r}_j)\mathcal{Q}_j$. In any
case, averaging allows us to reduce a whole manifold of possible encounters to a
single ``effective'' type of encounter.
\subsubsection{Limit of weak encounters}
Recovering the master equation approach from the encounter model in the limit of
weak encounters has already been discussed in Sec.~\ref{secCM} in the general
case involving triplet dephasing. We therefore here restrict to summarizing the
results for the simplified case considered in this section. In the limit
$\kappa$ $\!\ll$ $\!1$, the approximations
\begin{eqnarray}
  \tilde{\eta}&\approx&\frac{\kappa^2}{2}
  \sum_{j={\mathrm{S}},{\mathrm{T}}}(|\pi_j|^2+|\delta_j|^2)\ll1,
  \\
  \tilde{\eta}_j&\approx&2
  \frac{|\pi_j|^2}{\sum_{j={\mathrm{S}},{\mathrm{T}}}(|\pi_j|^2+|\delta_j|^2)}
  \in[0,2],
  \\
  \tilde{r}_j&\approx&\kappa^2|\pi_j|^2\ll1,
\end{eqnarray}
hold. If we set in (\ref{cMEST}) $\kappa^2r$ $\!=$ $\!2t$, cf. the comments
following (\ref{HI}), then $r\tilde{\eta}$ $\!\to$ $\!\eta$, $\tilde{\eta}_j$
$\!\to$ $\!\eta_j$, $r\tilde{r}_j$ $\!\to$ $\!r_j$, and (\ref{cMEST}) reproduces
(\ref{genmeRan}) which has been derived for a time-independent interaction
between the chemical system and its environment. Note that we can in this limit
also simplify the expressions for the maps,
$r\mathcal{A}_j$ $\!\to$ $\!r_j\langle\mathcal{Q}_j\rangle\hat{Q}_{\mathrm{P}}$,
$(j={\mathrm{S}},{\mathrm{T}})$, and
$r(\mathcal{A}_0-1)$ $\!\to$ $\!\sum_{j={\mathrm{S}},{\mathrm{T}}}\Bigl[
d_j\mathcal{L}(\hat{Q}_j)-\frac{r_j}{2}\{\hat{Q}_j,\bm{\cdot}\}\Bigr]$,
cf. (\ref{Ajwls}) and (\ref{A0wls}).
\subsection{
\label{sec8.2}
Singlet yield}
In most RPM-based models for the chemical compass in birds,
it is assumed that the biological signal is determined by the singlet yield
\cite{Sch78,*Sch82,Rit00,*Rod09}. In our model, this quantity can be directly
derived from the normalization of the solution (\ref{lntes}) of the linear
non-trace preserving master equation, i.e., the probability
$p(t)$ $\!=$ $\!\mathrm{Tr}\left[\hat{\varrho}_{\mathrm{N}}(t)\right]$ in
(\ref{rho}). The generator of the evolution is given by (\ref{LofA}),
$\mathcal{L}$ $\!=$ $\!r(\mathcal{A}-1)$. Consider a single chemical system
experiencing occasional encounters. An encounter is described by a CPT-map 
$\mathcal{A}_{\mathrm{CPT}}$ $\!=$ $\!\mathcal{A}_{\mathrm{S}}$ $\!+$
$\!\mathcal{A}_{\mathrm{T}}$ $\!+$ $\!\mathcal{A}_0$. If we substitute for
$\mathcal{A}$ in (\ref{LofA}) the ``no singlet click'' event $\mathcal{A}$ $\!=$
$\!\mathcal{A}_{\mathrm{CPT}}$ $\!-$ $\!\mathcal{A}_{\mathrm{S}}$ $\!=$
$\!\mathcal{A}_0$ $\!+$ $\!\mathcal{A}_{\mathrm{T}}$, then $p(t)$ is the
probability that no singlet click has been obtained within $(0,t]$.
Consequently, $p_{\mathrm{S}}(t)$ $\!=$ $\!1$ $\!-$ $\!p(t)$ is the probability
that a singlet click has been obtained up to time $t$. Consider now $n$ such
chemical systems in a cloud which are assumed to be independent. The singlet
yield $\Phi_{\mathrm{S}}$ of the cloud is given by the expectation value of a
binomial distribution $\mathrm{b}_{ln}(p_{\mathrm{S}})$ as defined in
(\ref{binomdist}),
\begin{eqnarray}
  \Phi_{\mathrm{S}}&=&np_{\mathrm{S}}(\infty),
  \\
\label{pS}
  p_{\mathrm{S}}(t)&=&1-\mathrm{Tr}\left[\hat{\varrho}_{\mathrm{N}}(t)\right],
  \\
\label{rhoR}
  \hat{\varrho}_{\mathrm{N}}(t)
  &=&\mathcal{T}_+\mathrm{e}^{r\int_0^t\mathrm{d}\tau
  (\mathcal{A}_{\mathrm{CPT}}-\mathcal{A}_{\mathrm{S}}-1)(\tau)
  }\hat{\varrho}(0),
\end{eqnarray}
where $\mathcal{A}_j$ and $\mathcal{A}_0$ are given by
(\ref{Ajrho}) and (\ref{A0rho}). It is hence sufficient to consider the single
system yield $p_{\mathrm{S}}$. In what follows, we calculate this quantity in
the special case of the von Neumann encounter defined in (\ref{vNe}).

\begin{figure*}[ht]
\includegraphics[width=16cm]{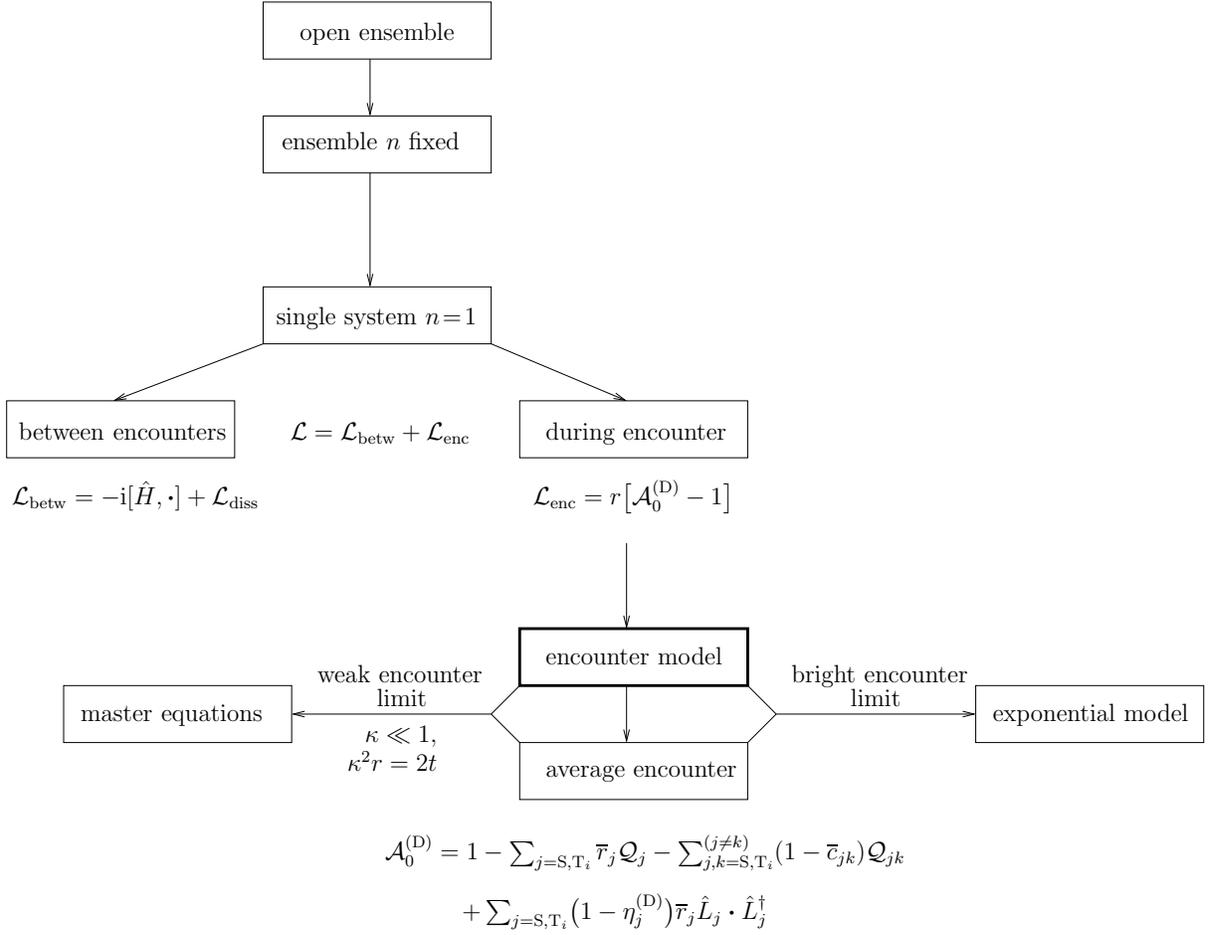}
\caption{\label{fig6a}
Summary of the different generators $\mathcal{L}$ of state evolution of the
chemical system (i.e., geminate radical pairs including recombination products
and nuclear spins) as discussed in the text. Between encounters, the system
state evolves due to internal (hyperfine) and external (Zeeman) interactions,
captured in $\hat{H}$, as well as potential other decoherence effects
$\mathcal{L}_{\mathrm{diss}}$ (not discussed). Encounters occurring at a rate
$r$ are accompanied by strong transient direct (dipolar and exchange)
interactions between the electrons, whose details are unknown, giving rise to
electron spin dephasing. An encounter may be followed by a recombination, i.e.,
transitions $\hat{L}_j$ to the reaction product state, which can be detected
with efficiencies $\eta_j^{(\mathrm{D})}$. All these effects are included in the
description of an encounter as a generalized measurement
$\{\mathcal{A}_j^{(\mathrm{D})},\mathcal{A}_0^{(\mathrm{D})}\}$, whose outcomes
are either $j$ $\!=$ $\!{\mathrm{S}},{\mathrm{T}_i}$ recombination fluorescence
signals or absence of fluorescence ($\mathcal{A}_0^{(\mathrm{D})}$). The latter
gives a contribution $\mathcal{L}_{\mathrm{enc}}$ to the dark state evolution
of a single chemical system that holds until a fluorescence confirms its
recombination.
For an ensemble of $n$ independent chemical systems, the dark evolution can be
generalized to an evolution conditioned on a given record of fluorescence
signals. The more general case of an open ensemble of chemical systems subject
to continuous excitation and recombination is beyond the scope of this work.
The description of the encounters depends on free parameters. Averaging over the
latter leads to the notion of an average encounter. While the limit of frequent
weak encounters recovers the master equation approach, the opposite limit of
strong encounters leads to what we call exponential model [cf. Eq.~(\ref{Sy})],
where a radical pair evolves entirely due to $\mathcal{L}_{\mathrm{betw}}$
unless a single re-encounter leads to its recombination. Each $\mathcal{L}$
obtained with the possible $\mathcal{A}_0^{(\mathrm{D})}$ and its limits can in
turn be used to construct the various master equations as presented in
Fig.~\ref{fig6b}.
}
\end{figure*}

Since the trace in (\ref{pS}) is picture invariant, we first calculate
(\ref{rhoR}) for the case (\ref{vNe}) in the Heisenberg picture, analogous to
(\ref{Heisexp}) (omitting the picture index H for the maps),
\begin{eqnarray}
  \hat{\varrho}_{\mathrm{N,H}}(t)
  &=&\mathcal{T}_+\mathrm{e}^{r\int_0^t\mathrm{d}\tau
  (\mathcal{A}_{\mathrm{T}}+\mathcal{Q}_{\mathrm{P}}-1)(\tau)
  }\hat{\varrho}(0)
  \nonumber\\
  &=&\mathrm{e}^{-rt}\sum_{k=0}^\infty
  \frac{r^k}{k!}\mathcal{T}_+\!\!\int_0^{t}\!\!\mathrm{d}t_k
  \cdots\!\!\int_0^{t}\!\!\mathrm{d}t_2\!\!\int_0^{t}\!\!\mathrm{d}t_1
  \nonumber\\
  &\times&(\mathcal{A}_{\mathrm{T}}\!+\!\mathcal{Q}_{\mathrm{P}})(t_k)
  \cdots(\mathcal{A}_{\mathrm{T}}\!+\!\mathcal{Q}_{\mathrm{P}})(t_1)
  \hat{\varrho}(0).\quad
\end{eqnarray}
Using that $\mathcal{A}_{\mathrm{T}}(t^\prime)\mathcal{A}_{\mathrm{T}}(t)$ $\!=$
$\!\mathcal{A}_{\mathrm{T}}(t^\prime)\mathcal{Q}_{\mathrm{P}}(t)$ $\!=$ $\!0$
together with $\mathcal{Q}_{\mathrm{P}}(t^\prime)\mathcal{A}_{\mathrm{T}}(t)$
$\!=$ $\!\mathcal{A}_{\mathrm{T}}(t)$ and $\mathcal{Q}_{\mathrm{P}}(t)$ $\!=$
$\!\mathcal{Q}_{\mathrm{P}}$, we obtain
\begin{eqnarray}
  \hat{\varrho}_{\mathrm{N,H}}(t)&=&
  \mathrm{e}^{-rt}\Biggl[\sum_{k=0}^\infty
  \frac{r^k}{k!}\mathcal{T}_+\!\!\int_0^{t}\!\!\mathrm{d}t_k
  \cdots\!\!\int_0^{t}\!\!\mathrm{d}t_1
  \mathcal{Q}_{\mathrm{P}}^k
  \nonumber\\
  &+&\sum_{k=1}^\infty
  \frac{r^k}{k!}\mathcal{T}_+\!\!\int_0^{t}\!\!\mathrm{d}t_k
  \cdots\!\!\int_0^{t}\!\!\mathrm{d}t_1
  \mathcal{A}_{\mathrm{T}}(t_1)\Biggr]\hat{\varrho}(0).\quad\quad
\end{eqnarray}
Using further
\begin{equation}
  \mathcal{T}_+\!\!\int_0^{t}\!\!\mathrm{d}t_k
  \cdots\!\!\int_0^{t}\!\!\mathrm{d}t_1\,f(t_1)
  =\int_0^{t}\!\!\mathrm{d}t_1\,k(t-t_1)^{k-1}f(t_1),
\end{equation}
we obtain
\begin{equation}
  \hat{\varrho}_{\mathrm{N,H}}(t)=
  \Bigl[\mathrm{e}^{r(\mathcal{Q}_{\mathrm{P}}-1)t}
  +\int_0^{t}\!\!\mathrm{d}t_1\,r\mathrm{e}^{-rt_1}
  \mathcal{A}_{\mathrm{T}}(t_1)\Bigr]\hat{\varrho}(0).
\end{equation}
Performing the trace and replacing
$\langle\mathcal{A}_{\mathrm{T}}(t_1)\rangle$ $\!=$ $\!1$
$\!-$ $\!\langle\mathcal{A}_{\mathrm{S}}(t_1)\rangle$ $\!-$
$\!\langle\mathcal{Q}_{\mathrm{P}}(0)\rangle$ , we obtain for (\ref{pS})
\begin{equation}
\label{Sy}
  p_{\mathrm{S}}(t)=\int_0^{t}\!\!\mathrm{d}t_1\,r\mathrm{e}^{-rt_1}
  \langle\hat{Q}_{\mathrm{S}}(t_1)\rangle
\end{equation}
in agreement with the exponential model. A summary of the different limits of
the encounter model is given in Fig.~\ref{fig6a}.
\section{
\label{sec9}
Summary}
\subsection{
\label{sec9.1}
Results}
We have formulated a conceptually simple encounter model describing the
recombination of RPs as an interaction with a minimal phenomenological
environment, cf. (\ref{HI}). By speaking of a ``chemical system'', we want to
stress (apart from including the nuclear spins) that we treat the
RP-recombination as transition to a reaction product (P) subspace that is
orthogonal to the (R) subspace corresponding to separated radicals.

The strength of our model interaction increases with the proximity of the
radicals and thus undergoes random fluctuations in general. Encounters of a
moving RP can then be described in an idealized way by switching on the
interaction for short random periods of time, whereas it is switched off when
the radicals are sufficiently separated. While the temporal statistics of the
encounters is assumed to be given classically and independent of the internal
quantum state of the chemical system, the effect of an encounter on its state
is described as a measurement, which is a natural consequence of tracing out the
model environment.

With this approach we achieve three main goals:

\emph{(1) Statistics of encounters}: We connect two complementary types of
approaches to the RP-evolution as limiting cases of our model. Encounter-based
approaches assume that the two electron spins (and their local nuclear spins)
evolve freely until a RP-encounter terminates this evolution and can here be
recovered by assuming a sudden strong model interaction, which inevitably leads
to a RP-recombination. In contrast, master equation-based approaches describe a
continuous conversion to recombination products by means of reaction operators
and are here recovered as the limit of a constant weak model interaction. Both
approaches are equivalent if applied to a large ensemble of independent RPs.

\emph{(2) Effect of encounters}: In our model, encounters can happen by chance
at any time. By choosing the parameters of this interaction, we can alter the
effect that an encounter has on the internal spin state of the RP, allowing for
a large variety of differently strong encounters. In encounter-based approaches
this variety allows for multiple encounters, since an encounter does not
necessarily lead to a RP recombination. In master equation-based approaches,
this variety provides a natural incorporation of a varying degree of dephasing
of the RP-spins that may occur in addition to a mere recombination.
We demonstrate that particular encounters may even deterministically flip the
RP spins without any dephasing or recombination
[Grover reflections, cf. (\ref{darkodd})].
Moreover we predict that the RP state will also undergo a separate dephasing in
the triplet subspace in general, cf. (\ref{30neu}) and analogous (\ref{Amj1}),
(\ref{Am01}). The standard approach of Haberkorn and its measurement-based
extensions can be recovered by setting the individual triplet dephasing
coefficients to zero, which is done in the main part of this work, to facilitate
comparison of the results with existing work. A classical averaging over all
possible types of encounters with a suitable probability distribution introduces
the notion of an average encounter that determines the state evolution as seen
by an experimenter lacking knowledge of the details of the actual encounters as
they happen.

\emph{(3) Conditional evolution and reaction operator controversy}: Our model
demonstrates, in agreement with previous treatments, that the RP spin state can
undergo dephasing prior to its recombination, at a rate that depends on the
experimental details. Independent of that, recently a nonlinear master equation
has been proposed for the evolution of the RP \cite{Kom09,Kom11}.
We explain why we regard this equation as untenable. If it refers to an
unconditional state evolution, it should be linear and when considered as a
conditional evolution, it is of wrong form. Nevertheless, nonlinear master
equations (of a different type) are legitimate in the description of state
evolutions \emph{conditioned} on additional assumptions such as a given
measurement record or subspace projection. We draw an analogy to the micromaser
and explain how nonlinear master equations can serve to describe such
conditional state evolutions.
As a byproduct, this allows us to reconsider Haberkorn's discussion of the
physicality of reaction operators from the viewpoint of quantum measurements,
cf. the paragraph following (\ref{Pex}). In the context of the RPM, we show how
nonlinear master equations must be properly applied to describe the evolution of
the chemical system prior to the detection of its recombination. This analysis
of the ``dark evolution'' shows that for imperfect detection of recombination,
e.g. due to lossy fluorescence detectors or radiationless transitions, the
conditional dark RP-state must be distinguished from the non-normalized
RP-component of the state of the chemical system, cf. Sec.~\ref{sec6}.

\emph{(4) Ensemble fluorescence}: We generalize our results to the evolution of
the RP state conditioned on the fluorescence emerging from a fixed number of
initially prepared and independently evolving RPs. A large number of recombining
RPs emit a continuous fluorescence signal in contrast to a single detector
click obtainable from an individual RP. The nonlinearity of the resulting master
equation (\ref{sME}) is then given by the deviation of the fluorescence signal
from its mean value.

Our analysis is phenomenological in the sense that we do not aim to derive it
from a realistic microscopic model, which would tie it to a particular setting
and introduce uncertainties into estimated parameters due to a lack of knowledge
of the microscopic details. Instead we choose a rigorous and general formulation
from an open quantum systems point of view and leave the free parameters open to
be fit with experiments.
\subsection{
\label{sec9.2}
Outlook}
\emph{(1) Continuous generation of RPs}:
We consider the evolution of RPs as an initial value problem, i.e., they are
created only initially. A continuous generation of new RPs can be accounted for
by a Lindblad term $\mathcal{L}(\hat{L}_j^\dagger)$ describing incoherent
transitions from the P- to the R-subspace. Note that a coherent excitation of
RPs by an auxiliary classical field with complex amplitude $\alpha$ can be
described by adding a driving term $\alpha\hat{L}_j^\dagger+\alpha^*\hat{L}_j$
to the Hamiltonian of the chemical system. This would however generate
coherences between the R and P spaces and be inconsistent with our model, which
(phenomenologically) assumes that the state of the chemical system is always in
a mixture of its separated and recombined forms. Although there is no
fundamental law that forbids superpositions between these forms, we assume that
because of the energy required to excite a RP, such superpositions constitute
highly nonclassical states, which are converted to a classical mixture on a
very short time scale.

\emph{(2) Feedback and control schemes}:
Our extension to ensemble fluorescence provides the basis for novel feedback
and control setups. We provide a simple formal framework for experimental
applications that may suggest gateways towards learning algorithms for RP
state reconstruction or the design of autonomous devices.
In particular, controlling the time and strength of the encounters would allow
one to verify and possibly apply the predicted effects such as the mentioned
separate dephasing in the triplet subspace and the Grover reflections.
\begin{acknowledgments}
The research was funded by the Austrian Science Fund (FWF) through the SFB
FoQuS: F4012.
\end{acknowledgments}
%
\newpage
\appendix
\section{
\label{secME}
Master equation}
\subsection{Time-convolutionless equation}
We first resume the derivation of the time-convolutionless equation as done in
\cite{bookBreuer}. The total evolution is described by
\begin{equation}
\label{me}
  \frac{\partial}{\partial{t}}\hat{\rho}=\mathcal{L}(t)\hat{\rho},
\end{equation}
where $\mathcal{L}(t)\hat{\rho}$ $\!=$
$\!-\mathrm{i}[\hat{H}_{\mathrm{I}}(t),\hat{\rho}]$ may refer to the
interaction Hamiltonian in the interaction picture,
$\hat{H}$ $\!=$ $\!\hat{H}_0$ $\!+$ $\!\hat{H}_{\mathrm{I}}$,
$\hat{H}_{\mathrm{I}}(t)$ $\!=$ $\!\mathrm{e}^{\mathrm{i}t\hat{H}_0}
\hat{H}_{\mathrm{I}}\mathrm{e}^{-\mathrm{i}t\hat{H}_0}$
(setting $\hbar$ $\!=$ $\!1$). Defining two orthogonal
time-independent (super)projectors $\mathcal{P}$ and $\mathcal{Q}$ which resolve
the identity, $\mathcal{P}$ $\!+$ $\!\mathcal{Q}$ $\!=$ $\!1$,
allows for the decomposition of (\ref{me}) into a relevant and an irrelevant
part,
\begin{eqnarray}
\label{meP}
  \frac{\partial}{\partial{t}}\mathcal{P}\hat{\rho}
  &=&(\mathcal{P}\mathcal{L}\mathcal{P}
  +\mathcal{P}\mathcal{L}\mathcal{Q})\hat{\rho},
  \\
\label{meQ}
  \frac{\partial}{\partial{t}}\mathcal{Q}\hat{\rho}
  &=&(\mathcal{Q}\mathcal{L}\mathcal{P}
  +\mathcal{Q}\mathcal{L}\mathcal{Q})\hat{\rho}.
\end{eqnarray}
The irrelevant part (\ref{meQ}) can be formally solved,
\begin{equation}
\label{Qs1}
  \mathcal{Q}\hat{\rho}=\mathcal{G}(t,0)\mathcal{Q}\hat{\rho}_0
  +\Sigma(t,0)(\mathcal{P}+\mathcal{Q})\hat{\rho},
\end{equation}
with
\begin{eqnarray}
   \Sigma(t,0)&=&\int_0^t\mathrm{d}s\,\mathcal{G}(t,s)\mathcal{Q}\mathcal{L}(s)
  \mathcal{P}\mathcal{U}^{-1}(t,s),
  \\
  \mathcal{G}(t,s)
  &=&\mathcal{T}_+
  \mathrm{e}^{\int_s^t\mathrm{d}\tau\mathcal{Q}\mathcal{L}(\tau)},
  \\
\label{meS}
  \mathcal{U}^{-1}(t,s)
  &=&\mathcal{T}_-\mathrm{e}^{-\int_s^t\mathrm{d}\tau\mathcal{L}(\tau)},
\end{eqnarray}
where the backward propagator $\mathcal{U}^{-1}$ describes the inverse of the
unitary evolution of the total system (\ref{me}) using negative time ordering
$\mathcal{T}_-$. (If an inverse does not exist, because the evolution of the
total system is not unitary, then we may aim at the Nakajima-Zwanzig-equation
instead.) For sufficiently small $\mathcal{L}$ the inverse $(1-\Sigma)^{-1}$
exists and (\ref{Qs1}) can be rewritten as
\begin{equation}
\label{Qs2}
  \mathcal{Q}\hat{\rho}=(1-\Sigma)^{-1}\Sigma\,\mathcal{P}\hat{\rho}
  +(1-\Sigma)^{-1}\mathcal{G}(t,0)\mathcal{Q}\hat{\rho}_0,
\end{equation}
which shows that the irrelevant part is given by the relevant part and the
initial condition for the irrelevant part. Inserting (\ref{Qs2}) into
(\ref{meP}) gives the equation for the relevant part
\begin{equation}
\label{TCL}
  \frac{\partial}{\partial{t}}\mathcal{P}\hat{\rho}
  =\mathcal{K}\mathcal{P}\hat{\rho}
  +\mathcal{I}\mathcal{Q}\hat{\rho}_0,
\end{equation}
where
\begin{eqnarray}
  \mathcal{K}&=&\mathcal{P}\mathcal{L}(1-\Sigma)^{-1}\mathcal{P},
  \\
  \mathcal{I}
  &=&\mathcal{P}\mathcal{L}(1-\Sigma)^{-1}\mathcal{G}(t,0)\mathcal{Q}.
\end{eqnarray}
Expanding $(1-\Sigma)^{-1}$ $\!=$ $\!\sum_{n=0}^\infty\Sigma^n$ and 
$\mathcal{G}$, $\mathcal{U}^{-1}$, and $\Sigma$ in powers of $\mathcal{L}$ gives
an expansion
\begin{equation}
\label{leme}
  \mathcal{K}\!=\!\mathcal{P}\mathcal{L}(t)\mathcal{P}\!+\!\int_0^t\mathrm{d}s
  \bigl[\mathcal{P}\mathcal{L}(t)\mathcal{L}(s)\mathcal{P}
  \!-\!\mathcal{P}\mathcal{L}(t)\mathcal{P}\mathcal{L}(s)\mathcal{P}\bigr]
  \!+\!\ldots.
\end{equation}
and a similar one can be given for $\mathcal{I}$. The latter is not needed if
$\mathcal{Q}\hat{\rho}_0$ $\!=$ $\!0$. In that case, it is also possible to use
(\ref{meS}) for the formal solution
$\hat{\rho}$ $\!=$ $\!\mathcal{U}\hat{\rho}_0$ of (\ref{me}) and
project $\mathcal{P}\hat{\rho}$ $\!=$
$\!\mathcal{P}\mathcal{U}\mathcal{P}\hat{\rho}_0$ for which
$\frac{\partial}{\partial{t}}\mathcal{P}\hat{\rho}$ $\!=$
$\!(\mathcal{P}\dot{\mathcal{U}}\mathcal{P})
(\mathcal{P}\mathcal{U}\mathcal{P})^{-1}\mathcal{P}\hat{\rho}$.
Further simplifications are possible if
$\mathcal{P}\mathcal{L}(t_{2n+1})\cdots\mathcal{L}(t_1)\mathcal{P}$ $\!=$ $\!0$
or if $\frac{\partial}{\partial{t}}\mathcal{L}$ $\!=$ $\!0$.
\subsection{
\label{secPES}
Pure environment state}
The considerations so far are independent of a concrete definition of
$\mathcal{P}$. A projection onto a factorized state $\mathcal{P}\hat{\rho}$
$\!=$ $\!\mathrm{Tr}_{\mathrm{E}}(\hat{\rho})\otimes\hat{\rho}_{\mathrm{E}}$
with a time-independent reference state $\hat{\rho}_{\mathrm{E}}$ leads to
a trace preserving master equation.
As an example, we consider (\ref{TCL}) with the two terms given in (\ref{leme})
for the case of a projection with a pure environment state
$\hat{\rho}_{\mathrm{E}}$ $\!=$ $\!|0\rangle\langle0|$,
$\mathcal{Q}\hat{\rho}_0$ $\!=$ $\!0$, and a time-independent
$\hat{H}_{\mathrm{I}}$. (Note that such time-independence in the interaction
picture can be achieved by the rotating wave approximation, i.e., neglecting
oscillating terms.) Defining
\begin{eqnarray}
\label{Lk}
  \hat{c}_k&=&\langle{k}|\hat{H}_{\mathrm{I}}|0\rangle,
  \\
\label{Lc}
  \mathcal{L}(\hat{c})\hat{\varrho}&=&\hat{c}\hat{\varrho}\hat{c}^\dagger
  -\frac{1}{2}\{\hat{c}^\dagger\hat{c},\hat{\varrho}\},
\end{eqnarray}
with orthonormal environment states $|k\rangle$ and $\{,\}$ denoting the
anti-commutator, (\ref{TCL}) gives for the system state
$\hat{\varrho}$ $\!=$ $\!\mathrm{Tr}_{\mathrm{E}}\hat{\rho}$ a Lindblad equation
\begin{equation}
\label{TCLpe}
  \frac{\partial}{\partial{t}}\hat{\varrho}=-\mathrm{i}[\hat{c}_0,\hat{\varrho}]
  +2t\sum_{k\neq0}\mathcal{L}(\hat{c}_k)\hat{\varrho}.
\end{equation}
In the limit $t\to0$, (\ref{TCLpe}) gives the unitary dynamics
$\dot{\hat{\varrho}}$ $\!=$ $\!-\mathrm{i}[\hat{c}_0,\hat{\varrho}]$ 
determined by the Hermitian $\hat{c}_0$.
In the special case $\hat{c}_0$ $\!=$ $\!0$, we obtain
\begin{equation}
\label{TCLpe0}
  \frac{\partial}{\partial{t}}\hat{\varrho}
  =2t\sum_k\mathcal{L}(\hat{c}_k)\hat{\varrho}.
\end{equation}

Directly expanding the unitary state evolution of the total system
\begin{eqnarray}
  \hat{\rho}&=&\mathrm{e}^{-\mathrm{i}t\hat{H}_{\mathrm{I}}}
  \,\hat{\rho}_0\,\mathrm{e}^{\mathrm{i}t\hat{H}_{\mathrm{I}}}
  \nonumber\\
  &=&\hat{\rho}_0-\mathrm{i}t[\hat{H}_{\mathrm{I}},\hat{\rho}_0]
  +t^2\mathcal{L}(\hat{H}_{\mathrm{I}})\hat{\rho}_0+\mathcal{O}(t^3),
\label{totexpansion}
\end{eqnarray}
assuming a pure initial environment state $|0\rangle$, so that
$\hat{\rho}_0$ $\!=$ $\!\hat{\varrho}_0\otimes|0\rangle\langle0|$, 
truncating $\mathcal{O}(t^3)$, and performing the trace over the environment
approximates the system state as
\begin{eqnarray}
  \hat{\varrho}&=&\hat{\varrho}_0-\mathrm{i}t[\hat{c}_0,\hat{\varrho}_0]
  +t^2\sum_k\mathcal{L}(\hat{c}_k)\hat{\varrho}_0,
  \\
\label{uc1}
  \frac{\partial}{\partial{t}}\hat{\varrho}&=&
  -\mathrm{i}[\hat{c}_0,\hat{\varrho}_0]
  +2t\sum_k\mathcal{L}(\hat{c}_k)\hat{\varrho}_0,
\end{eqnarray}
where, in contrast to (\ref{TCLpe}), the sums here include $k$ $\!=$ $\!0$.
Assuming that after sufficiently short time intervals, the environment state is
``reset'' to the initial $|0\rangle$, we may replace in (\ref{uc1})
$\hat{\varrho}_0$ with $\hat{\varrho}$. This gives the same limit
$-\mathrm{i}[\hat{c}_0,\hat{\varrho}]$ as (\ref{TCLpe}), and it reduces to
(\ref{TCLpe0}) for $\hat{c}_0$ $\!=$ $\!0$, but in general, the additional
$k$ $\!=$ $\!0$ second order terms may pretend an incoherent evolution due to
the truncation error. This can be seen by choosing
$\hat{H}_{\mathrm{I}}$ $\!=$ $\!\hat{H}_{\mathrm{S}}\otimes|0\rangle\langle0|$
with some system Hamiltonian $\hat{H}_{\mathrm{S}}$ and comparing
(\ref{uc1}) with (\ref{TCLpe}).
\subsection{Trace-reducing equation}
An additional projection $\mathcal{R}$ onto a system subspace denoted by
$(\bm{\cdot})_{\mathrm{R}}$ $\!=$ $\!\mathcal{R}(\bm{\cdot})$ $\!=$
$\!\hat{R}(\bm{\cdot})\hat{R}$
with $\hat{R}$ $\!=$ $\!\hat{R}^2$ $\!=$ $\!\hat{R}^\dagger$, hence
$\mathcal{P}\hat{\rho}$ $\!=$ $\!\hat{R}[\mathrm{Tr}_{\mathrm{E}}
(\hat{\rho})\otimes\hat{\rho}_{\mathrm{E}}]\hat{R}$ $\!=$
$\!\hat{\varrho}_{\mathrm{R}}\otimes\hat{\rho}_{\mathrm{E}}$, gives a master
equation for which (\ref{leme}) is linear but does not preserve the trace
\cite{WKB09}. For the case considered in App.~\ref{secPES}, this yields
\begin{eqnarray}
  \frac{\partial}{\partial{t}}\hat{\varrho}_{\mathrm{R}}&=&
  -\mathrm{i}[\hat{c}_{0\mathrm{R}},\hat{\varrho}_{\mathrm{R}}]
  \nonumber\\
  &+&2t\sum_{k}\bigl(\hat{c}_{k\mathrm{R}}\hat{\varrho}_{\mathrm{R}}
  \hat{c}_{k\mathrm{R}}^\dagger
  -\frac{1}{2}\{(\hat{c}_k^\dagger\hat{c}_k)_{\mathrm{R}},
  \hat{\varrho}_{\mathrm{R}}\}\bigr)
  \nonumber\\
  &-&2t\bigl(\hat{c}_{0\mathrm{R}}\hat{\varrho}_{\mathrm{R}}
  \hat{c}_{0\mathrm{R}}^\dagger
  -\frac{1}{2}\{\hat{c}_{0\mathrm{R}}^\dagger\hat{c}_{0\mathrm{R}},
  \hat{\varrho}_{\mathrm{R}}\}\bigr).
\label{TCLpeR}
\end{eqnarray}
The first, second and third line in (\ref{TCLpeR}) correspond to the first,
second and third term in (\ref{leme}). For $\hat{R}$ $\!=$ $\!1$,
(\ref{TCLpeR}) reduces to (\ref{TCLpe}).
In the special case $\hat{c}_0$ $\!=$ $\!0$, we now obtain instead of
(\ref{TCLpe0}) 
\begin{equation}
\label{TCLpe0R}
  \frac{\partial}{\partial{t}}\hat{\varrho}_{\mathrm{R}}
  =2t\sum_{k}\bigl(\hat{c}_{k\mathrm{R}}\hat{\varrho}_{\mathrm{R}}
  \hat{c}_{k\mathrm{R}}^\dagger
  -\frac{1}{2}\{(\hat{c}_k^\dagger\hat{c}_k)_{\mathrm{R}},
  \hat{\varrho}_{\mathrm{R}}\}\bigr).
\end{equation}
If we express (\ref{TCLpe0}) in the form of  
$\frac{\partial}{\partial{t}}\hat{\varrho}$ $\!=$ $\!\mathcal{L}\hat{\varrho}$,
then (\ref{TCLpe0R}) can be written as
$\frac{\partial}{\partial{t}}\hat{\varrho}_{\mathrm{R}}$ $\!=$
$\!\mathcal{L}_{\mathrm{R}}\hat{\varrho}_{\mathrm{R}}$, where
$\mathcal{L}_{\mathrm{R}}$ $\!=$ $\!\mathcal{R}\mathcal{L}\mathcal{R}$.
\section{
\label{secJO}
Jump operators}
We here briefly resume the notation used within this work.
Let $\hat{Q}_j$ $\!=$ $\!|j\rangle\langle{j}|$ and
$\hat{Q}_{\mathrm{P}}$ $\!=$ $\!|\mathrm{P}\rangle\langle\mathrm{P}|$
be the projectors onto the RP states
$j$ $\!=$ $\!\{\mathrm{S},\mathrm{T}_0,\mathrm{T}_+,\mathrm{T}_-\}$, and the
orthogonal reaction product state P of the chemical system,
\begin{equation}
\label{STPdec}
  \hat{Q}_{\mathrm{R}}+\hat{Q}_{\mathrm{P}}=1,
  \quad
  \hat{Q}_{\mathrm{R}}=\hat{Q}_{\mathrm{S}}+\hat{Q}_{\mathrm{T}},
  \quad
  \hat{Q}_{\mathrm{T}}=\sum_{j=\mathrm{T}_i}\hat{Q}_j,
\end{equation}
where the S and T subspaces have been combined to the RP subspace R.

The jump operators
\begin{equation}
\label{Lj}
  \hat{L}_j=|P\rangle\langle{j}|
\end{equation}
describe transitions from the
$j$ $\!=$ $\!\{\mathrm{S},\mathrm{T}_0,\mathrm{T}_+,\mathrm{T}_-\}$ states
to the product state P. We thus have $\hat{L}_j\hat{L}_{j^\prime}$ $\!=$
$\!\hat{L}_j^\dagger\hat{L}_{j^\prime}^\dagger$ $\!=$ $\!0$,
$\hat{L}_j^\dagger\hat{L}_j^{}$ $\!=$ $\!\hat{Q}_j$,
$\hat{L}_j\hat{L}_{j^\prime}^\dagger$ $\!=$
$\!\delta_{jj^\prime}\hat{Q}_{\mathrm{P}}$,
from which we can deduce the relations
$\hat{Q}_j\hat{Q}_{j^\prime}$ $\!=$ $\!\delta_{jj^\prime}\hat{Q}_j$,
$\hat{L}_j\hat{Q}_{j^\prime}$ $\!=$ $\!\delta_{jj^\prime}\hat{L}_j$,
$\hat{Q}_{j^\prime}\hat{L}_j$ $\!=$ $\!\hat{L}_j\hat{Q}_{\mathrm{P}}$ $\!=$
$\!0$, and $\hat{Q}_{\mathrm{P}}\hat{L}_j$ $\!=$ $\!\hat{L}_j$.
If we use the shorthand notation
\begin{equation}
\label{projs}
  \mathcal{Q}\hat{\varrho}\equiv\hat{Q}\hat{\varrho}\hat{Q},
  \quad
  \mathcal{Q}_{jk}\hat{\varrho}\equiv\hat{Q}_j\hat{\varrho}\hat{Q}_k,
\end{equation}
where the $\hat{Q}$ and $\hat{Q}_j$ can be any projectors, then
for a state $\hat{\varrho}$ that obeys (\ref{inicon}), i.e., if
$\mathcal{Q}_{\mathrm{RP}}\hat{\varrho}$ $\!=$ $\!0$, we obtain
\begin{eqnarray}
\label{substjk}
  \sum_{j,k={\mathrm{S}},{\mathrm{T}_i}}^{(j\neq{k})}\mathcal{Q}_{jk}&=&
  1-\mathcal{Q}_{\mathrm{P}}-\sum_{j=\mathrm{S},\mathrm{T}_i}\mathcal{Q}_j,
  \\
\label{subst}
  \mathcal{Q}_{\mathrm{ST}}+\mathcal{Q}_{\mathrm{TS}}
  &=&1-\mathcal{Q}_{\mathrm{P}}-\sum_{j={\mathrm{S}},{\mathrm{T}}}\mathcal{Q}_j,
  \\
\label{STid}
  \mathcal{L}(\hat{Q}_{\mathrm{S}})=\mathcal{L}(\hat{Q}_{\mathrm{T}})
  &=&-\frac{1}{2}(\mathcal{Q}_{\mathrm{ST}}+\mathcal{Q}_{\mathrm{TS}}),
  \\
  \mathcal{Q}_{\mathrm{R}}+\mathcal{Q}_{\mathrm{P}}&=&1.
\end{eqnarray}
\section{
\label{secNHC}
Non-Hermitian coherent evolution}
An evolution which keeps a pure state pure, i.e., the purity of the normalized
state $\mathrm{Tr}(\hat{\varrho}^2)$ $\!=$ $\!\mathrm{Tr}(
\hat{\varrho}_{\mathrm{R}}^2)/(\mathrm{Tr}\hat{\varrho}_{\mathrm{R}})^2$
\cite{Hab76} remains unity, can be written as
$\frac{\partial}{\partial{t}}|\Psi\rangle_{\mathrm{R}}$ $\!=$
$\!\hat{A}|\Psi\rangle_{\mathrm{R}}$, which for
$\hat{\varrho}_{\mathrm{R}}$ $\!=$ $\!|\Psi\rangle_{\mathrm{R}}\langle\Psi|$
reads $\frac{\partial}{\partial{t}}\hat{\varrho}_{\mathrm{R}}$ $\!=$
$\!\hat{A}\hat{\varrho}_{\mathrm{R}}$ $\!+$
$\!\hat{\varrho}_{\mathrm{R}}\hat{A}^\dagger$ $\!=$
$\!\{\hat{A}_+,\hat{\varrho}_{\mathrm{R}}\}$ $\!+$
$\![\hat{A}_-,\hat{\varrho}_{\mathrm{R}}]$ $\!=$
$\!2(\hat{A}\hat{\varrho}_{\mathrm{R}})_+$,
where $\hat{A}_\pm$ $\!=$ $\!(\hat{A}\pm\hat{A}^\dagger)/2$,
and which has the solution $|\Psi(t)\rangle_{\mathrm{R}}$ $\!=$
$\!\mathcal{T}_+\mathrm{e}^{\int_0^t\mathrm{d}\tau\hat{A}(\tau)}
|\Psi(0)\rangle_{\mathrm{R}}$.
Equivalently, we may write in a more convenient form $\hat{H}_{\mathrm{eff}}$
$\!\equiv$ $\!\mathrm{i}\hat{A}$ $\!=$ $\!\hat{H}$ $\!+$
$\!\mathrm{i}\hat{Q}$, where $\hat{H}$ $\!=$
$\!(\hat{H}_{\mathrm{eff}})_+$ and $\hat{Q}$ $\!=$
$\!-\mathrm{i}(\hat{H}_{\mathrm{eff}})_-$ are Hermitian. This gives
\begin{eqnarray}
  \frac{\partial}{\partial{t}}|\Psi\rangle_{\mathrm{R}}
  &=&-\mathrm{i}\hat{H}_{\mathrm{eff}}|\Psi\rangle_{\mathrm{R}},
  \\
\label{ppe}
  \frac{\partial}{\partial{t}}\hat{\varrho}_{\mathrm{R}}
  &=&-\mathrm{i}[\hat{H},\hat{\varrho}_{\mathrm{R}}]
  +\{\hat{Q},\hat{\varrho}_{\mathrm{R}}\}.
\end{eqnarray}
An example in our context is the Haberkorn equation \cite{Hab76}, which follows
from identifying $\hat{H}$ with the spin Hamiltonian of the electrons
constituting the RP and setting $\hat{Q}$ $\!=$
$\!-\frac{1}{2}\sum_{j={\mathrm{S}},{\mathrm{T}}}r_j\hat{Q}_j$, cf. 
(\ref{gensolh}).
\section{
\label{sec8.3}
Comparison with one-atom-maser}
The encounter model bears close analogy to the micromaser considered in
\cite{Bri94} as described in Sec.~\ref{sec:one-atom-maser}. The cavity field of
the micromaser corresponds to the chemical system, that can have transitions
from the R- to the P-subspace, whereas the 2-level diagnosis atoms of
the micromaser correspond to the 9-level model environment that interacts with
the chemical system via the interaction (\ref{HI}). A diagnosis atom passing the
micromaser cavity corresponds to an encounter in our model, whereas each
detector passage of a diagnosis atom corresponds to a measurement of our model
environment by means of a detector. Finally, the A-B-0 outcomes of the
measurement of the diagnosis atoms in the micromaser correspond to the
S-T-0 outcomes of a singlet or triplet fluorescence measurement on our model
environment. In the remainder of this section we discuss the peculiarities
that arise, however,  if we entirely restrict to the R-subspace, that is, if we
try to associate the RP (instead of the chemical system including the reaction
products) with the cavity field of the micromaser.

The maps $\mathcal{A}_j^{(\mathrm{D})}$ and $\mathcal{A}_0^{(\mathrm{D})}$ in
(\ref{AjrhoD}) and (\ref{A0rhoD}) act on the total state of the chemical system,
which includes the reaction products. We now ask whether it is possible to
define effective transformations $\mathcal{A}_j^{\mathrm{eff}}$ and
$\mathcal{A}_0^{\mathrm{eff}}$ that again correspond to a
$j$ $\!=$ $\!{\mathrm{S}},{\mathrm{T}}$ or no fluorescence signal, but solely
act on the R-subspace. We start with identifying $\mathcal{A}_0^{\mathrm{eff}}$
$\!\equiv$ $\!\tilde{\mathcal{A}}_{0\mathrm{R}}$ $\!=$
$\!\sum_{j={\mathrm{S}},{\mathrm{T}}}(1-\tilde{\eta}_j)\mathcal{Q}_j$, since
this map determines the linear (non-trace preserving) part of the evolution
(\ref{cMEST}) -- (\ref{genmeRanME}) in the R-subspace
(where $\mathcal{Q}_{\mathrm{R}}$ acts like the identity), i.e.,
$\frac{\partial}{\partial{t}}\hat{\varrho}_{\mathrm{N,R}}$ $\!=$
$\!(\mathcal{L}_{\mathrm{betw,R}}+\mathcal{L}_{\mathrm{enc,R}})
\hat{\varrho}_{\mathrm{N,R}}$, with $\mathcal{L}_{\mathrm{enc,R}}$ $\!=$
$\!\tilde{r}(\tilde{\mathcal{A}}_{0\mathrm{R}}-1)$, and $\tilde{r}$ is defined
in (\ref{rtilde}).

We now demand that $\mathcal{A}_0^{\mathrm{eff}}$ $\!+$
$\!\sum_{j={\mathrm{S}},{\mathrm{T}}}\mathcal{A}_j^{\mathrm{eff}}$ $\!=$
$\!\mathcal{A}_{\mathrm{CPT}}^{\mathrm{eff}}$ $\!=$
$\!\sum_{j={\mathrm{S}},{\mathrm{T}}}\mathcal{Q}_j$ and try to find the
$\mathcal{A}_j^{\mathrm{eff}}$, for which we have
\begin{equation}
\label{sumeff}
  \mathcal{A}_{\mathrm{S}}^{\mathrm{eff}} 
  +\mathcal{A}_{\mathrm{T}}^{\mathrm{eff}}
  =\tilde{\eta}_{\mathrm{S}}\mathcal{Q}_{\mathrm{S}}
  +\tilde{\eta}_{\mathrm{T}}\mathcal{Q}_{\mathrm{T}}.
\end{equation}
The similarity of (\ref{cMEST}) with the corresponding equation for the
micromaser suggests to identify $\mathcal{A}_j^{\mathrm{eff}}$ $\!=$
$\!\tilde{\eta}_j\mathcal{Q}_j$ as transformations of the RP state
which correspond to an imperfect singlet or triplet projection, just as
the respective transformations $p_{\mathrm{A}}\mathcal{A}_{\mathrm{A}}$ and
$p_{\mathrm{B}}\mathcal{A}_{\mathrm{B}}$ of the radiation field in the
micromaser associated with an A or B click, produced by a detector with
efficiencies $p_{\mathrm{A}}$ and $p_{\mathrm{B}}$.
That is, we identify A and B with S and T, $\mathcal{A}$ with
$\mathcal{Q}_{\mathrm{S}}\hat{\varrho}_{\mathrm{R}}$ and $\mathcal{B}$ with
$\mathcal{Q}_{\mathrm{T}}\hat{\varrho}_{\mathrm{R}}$,
respectively, so that $\mathcal{A}_{\mathrm{CPT}}$ $\!=$
$\!\mathcal{Q}_{\mathrm{S}}$ $\!+$ $\!\mathcal{Q}_{\mathrm{T}}$ describes the
transformation of the RP if the fluorescence is ignored. $r$ in the micromaser
is identified with $\tilde{\eta}r$, whereas $p_{\mathrm{A}}$ and
$p_{\mathrm{B}}$ are identified with $\tilde{\eta}_{\mathrm{S}}$ and
$\tilde{\eta}_{\mathrm{T}}$. An interpretation of the $\tilde{\eta}_j$ as
detection efficiencies is however not possible if $\tilde{\eta}_j>1$, cf. the
comments at the end of Sec.~\ref{sec5.2}.

Similarly, the effective transformations are not the projected full space
transformations, e.g.,
$\mathcal{A}_0^{\mathrm{eff}}$ $\!\neq$ $\!\mathcal{A}_{0\mathrm{R}}$, since in
(\ref{117}) we have factored out $\tilde{\eta}$, which led to the
introduction of the $\tilde{\eta}_j$. The map $\mathcal{A}_0$ in (\ref{A0rho})
and its projection $\mathcal{A}_{\mathrm{R}}$ $\!=$
$\!(1-\tilde{\eta})\mathcal{Q}_{\mathrm{R}}$ $\!+$
$\!\tilde{\eta}\mathcal{A}_0^{\mathrm{eff}}$ can hence be interpreted as an
``ensemble partition'' into an unaffected part with weight $(1-\tilde{\eta})$
and an ``observed but not detected'' (or encountered but not recombined) part
with weight $\tilde{\eta}$ only if $\tilde{\eta}<1$. In the latter case,
$\tilde{\eta}$ can be interpreted as an efficiency allowing in turn to interpret
(\ref{rtilde}) as an effective encounter rate $\tilde{\eta}r$ in (\ref{117}).

At this point it should be said that the evolution (\ref{cMEST}) just
contains the sum (\ref{sumeff}), and a decomposition of this sum defining
individual $\mathcal{A}_j^{\mathrm{eff}}$ is arbitrary.
A different possible decomposition of the sum (\ref{sumeff}) that avoids this
misinterpretation for $\tilde{\eta}_j>1$, but reduces to the 
$\mathcal{A}_j^{\mathrm{eff}}$ suggested above if $\tilde{\eta}_j\le1$ is
\begin{eqnarray}
\label{chosendec}
  &&\hspace{-0.6cm}\tilde{\eta}_j=\nu_j+\mu_j,
  \\
  &&\hspace{-0.6cm}\nu_j\!=\!\min(\tilde{\eta}_j,1),
  \quad\mu_j\!=\!\max(\tilde{\eta}_j,1)\!-\!1,
\end{eqnarray}
so that $\nu_j$, $\mu_j$ $\in$ $\![0,1]$ may now be interpreted as efficiencies,
and identify
\begin{eqnarray}
  \mathcal{A}_{\mathrm{S}}^{\mathrm{eff}}&=&
  \nu_{\mathrm{S}}\mathcal{Q}_{\mathrm{S}}+
  \mu_{\mathrm{T}}\mathcal{Q}_{\mathrm{T}},
  \\
  \mathcal{A}_{\mathrm{T}}^{\mathrm{eff}}&=&
  \nu_{\mathrm{T}}\mathcal{Q}_{\mathrm{T}}+
  \mu_{\mathrm{S}}\mathcal{Q}_{\mathrm{S}}.
\end{eqnarray}
This shows that a perfect fluorescence detection ($\hat{\Pi}_{\mathrm{S}}$ $\!=$
$\!|\pi_{\mathrm{S}}\rangle\langle\pi_{\mathrm{S}}|$, $\hat{\Pi}_{\mathrm{T}}$
$\!=$ $\!\sum_{j={\mathrm{T}_i}}|\pi_j\rangle\langle\pi_j|$, $\hat{\Pi}_0$ $\!=$
$\!1-\sum_{j={\mathrm{S}},{\mathrm{T}}}\hat{\Pi}_j$) appears on
the R-subspace indeed as an imperfect detection, but in general as an
``unbalanced dephasing'' rather than a projection multiplied with an efficiency.
This effective detection corresponds to a POVM
\begin{eqnarray}
  \hat{\Pi}_{\mathrm{S}}^{\mathrm{eff}}&=&
  \nu_{\mathrm{S}}\hat{Q}_{\mathrm{S}}+
  \mu_{\mathrm{T}}\hat{Q}_{\mathrm{T}},
  \\
  \hat{\Pi}_{\mathrm{T}}^{\mathrm{eff}}&=&
  \nu_{\mathrm{T}}\hat{Q}_{\mathrm{T}}+
  \mu_{\mathrm{S}}\hat{Q}_{\mathrm{S}},
  \\
  \hat{\Pi}_0^{\mathrm{eff}}&=&
  \sum_{j={\mathrm{S}},{\mathrm{T}}}(1-\tilde{\eta}_j)\hat{Q}_j.
\end{eqnarray}
 (\emph{Note}:
The effective POVM and the corresponding effective transformations disregard the
recombination similar to the description of photodetection in terms of photon
number state projectors, where the detection leads to a photon absorption
by the detector rather than preparing the radiation field in a photon number
state.)

The analysis also holds for (\ref{genmeRan}), which is based on the master
equation approach and corresponds to the weak encounter limit of the encounter
model. There, the $\eta_j$ define the ratio of the decay rates $r_j$ with
respect to the sum of the \emph{average} decay and dephasing rates $r$ and $d$.
Since $\sum\eta_j$ $\!\le$ $\!2$, we have $0$ $\!\le$ $\!\eta_j$ $\!\le$ $\!2$
in general. In particular, if $r_j$ $\!=$ $\!r$, then
$\eta_j$ $\!=$ $\!r/(r+d)$ $\!\le$ $\!1$.
Similarly, if $r_j$ $\!\le$ $\!d_j$, then
$\eta_j$ $\!\le$ $\!{r}_j/\sum{r}_j$ $\!\le$ $\!1$. Such cases with
$\eta_j$ $\!\le$ $\!1$ suggest to interpret the $\eta_j$ as efficiencies of a
singlet or triplet state detection. In particular, if $r_j$ $\!\le$ $\!d_j$,
we may think of a recombination as a two-step process where the recombination
itself must be proceeded by a reaction, that is, a dephasing (caused e.g., by
an encounter of the RP diffusing in a solvent). This includes pure
dephasing as limiting case for which $\eta_j$ $\!=$ $\!0$ and the Jones-Hore
case, where $\eta_j$ $\!=$ $\!{r}_j/\sum{r}_j$, but not the Haberkorn situation,
for which $\sum\eta_j$ $\!=$ $\!2$. Only if additionally $r_j$ $\!=$ $\!r$, the
Haberkorn case describes perfect detection $\eta_j$ $\!=$ $\!1$. In cases, which
allow $\eta_j>1$, such interpretation of $\eta_j$ as detection efficiencies is
not possible. 
\end{document}